\documentclass[12pt]{iopart}
\usepackage{times}
\usepackage{geometry}
\usepackage{titlesec}
\usepackage{setspace} % For adjusting line spacing
\usepackage{ragged2e}
\usepackage{hyperref}
\usepackage{graphicx}
\usepackage{caption}
\usepackage{hhline}
\usepackage{xcolor}
\usepackage{ifthen}
\usepackage{url}
\usepackage{footmisc}
\usepackage{xcolor}
\usepackage{soul}
\usepackage{subfigure}
\usepackage{etoolbox}
\usepackage{adjustbox}
\usepackage[style=numeric, sorting=none]{biblatex}
\usepackage{textcase}
\usepackage{tabularx} % For better control over table column widths
\usepackage{comment}
\usepackage{booktabs}
\usepackage{mathptmx}
\usepackage{wrapfig}

\begin{document}

\title[N.Carey et al., PPCF/NF-XX (2025)]{Neural operator surrogate models of plasma edge simulations: feasibility and data efficiency}

\author{N. Carey$\textsuperscript{1}$, L. Zanisi$\textsuperscript{1}$, S. Pamela$\textsuperscript{1}$, V. Gopakumar$\textsuperscript{1,4}$, J. Omotani$\textsuperscript{1}$, J. Buchanan$\textsuperscript{1}$, J. Brandstetter$\textsuperscript{2,3}$, Fabian Paischer$\textsuperscript{2}$, Gianluca Galletti$\textsuperscript{2}$, Paul Setinek$\textsuperscript{2}$}

\address{UKAEA, Culham Centre for Fusion Energy, Abingdon, UK$\textsuperscript{1}$}
\address{ELLIS Unit, Linz, LIT AI Lab, Institute for Machine Learning, Johannes Kepler University, Linz, Austria$\textsuperscript{2}$}
\address{NXAI GmbH, Austria$\textsuperscript{3}$}
\address{Centre for Artificial Intelligence, UCL, London, UK$\textsuperscript{4}$}
\ead{Naomi.Carey@ukaea.uk}
\vspace{10pt}
\begin{indented}
\item December 2024
\end{indented}

\begin{abstract}

Simulation-based plasma scenario development plays a crucial role in designing next-generation tokamaks and fusion power plants. However, the inclusion of high-fidelity simulations of Scrape-Off Layer (SOL) turbulence and transient MHD events such as Edge Localized Modes (ELMs) in highly iterative applications remains computationally prohibitive, limiting their use in design and control workflows. Understanding these phenomena is vital, as they govern heat flux on plasma-facing components, influencing reactor performance and material lifetime. This study explored Fourier Neural Operators (FNOs) as surrogate models to accelerate plasma simulations from the JOREK MHD and STORM turbulence codes.

 FNOs were trained on single-step rollouts and evaluated in terms of long-term predictive accuracy in an auto-regressive manner. To mitigate the computational burden of dataset generation, a transfer learning strategy was explored, leveraging low-fidelity simulations to improve performance on high-fidelity datasets. These results showed that FNOs effectively captured initial plasma evolution, including blob movement and density source localization for JOREK and STORM, respectively. However, long rollouts accumulated errors and exhibited sensitivity to certain physical phenomena, leading to non-monotonic error spikes. Transfer learning significantly reduced errors for small dataset sizes and short rollouts, achieving an order-of-magnitude reduction when transfering from low- to high-fidelity datasets. However, its effectiveness diminished with longer rollouts and larger dataset sizes, especially when applied to datasets with significantly different dynamics. Attempts to transfer models to previously unseen variables in simulations were unsuccessful, underscoring the limitations of transfer learning in this context.

 These findings demonstrate the promise of neural operators for accelerating fusion-relevant PDE simulations. However, they also highlight key challenges: improving long-term accuracy to mitigate error accumulation, capturing critical physical behaviors, and developing robust surrogates that effectively leverage multi-fidelity, multi-physics datasets.

\end{abstract}

%
% Uncomment for keywords
%\vspace{2pc}
%\noindent{\it Keywords}: XXXXXX, YYYYYYYY, ZZZZZZZZZ
%
% Uncomment for Submitted to journal title message
%\submitto{\JPA}
%
% Uncomment if a separate title page is required
%\maketitle
% 
% For two-column output uncomment the next line and choose [10pt] rather than [12pt] in the \documentclass declaration
%\ioptwocol
%

\section{Introduction}

\subsection{Motivation} \label{sec:motivation}

Modelling of the edge plasma is fundamental to ensure appropriate divertor and core performance, and their reliable characterisation is desirable for not just designing and interpreting the current generation of experiments, but to inform the next generation devices like ITER and STEP \cite{Loarte2007Chapter4P,Eich2013ScalingOT}. However, modelling plasma is computationally intensive, requiring the solution of strongly coupled partial differential equations (PDEs), which are often analytically intractable. Numerical approximation schemes provide a practical approach, balancing accuracy and computational efficiency, as implemented in frameworks such as BOUT++ \cite{Riva2019-fh}, JOREK \cite{Hoelzl_2021}, JINTRAC \cite{romanelli2014jintrac}, and SOLPS \cite{WIESEN2015480}.

However, even for reduced-order models, which are often 1 dimensional (e.g., DIV1D \cite{Derks2022}), the computational cost is still high or even prohibitive for iterative applications, such as optimisation, or real-time control. The MHD code JOREK, which simulates transient MHD events such as Edge Localized Modes (ELMs), is a prominent example of this issue. While ELM simulations are crucial for designing tokamaks and their control systems \cite{Huijsmans2015}, accurately simulating ELM behavior in DEMO-class devices is slow and takes up large computational resources.

Beyond transient MHD events, turbulence in the Scrape-Off Layer (SOL) drives fine-scale, continuous heat, momentum and particle transport. The SOL dynamics is known to be dominated by filaments, i.e. coherent structures which play a crucial role in transport plasma particles over large distances from the separatrix. These filaments contribute to shaping the time-averaged density profile in the SOL and simulating them poses a major computational challenge \cite{10.1063/1.4940330}. The contrast in time scales between parallel and perpendicular transport imposes strict time-step limitations, making it difficult to balance accuracy and efficiency. Consequently, current solvers cannot yet routinely simulate turbulence at the scale of DEMO-class reactors, focusing instead on smaller experimental devices like TCV or ASDEX Upgrade \cite{STEGMEIR2023108801}.

Additionally, numerical solvers often have to be specifically tailored to the equations and spatial geometry they address \cite{ODEbook}, are challenging to deploy quickly due to their complex software infrastructure and library dependencies, and these solvers may fail to converge when integrated into larger simulation frameworks, often due to model mismatches or the need for expert manual intervention. 

Neural network (NN) based surrogate models offer a promising avenue to enable the fast, inexpensive estimation of plasma states given the initial and boundary conditions for a physical model of choice. For example, Convolutional Neural Networks (CNNs) have been applied to emulate the behaviour of SOLPS \cite{WIESEN2015480,Gopakumar2020,DASBACH2023101396} over a restricted area of its parameter space, demonstrating orders of magnitude speedup. Nevertheless, solutions derived from CNNs lack discretization invariance (both spatially and temporally), and therefore are applicable only to the specific discretization of the numerical PDE solution that they were trained on. Recent work has introduced Neural Operators (NOs) \cite{kovachki2023neural}, a family of surrogate models that learn infinite-dimensional mappings between function spaces, and are thus capable of learning a continuous, discretization-invariant representation of PDE solutions \cite{li2021fourier, alkin2024universalphysicstransformersframework}. The Fourier Neural Operator (FNO), where convolutional filters are learned in Fourier space, has proved a particularly successful architecture for modelling PDEs with NNs \cite{li2021fourier}. Although more recent neural operator architectures have demonstrated superior performance in certain tasks, they often come with increased computational overhead and implementation complexity. Given the relatively moderate complexity of the datasets used here and the exploratory nature of this study, FNO offers a practical balance between performance, simplicity, and availability. It has been successfully applied to a range of PDE problems, including in simplified fusion-relevant cases \cite{Gopakumar2024,Poels_2023} making it a strong candidate for further investigations in more complex use cases, such as SOL turbulence and reduced-fidelity MHD ELM simulations.

Future work will extend this approach by incorporating and benchmarking state-of-the-art models.

\subsection{Similar works}

Machine learning-based surrogate models, including neural operators (NOs), are being applied in plasma physics to provide rapid approximations for various problems. In fusion applications, these include emulating turbulent transport \cite{van_de_Plassche_2020, Ho_2021}, estimating the Tritium breeding ratio \cite{M_nek_2023}, and modelling edge plasma behavior \cite{Gopakumar_2024, DASBACH2023101396}.

Recent approaches to neural PDE surrogates can be split into thee main groups \cite{lippe2023pderefiner}: (1) approaches that directly learn to approximate the PDE \cite{raissi2017physicsinformeddeeplearning} (2) data driven approaches that map the current state of a system to its future state using a learned evolution operator \cite{gopakumar2023plasma, brandstetter2023message} and (3) hybrid approaches that combine neural networks with traditional numerical solvers \cite{pamela2024neuralparareal, StylES}. For time-dependent PDEs, many approaches rely on autoregressive rollouts, where outputs from prior steps were fed back as inputs to compute solutions over long time horizons. Since directly predicting long-term evolution is highly challenging, autoregressive modeling allows the system to approximate future states incrementally. However, a key drawback of autoregressive rollouts is the accumulation of errors at each prediction step, which can cause the predicted trajectories to diverge from the ground truth over time \cite{lippe2023pderefiner, brandstetter2023message, mikhaeil2022difficultylearningchaoticdynamics}. Traditional numerical methods mitigate such instabilities using implicit schemes or explicit methods such as adaptive step sizes to improve accuracy and stability \cite{ODEbook, Iserles_2008}. In contrast, neural operators deployed autoregressively operate in a Markovian fashion, lacking a forcing function to correct errors. Consequently, evaluating their performance over longer time horizons is crucial for assessing their reliability in fusion applications.

Despite these advances, the application of NOs to fusion-related time-dependent problems remains limited as the use of machine learning, particularly neural operators, in fusion research is still a relatively new and emerging area. A first exploration of FNOs for emulation of MHD was presented in \cite{Gopakumar_2024}, where it was shown that the FNO outperforms non discretization-invariant, CNN-based architectures. In the same study, the FNO was shown to be capable of accurately predicting the short-term evolution of plasma in the Mega Ampere Spherical Tokamak visible camera images. In another study, \cite{Poels_2023} benchmarked a range of NOs for reproducing DIV1D simulations, with FNOs serving as a competitive baseline.

Building on these advances, previous work by the authors introduced the Neural-Parareal framework \cite{pamela2024neuralparareal}, which focuses on integrating NOs as coarse solvers within the Parareal method - a time-parallelization technique for High-Performance Computing (HPC) simulations. This approach focused on dynamically training NOs using data generated by the framework to improve surrogate model accuracy and speed-up. In contrast, this work examines the performance of NOs over longer time periods independently.

Additionally, creating simulation data to train surrogate models is often prohibitively expensive, requiring computationally-expensive iterative solvers such as finite-volume schemes and finite-element methods to be run on supercomputers. Strategies that focus surrogate model training (and thus generation of simulation data) on the most informative regions of parameter space have been proposed in \cite{Zanisi2024,Hornsby2024, RodriguezFernandez2024}, showing significant data efficiency gains.

In contrast, this work focuses on transfer learning, which leverages pre-trained neural networks to improve the performance of surrogate models, offering a different approach to enhancing data efficiency (e.g., \cite{Weiss2016ASO}) as demonstrated in natural language processing and computer vision tasks \cite{houlsby2019parameterefficienttransferlearningnlp, raffel2023exploringlimitstransferlearning, saito2018maximumclassifierdiscrepancyunsupervised, Weiss2016ASO, NIPS2014_375c7134}. By leveraging knowledge acquired from a source domain, transfer learning can improve learning efficiency and performance in target domains where data is scarce. The hierarchical nature of physics models, where different tradeoffs between computation speed and physics accuracy are available, is therefore particularly amenable to transfer learning. Traditionally, multi-fidelity modelling leverages many fast low-fidelity simulations to explore the data manifold and only a few high-fidelity simulations to refine the information gathered from the low-fidelity runs. In this work, multifidelity modelling is achieved by training surrogate models on a source domain with low physics fidelity and where data is abundant and inexpensive are fine tuned to a target domain with higher physics fidelity where simulation data is costly. Similarly, transfer learning between physically adjacent but distinct domains, such as SOL turbulence and ELM simulations, is also explored. Although these regimes differ in detail and governing equations, they often share fundamental structures that transfer learning may exploit to improve data efficiency and model generalization. Prior work has explored transfer learning in neural operator surrogates, demonstrating its potential for data efficiency improvements in simulation-based tasks \cite{subramanian2023foundationmodelsscientificmachine}. However, the applicability and scalability of this technique across distinct simulation frameworks and fidelity levels remains an open question, as does its capability to handle long-term trajectory predictions.

\subsection{Overview of this work}

This work presents Fourier Neural Operator (FNO)-based surrogate models, implemented using the PDEArena library \cite{gupta2022multispatiotemporalscale}, for two widely used plasma simulation codes: STORM \cite{Walkden2016-ys} and JOREK \cite{Hoelzl_2021}. STORM, built on the BOUT++ framework, is a fluid code focused on modelling turbulence and transport processes in the SOL region. JOREK, on the other hand, is a widely adopted code for studying large-scale
MHD instabilities in both core and edge regions of tokamaks \cite{Smith2020-cv}. As discussed in \ref{sec:motivation}, this paper focuses on the FNO due to its demonstrated efficiency on medium-scale PDE problems like the datasets utilized in the present work \cite{Azizzadenesheli2024, dehoop2022costaccuracytradeoffoperatorlearning}.

The investigation focused on the following objectives:
\begin{itemize}
    \item \textbf{Assessing long-term prediction accuracy in autoregressive rollouts for neural operator surrogate models.} Errors in autoregressive rollouts compound over time, leading to instabilities. This study also examined how trajectory-specific dynamics contribute to error accumulation. Particular attention was given to transitions or significant events within the trajectory, as these could exacerbate error growth.
    \item \textbf{Exploring the potential of transfer learning in neural operators.} This study investigated transfer learning as a strategy to enhance data efficiency by leveraging knowledge from related datasets. Models trained from scratch were compared with those fine-tuned using transfer learning, and their performance was evaluated in terms of how it scaled with dataset size and rollout length. The objective was to assess the feasibility of using transfer learning across diverse scenarios relevant to fusion modelling, providing insights into its strengths and limitations. Specifically three scenarios were explored:
    \begin{enumerate}
        \item Cross-Simulation-Code Transfer (between STORM and JOREK): Transfer learning was tested between two independently developed simulation codes with distinct applications and physics models. This scenario evaluated the adaptability of neural operator surrogates to handle entirely different simulation frameworks.
        \item Cross-Fidelity Transfer (within JOREK): The transfer of knowledge was explored between two datasets of differing physics fidelity, specifically from low-fidelity-physics to high-fidelity-physics simulations from JOREK. This approach evaluated the potential of transfer learning to reduce the computational burden of high-fidelity-physics modelling by using low-cost, lower-fidelity-physics simulations as a foundation.
        \item Cross-Variable Transfer (within JOREK): A key challenge encountered when transferring from low- to high-fidelity simulations was that the target dataset contained variables absent in the source dataset. Many deep learning architectures such as FNO lack the flexibility to seamlessly incorporate prior knowledge from learned variables into datasets with new variables. To address this, experiments with a two-step transfer learning approach were carried out: first, transferring knowledge from the low-fidelity dataset to the common variables in the high-fidelity dataset, followed by a second transfer step to extend the learned representations to the variables unique to the target dataset.
    \end{enumerate}
\end{itemize}

\section{Problem Setting}

\subsection{Dataset} \label{dataset}

In this work, training datasets created from STORM \cite{Walkden2016-ys} and JOREK \cite{Hoelzl_2021} simulations were used which have already been adopted in several studies \cite{pamela2024neuralparareal, gopakumar2023fourierneuraloperatorplasma, carey2024dataefficiencylongterm}. JOREK is designed to simulate large-scale MHD-instabilities in the core and edge plasma region, whereas STORM is a fluid code focused on modelling turbulance and transport processes in the SOL at much smaller scales. All simulations in these datasets were run in simplified 2D rectangular slab geometry (with toroidal curvature). The long term goal will be to expand to 3D cases with more complex simulations in full toroidal geometry, which is left for future work. Dimensions of the dataset are be found in table \ref{data_sourcetable}.

\begin{table}[h]
\centering 
\resizebox{\textwidth}{!}{
\begin{tabular}{@{}lllll@{}}
\toprule
Dataset name        & Dataset size              & Trajectory length (num timesteps)                 & Num variables & Dimensions \\ \midrule
STORM               & 1000 trajectories         & 1000 per traj                 & 2             & 384x256    \\
Electrostatic JOREK & 2000 trajectories         & 200 per traj                  & 2             & 100x100    \\
Reduced-MHD JOREK   & 11391 slices and 20 trajs & 10 per slice and 200 per traj & 4             & 100x100    \\ \bottomrule
\end{tabular}}
\caption{Table describing each dataset's size, variables and spatial dimensions.}
\label{data_sourcetable}
\end{table}

\subsubsection{JOREK} \label{sec:JOREK_dataset}is a continuously developed simulation code that is widely used to study large-scale MHD instabilities from the core and edge plasma regions \cite{Smith2020-cv}. The JOREK dataset specifically focuses on simulating filamentary blobs in the edge region of a tokamak. Throughout the simulation, the toroidal curvature combined with the pressure gradient of the blobs generates an electric field, resulting in the blobs moving radially outwards (away from the centre of the tokamak). These blobs are expelled out of the hot confined plasma region through turbulence or MHD instabilities, such as Edge-Localised-Modes (ELMs) \cite{Hoelzl_2021}. These are important to study as ELMs can eject hot plasma filaments onto the machine's first wall and damage it. The initial conditions of the blobs were generated from uniform distributions in positions (R,Z), amplitudes of density and temperature. The number of blobs were left to vary between 1 and 10 uniformly. Two datasets were generated from JOREK:
\begin{enumerate}
    \item \textbf{Electrostatic JOREK} simulations evolve three physical variables: density $\rho$, electric potential $\Phi$ and temperature $T$. One auxiliary variable, the toroidal vorticity $\omega$ was used for numerical stability. This model is equivalent to reduced-MHD JOREK dataset below but without the magnetic potential and current components in the model. This dataset is described in further detail in \cite{pamela2024neuralparareal}.
    \item \textbf{Reduced MHD JOREK} as implemented in routine JOREK studies, including a magnetic field but not the parallel velocity. These simulations evolve the four physical variables: $\rho$, $\Phi$, $T$ and poloidal magnetic flux $\Psi$. Two additional auxillary variables $\omega$ and toroidal plasma current density $zj$ were also used. These settings correspond to a higher-fidelity-physics compared to the electrostatic case, and required 2.25X more time to generate with similar hardware resources. This dataset was generated in batches for the application \cite{pamela2024neuralparareal}. The first batch of data was generated normally resulting in full trajectories (similar to electrostatic JOREK), whereas the three subsequent batches were generated as part of parareal (parallel-in-time) scheme resulting in chunked trajectories. As a consequence of the generation method, the beginning and end of each trajectory chunk does not line up exactly with each other, resulting in 20 individual sets of 10 timesteps referred to as timeslices. 
\end{enumerate}

Both JOREK dataset trajectories are evolved for 2000 timesteps (corresponding to a trajectory length of 2000) of approximately 0.15µs each on a high-order Bezier finite-element grid with uniform resolution of 200 by 200 elements. The poloidal 2D slab is centred at a toroidal major radius of 10m with a height and width of 1m. The boundary conditions around the domain are Dirichlet for all variables. High spatial and temporal resolution was chosen to ensure numerical stability during the simulations. An example of the simulations can be seen in figure \ref{fig:examples}. Additional information about generating the dataset can be found at \cite{pamela2024neuralparareal}

\subsubsection{STORM} is a fluid code built on the BOUT++ framework which is focused on modelling turbulance and transport processes in the SOL. It has been used to conduct nonlinear flux-driven simulations in double-null tokamak geometry under realistic parameters \cite{Riva2019-fh}. The physics model in the STORM simulations used here evolves two variables: the density $\rho$ and electric potential $\Phi$, with a constant background temperature level. A vertical band of density source at a fixed location and amplitude generates fluid turbulence in the radial direction due to the toroidal geometry. The dataset is obtained by varying the amplitude and width of the source which is fixed throughout the timesteps, resulting in the density increasing linearly with time. The simulations were all run in 2D regular slab geometry (384x256) with the dimensions being 150 and 100mm, located around the separatrix, allowing for analysis of much smaller scale turbulence. For STORM, a total of 1000 timesteps were run which evolve density and electric potential.

\begin{figure}
    \centering
    \subfigure[Elecrostatic JOREK]{
        \includegraphics[width=0.48\textwidth]{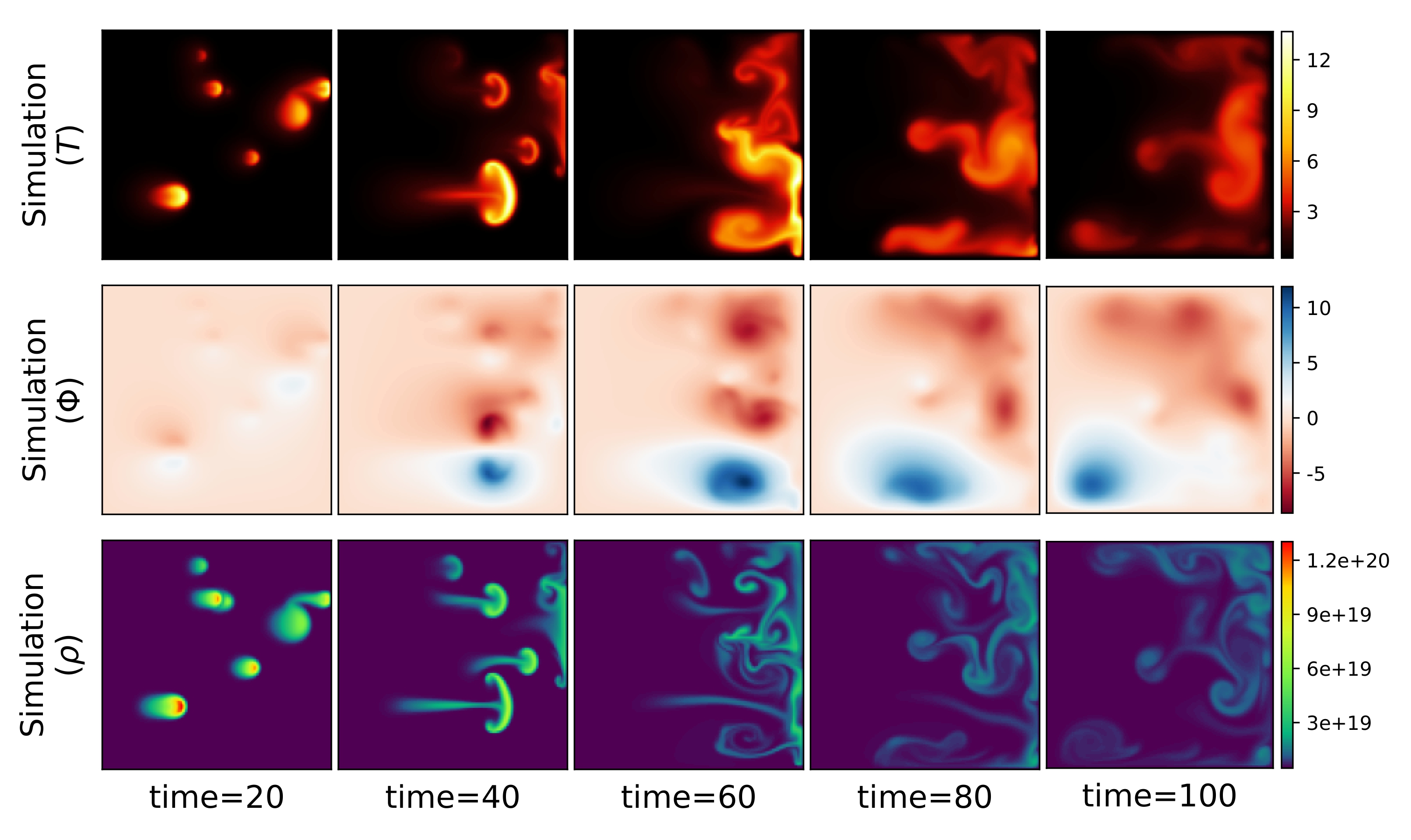}
    }
    \subfigure[Reduced-MHD JOREK]{
        \includegraphics[width=0.47\textwidth]{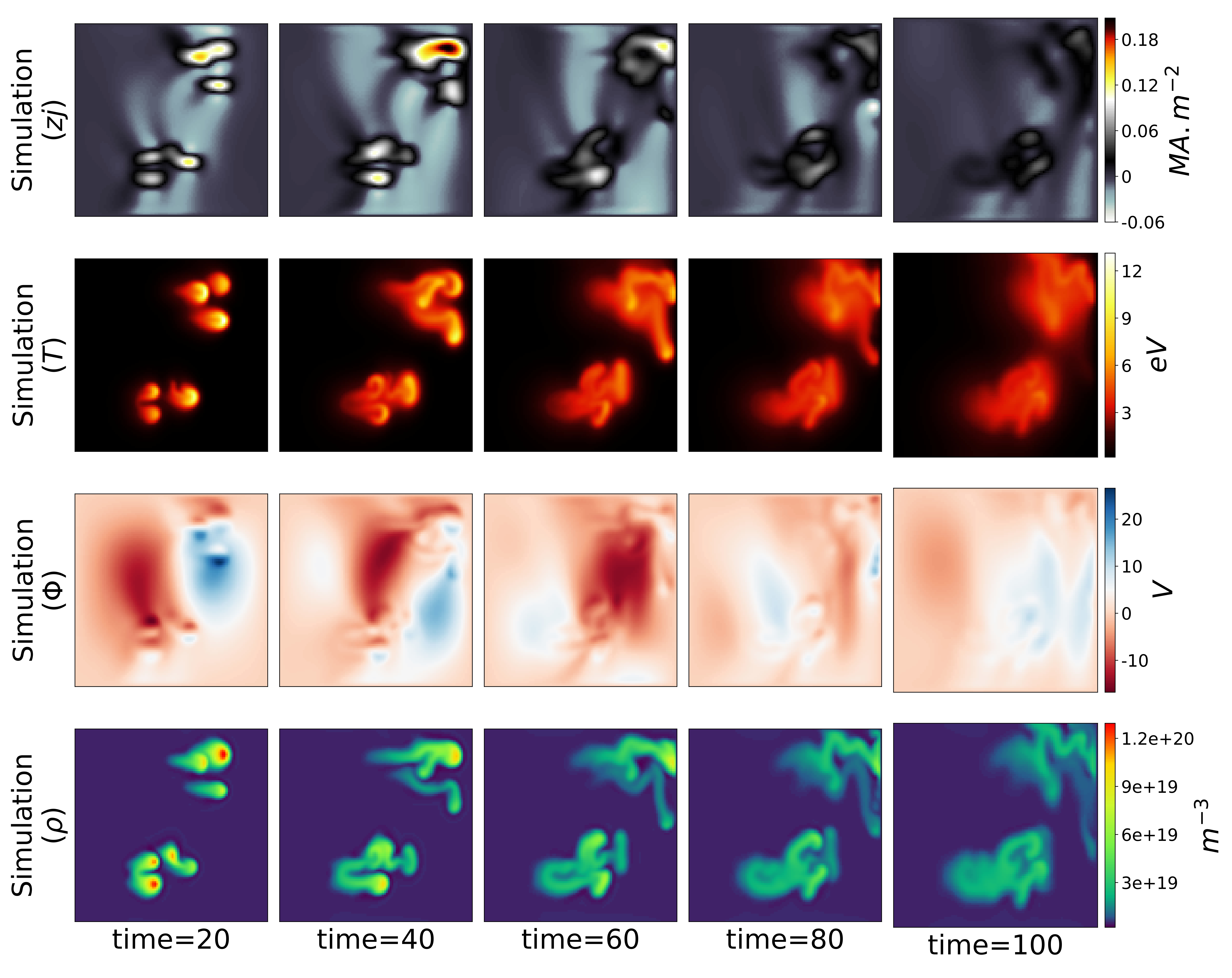}
    }    
    \subfigure[STORM]{
        \includegraphics[width=0.8\textwidth]{figures/2_1_2_example_traj_STORM.jpg}
    }
    \caption{Example rollout for each of the 3 datasets: (a) electrostatic JOREK dataset, (b) reduced-MHD JOREK dataset and (c) STORM dataset.}
    \label{fig:examples}
\end{figure}

The neural operator (NO) training focused only on the saturated turbulence phase in STORM when the root-mean-square (RMS) vorticity reached a statistically steady state and fluctuated around a constant value. For practical purposes, this was simplified by including only trajectory timesteps after $t=200$. This approach ensured that the dataset contained trajectories of consistent length and provided a buffer to account for any outliers. This was done as early experiments showed that the transition between the two regimes was extremely difficult to capture accurately due to the highly chaotic nature of turbulence onset. Unlike blobs in the JOREK datasets, which have a clear initial condition, the model found it extremely difficult to predict where the turbulence will start in the STORM cases, as it originated from the growth of very small numerical differences in the background noise. An example of the simulations can be seen in figure \ref{fig:examples}.

\subsection{Dataset pre-processing} \label{sec:data_preprocessing}

Given the high-resolution nature of the initial simulations, which were conservatively designed to ensure numerical stability, each dataset was spatially and temporally down-sampled. For example, the spatial resolution of the STORM dataset was reduced from $384\times256$ to $192\times128$, while the JOREK datasets (Electrostatic and Reduced MHD) were down-sampled from $200\times200$ to $100\times100$. Additionally, the temporal resolution of the JOREK datasets were reduced by a factor of 10, decreasing from 2000 timesteps to 200 by taking every tenth timestep. The dataset resolution that the model is trained on can be increased by reducing the downsampling, but this will result in longer training times and larger model sizes due to the increased data volume. For the current settings, training times are approximately 32, 72, and 144 hours on a single NVIDIA A100 GPU for the reduced-MHD JOREK, electrostatic JOREK, and STORM datasets, respectively, with the resulting model file size being 1.6 GB (after tracing with TorchScript). It is important to note that these specific down-sampling choices are arbitrary; a more extensive study would be required to determine the optimal down-sampling strategy. 

The datasets were partitioned into training and testing sets. From the two datasets containing only full trajectories, a roughly 90\%/10\% split into training/testing was performed; resulting in a dataset split of 799/100 trajectories for STORM and 1794/191 trajectories for electrostatic JOREK. Due to its unique generation method, as described in Section \ref{sec:JOREK_dataset}, the Reduced-MHD JOREK dataset was structured differently. Instead of full trajectories, it predominantly consisted of short partial trajectories, referred to as timeslices, along with a smaller number of complete trajectories. The 20 full trajectories were reserved for testing to allow for evaluation of longer rollouts, whilst the 12,000 shorter timeslices were used for training.

Due to each variable having widely different scales (e.g. density being a factor of $10^{20}$ whilst electric potential was around $10^1$), the data distribution within the spatial domain was severely imbalanced. To address this, min-max normalization was applied to the entire training dataset for each individual variable, ensuring that the data’s maximum and minimum values lay between -1 and 1. The same transformation was then applied to the evaluation dataset.

\section{Methods}
The PDEArena platform \cite{gupta2022multispatiotemporalscale} was utilized, as it provides a standardized code base that supports various neural PDE surrogates, allowing for streamlined implementation and ensuring that existing models can be easily adapted or new models implemented. The data loader was extended to include the datasets utilized for this work.

\subsection{Fourier Neural Operator}

Neural operators \cite{kovachki2023neural, wang2021learningsolutionoperatorparametric} are a powerful model class designed to learn mappings between continuous function spaces, making them ideal for approximating solutions to partial differential equations (PDEs) that describe continuous dynamics. Unlike traditional neural networks, which operate on finite-dimensional inputs and outputs, neural operators directly model the underlying continuous functions, offering a more natural and effective approach for PDE-driven phenomena.

In the Neural Operator (NO) setting, neural operator learns the mapping from an input function space A to the output function space U given example input and output functions $a\in A$ and $u\in U$. The NO function ($G_\theta$) can be written as a parameterized representation of this operator across function spaces:

\begin{equation}
    G_\theta: A \rightarrow U \; s.t.\; G_\theta(a) = u
\end{equation}

The basic structure of a neural operator is similar to that of a standard neural network, as visualised in figure 2 at \cite{li2021fourier} and can be written in the form:

\begin{equation}
u = G_\theta(a) = L_{lift} \circ K_1 \circ K_2 \circ ... \circ K_{T-1} \circ K_{T} \circ L_{proj}(a)
\end{equation}

where the components are:
\begin{enumerate}
    \item $L_{lift}$: a fully local, point-wise operation that projects the input domain to a higher dimensional latent representation ($a\in \textbf{R}^{A} \rightarrow v_0\in{\textbf{R}^{v_0}}$). Conventionally this is implemented in a NO as a standard linear layer.
    \item $K_t\;\forall t\in\{1,...,T\}$: the neural operator specific layer which is formulated as $K_t(v_{t-1})=\sigma(Wv_{t-1}(x)+\kappa(a;\phi)v_{t-1}(x))$ where the $\sigma$ denotes a non linear activation, $W$ is a local linear operator and $\kappa$ is a non-local integral kernel operator.
    \item $L_{proj}$: is a fully local, point-wise operation and is also conventionally implemented as a linear layer similar to $L_{lift}$. However, this operation acts in the opposite direction, projecting the latent representation back into the output domain ($v_0\in{\textbf{R}^{v_T}} \rightarrow u\in \textbf{R}^{U}$).
\end{enumerate}

For the FNO specifically, the non-linear integral kernal operator is derived as $(\kappa(a;\phi)v_t)(x)=\mathcal{F}^{-1}(\mathcal{F}(\kappa_\phi).\mathcal{F}(v_t))(x)$ where $\mathcal{F}$ is the Fourier transform. This equation is derived by representing the kernel integral operator as a computation in Fourier space and simplifying the resulting equation as explained in \cite{li2021fourier}.

In practice, the Fourier layer uses two learnable weight matrices, $R$ and $W$. The matrix $R$ operates in Fourier space, enabling convolution by parameterizing the mapping directly in the frequency domain, while $W$ performs a linear transformation in the input Euclidean space. The output of a Fourier layer is expressed as:
\begin{equation}
    y = \sigma\bigg(\mathcal{F}^{-1}\big(R\mathcal{F}(x)\big) + Wx + b\bigg)
\end{equation}
where the input $x$ is combined with a learned Fourier representation through $R$, a linear transformation through $W$, and a bias term $b$, followed by a nonlinearity $\sigma$. The addition of the linear transform W and bias term b is designed to allow the model to keep track of a non-periodic boundary \cite{li2021fourier}. However due to the addition of a linear transformation of the input Euclidean space, the grid must be fixed and the Fourier transform is commonly implemented via the Cooley–Tukey FFT algorithm. Fourier decomposition is performed across the two-dimensional spatial domain, although versions of the FNO also exist with Fourier transforms applied to both spatial and temporal dimensions \cite{li2021fourier}. A modified version of the FNO configuration ‘FNO-128-32m’ in PDEarena was used, where the number of Fourier blocks was increased to 3. A Fourier block as defined in PDEArena is equivalent to 2 Fourier layers as described above stacked on top of each other. Additionally, grid discretization was concatenated with field data along the same dimension, following the approach demonstrated in \cite{gopakumar2023plasma}.

To enforce non-negativity in physically constrained variables (such as density and temperature), we apply a modified ReLU activation at the output layer. This ensures that, after inverting the min-max normalization, the predicted values remain $\geq0$.

\subsection{U-Net mod64}

As a second baseline, we employed the U-Netmod-64 architecture from the PDEArena \cite{gupta2022multispatiotemporalscale}. U-Netmod is a modern variant of the classic U-Net, using convolutional downsampling layers instead of max-pooling and incorporating Wide ResNet-style blocks for improved feature extraction, while retaining skip connections that help preserve fine spatial details. This architecture was selected based on its strong performance in the PDEArena benchmarks and to match the parameter count of our FNO model for a fair comparison. However, the U-Net is fixed to a specific input resolution and cannot easily adapt to changes in spatial discretization. In contrast, the FNO is designed to learn mappings between continuous function spaces, making it particularly well-suited for approximating PDE solutions across varying resolutions. This resolution invariance was especially useful in some of the transfer learning experiments between datasets where the spatial resolution varied between the source and target domains. For succinctness, only the FNO results are explored in detail in the main text and the comparison between the FNO and U-Net is briefly discussed in appendix \ref{sec:model_unet}. U-Net surrogate models of STORM were not trained due to computational resource limitations. Transfer learning was not performed for the U-Net for similar reasons.

\subsection{Additional training details} \label{sec:additional_details}

\subsubsection{Further model hyperparameters}
The FNO architecture and training hyperparameters were based on the original FNO implementation \cite{li2021fourier} and the PDEArena framework defaults \cite{gupta2022multispatiotemporalscale}, with minor adjustments made to the learning rate and gradient clipping to improve training stability. No extensive hyperparameter optimization was performed, as this study focuses on demonstrating proof-of-concept performance; however, a more thorough tuning effort would be appropriate for future production use. A complete list of hyperparameters is provided in Section~\ref{sec:model_hyperparam}.

\subsubsection{Time bunding} The FNO operates on field values across a 2D grid and accepts $T_{in}$ consecutive input time steps and outputs a simultaneous set of $T_{out}$ subsequent time steps across the same grid. For the majority of this work, an input feed of $T_{in}=20$ time steps and a corresponding output of $T_{out}=5$ time steps were used, as it demonstrated the best performance during initial testing. However, for the reduced-MHD JOREK dataset, this input feed was reduced to $T_{in}=5$ time steps, due to only 10 consecutive timesteps being available for the majority of the dataset.

An experiment explored larger output steps ($T_{out}>5$, see \ref{sec:increasing t_out}), revealing that models trained to predict more $T_{out}$ steps performed similar or worse than those trained with smaller $T_{out}$ steps rolled out iteratively to the same total time length for both short and long rollouts. Larger $T_{out}$ models also consistently lost finer spatial details, aligning with the findings from \cite{lippe2023pderefiner}, which suggested that a common weakness of long rollouts was the inability to learn higher frequency components which seemed to be exacerbated when training on large rollouts.

\subsubsection{Training configuration} In PDEArena, the FNO was trained on the MSE loss between the output $T_{out}$ steps from the surrogate and simulation which are averaged across space and samples and summed across time and fields. To ensure comprehensive coverage of the simulations dynamics through the entire trajectory, the data the model was trained on was created by sampling timeslices starting at random points along the trajectory. Validation during training was performed similarly with an early stopping strategy utilised to determine when the model finished training.

\subsubsection{Testing on longer rollouts} 
\begin{figure}[h!]
    \centering
    \includegraphics[width=\textwidth]{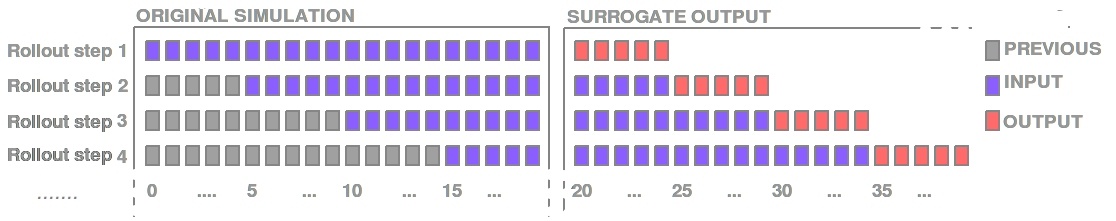}
    \caption{Example rollout using surrogate with $T_{in}=20$ and $T_{out}=5$. For the initial rollout step, the surrogate uses only timesteps from the original simulation, but rollout steps beyond that start using an increasing portion of generated rollout steps.}
    \label{fig:rollout}
\end{figure}

To generate extended temporal predictions, the model employs an autoregressive approach, where its output is recursively fed back into the FNO, alongside relevant input time steps, to predict subsequent field evolution over time (see Figure \ref{fig:rollout}). Notably, this autoregressive framework does not explicitly incorporate temporal dynamics into the model.

The starting point of a rollout refers to the timestep at which the predictions from the FNO start. For the results presented here, unless stated otherwise, rollouts begin at $t=20$ for models evaluated on the JOREK datasets and $t=200$ for models evaluated on the STORM dataset. This is to allow for the prior 20 timesteps that are fed to the FNO to generate the predictions and to avoid the turbulance buildup phase for STORM as explained in the STORM section. All reported errors, unless otherwise specified, are computed with min-max normalization applied, as errors without this transformation grow too large and lead to numerical overflow when calculated MSE for longer rollouts.

\subsubsection{Teacher-forcing testing} \label{sec:teacher-forcing} The autoregressive strategy introduced above may result in accumulating errors, since the predictions of the surrogate model are only approximation of the true solution and therefore the input manifold will progressively diverge from the training data distribution. A \emph{teacher-forcing} test  strategy consists in using as input the true PDE solution at any timestep. While this testing strategy does not reflect the typical downstream use case where the surrogate model is employed as a drop-in replacement of the PDE simulator, it learnt behaviours without accounting for accumulating rollout errors and it is therefore a useful diagnostic.

\subsubsection{Transfer learning} \label{sec:transfer_learning} Transfer learning was implemented via the fine-tuning approach. In the context of this work, the source domain ($D_S$) refers to the larger, well-sampled simulation dataset ($X_S,Y_S$), while the target domain 
($D_T$) corresponds to a smaller dataset representing a different but related simulation ($X_D,Y_D$). For example, for scenario one of the cross-simulation transfer, the source dataset was drawn from STORM and the target dataset was drawn from JOREK. More specifically, there are two steps in the fine-tuning framework: pre-training and fine-tuning. During pre-training, the model is trained on a labeled dataset from $D_S$. For all pre-training, the largest dataset available was used from the $D_S$. For finetuning, the model is first initialised with the pre-trained parameters, and all of the parameters are fine-tuned using labeled data from $D_T$. When the size of the training dataset was varied, this only affected the amount of data used for fine-tuning dataset. All hyperparameters remained unchanged during fine-tuning, except for gradient clipping, which was adjusted to match the target dataset, and the learning rate, which was set to a lower value. The learning rate was determined through a simple grid search, halving the value until performance ceased to improve, similar to the procedure used for scratch models (described in \ref{sec:model_hyperparam}).

\subsubsection{Transfer learning process to new variables} \label{sec:transfer_learning2} A challenge faced with many neural networks such as the FNO is that the number of variables that the model can be trained on is fixed. This means that it is not possible to perform the fine-tuning process above to a target dataset with more variables without changing the architecture. To address this, we adopted a two-stage fine-tuning strategy:
\begin{enumerate}
    \item Fine-tuning Stage 1 (Domain Adaptation): The pretrained model was fine-tuned on the target dataset using only the variables that were common to both source and target domains (electric potential and density). This step allowed the model to adjust to the physical and numerical context of the target simulation regime.
    \item Fine-tuning Stage 2 (New Variable Learning): The model was then further fine-tuned on the target dataset, this time predicting pair of the new variables (temperature and current) and computing the loss for those variables.
\end{enumerate}
This approach used the common variables in the target domain as a contextual bridge, allowing the model to adapt gradually rather than making a sudden leap from the source data to unrelated target outputs. Although more sophisticated methods exist in the literature for transfer learning, this approach was chosen for this case study for simplicity, with exploration of more advanced techniques left for future work.

\section{Results}

\subsection{Initial results} \label{sec:initial_results}

As a first experiment, FNOs were trained to learn the evolution for each of the three full simulation datasets (electrostatic JOREK, reduced-MHD JOREK and STORM), converging to the test single-step MSE in table \ref{table:initial_results:error_table}. Both the electrostatic JOREK and STORM surrogates were trained on all fields except the auxiliary variables, while reduced-MHD JOREK was trained on all available fields, including the auxiliary variables. The reduced-MHD JOREK model exhibited the largest error overall, and there was significant variation in the one-step error between variables for some models, which could suggest potential overfitting to specific variables over others in the dataset. A comparison against the U-Net \cite{ronneberger2015unetconvolutionalnetworksbiomedical}, is presented in Appendix \ref{sec:model_unet} and briefly discussed.

\begin{table}[h!]
\centering
\begin{adjustbox}{width=\textwidth}
\begin{tabular}{@{}lllllll@{}}
\toprule & \textbf{Electrostatic JOREK} & \textbf{STORM} & \textbf{Reduced-MHD JOREK} \\
\textbf{Variable} & \textbf{(MSE$\pm$STD)} & \textbf{(MSE$\pm$STD)} & \textbf{(MSE$\pm$STD)} \\\midrule
Temperature ($T$) & $4.34\pm 11.1\times10^{-8}$ & & $3.21 \pm 13.14\times10^{-4}$ \\
Electric potential ($\Phi$) & $2.95\pm 13.0 \times 10^{-6}$ & $1.71 \pm 2.36 \times 10^{-6}$ & $1.48 \pm 6.39\times 10^{-4}$
\\
Density ($\rho$) & $5.80 \pm 22.7\times10^{-6}$ & $7.73 \pm 13.1 \times 10^{-6}$ & $1.36 \pm 4.52\times 10^{-4}$ \\
Vorticity ($\omega$) & & & $5.00 \pm 20.0\times10^{-6}$ \\
Magnetic flux ($\Psi$) & & & $3.90 \pm 13.5\times10^{-5}$ \\
Current ($zj$) & & & $1.41 \pm 5.62\times 10^{-4}$ \\ \bottomrule
\end{tabular}
\end{adjustbox}
\caption{MSE on mininmum output length (meaning 5 timesteps) for each rescaled dataset variable averaged across different starting points for all dataset trajectories.}
\label{table:initial_results:error_table}
\end{table}

\begin{figure}[h!]
    \centering
    \subfigure[Particle density $\rho$ ($m^{-3}$)]{
    \includegraphics[width=0.67\textwidth]{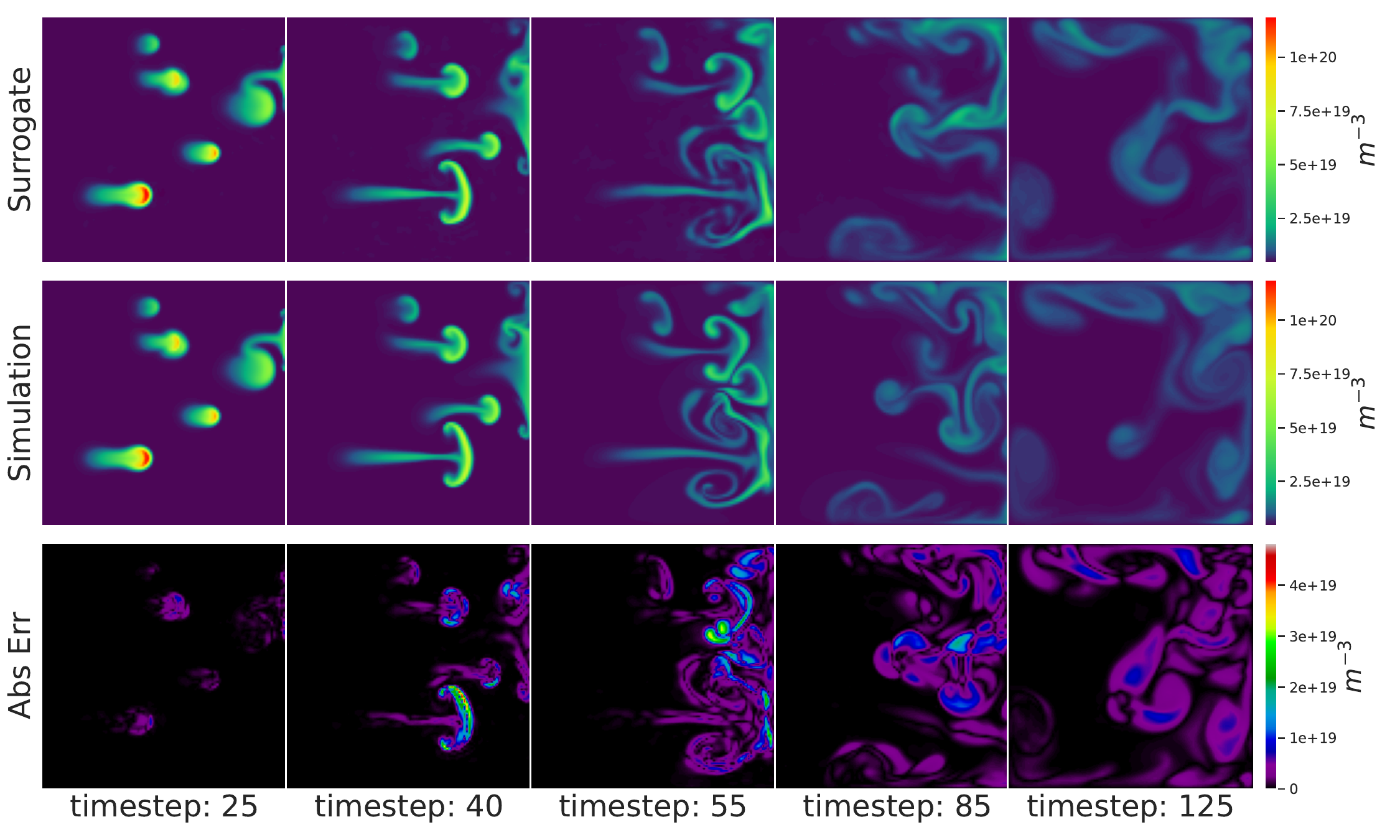}
    }
    \subfigure[Electric potential $\Phi$ ($V$)]{
    \includegraphics[width=0.67\textwidth]{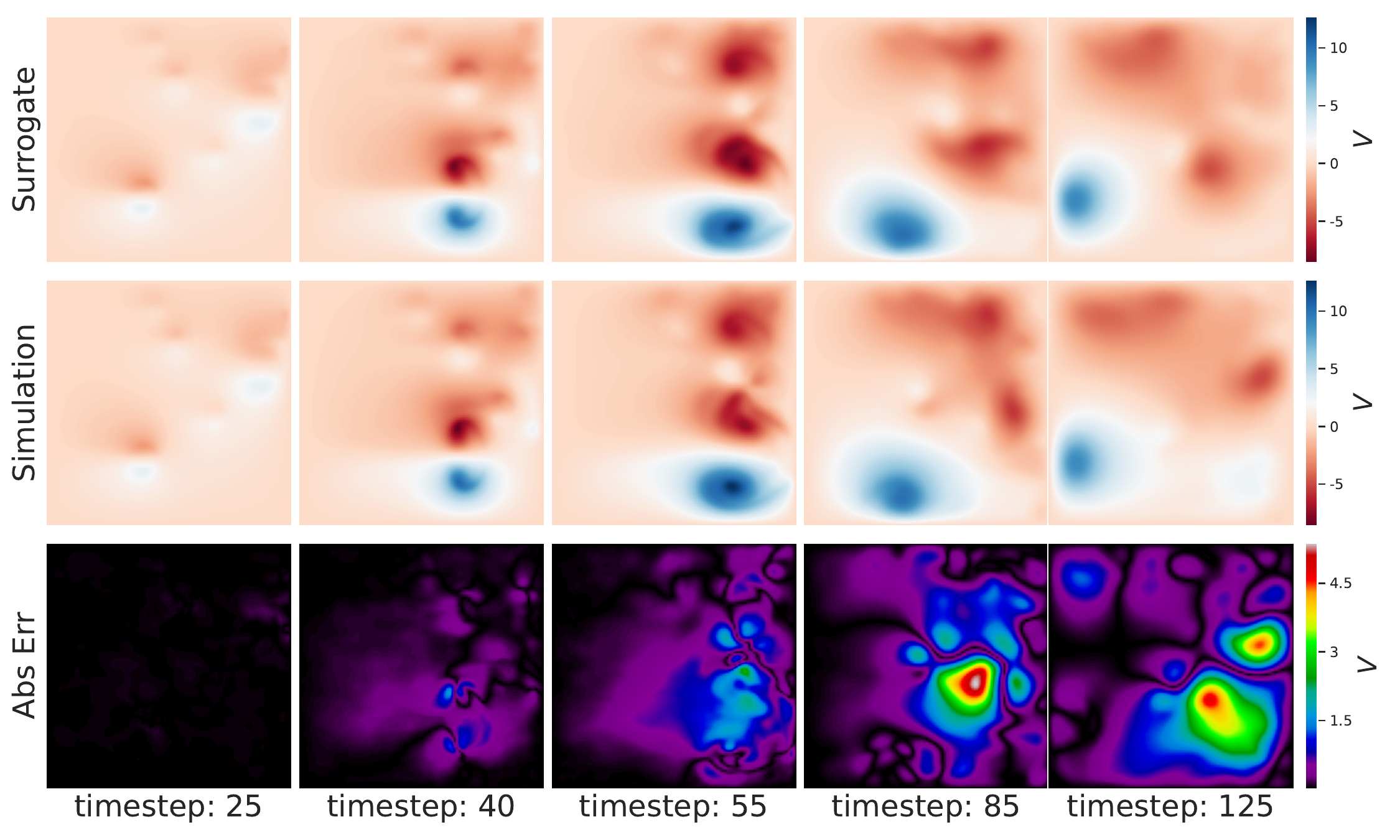}
    }
    \subfigure[Temperature $T$ ($eV$)]{
    \includegraphics[width=0.67\textwidth]{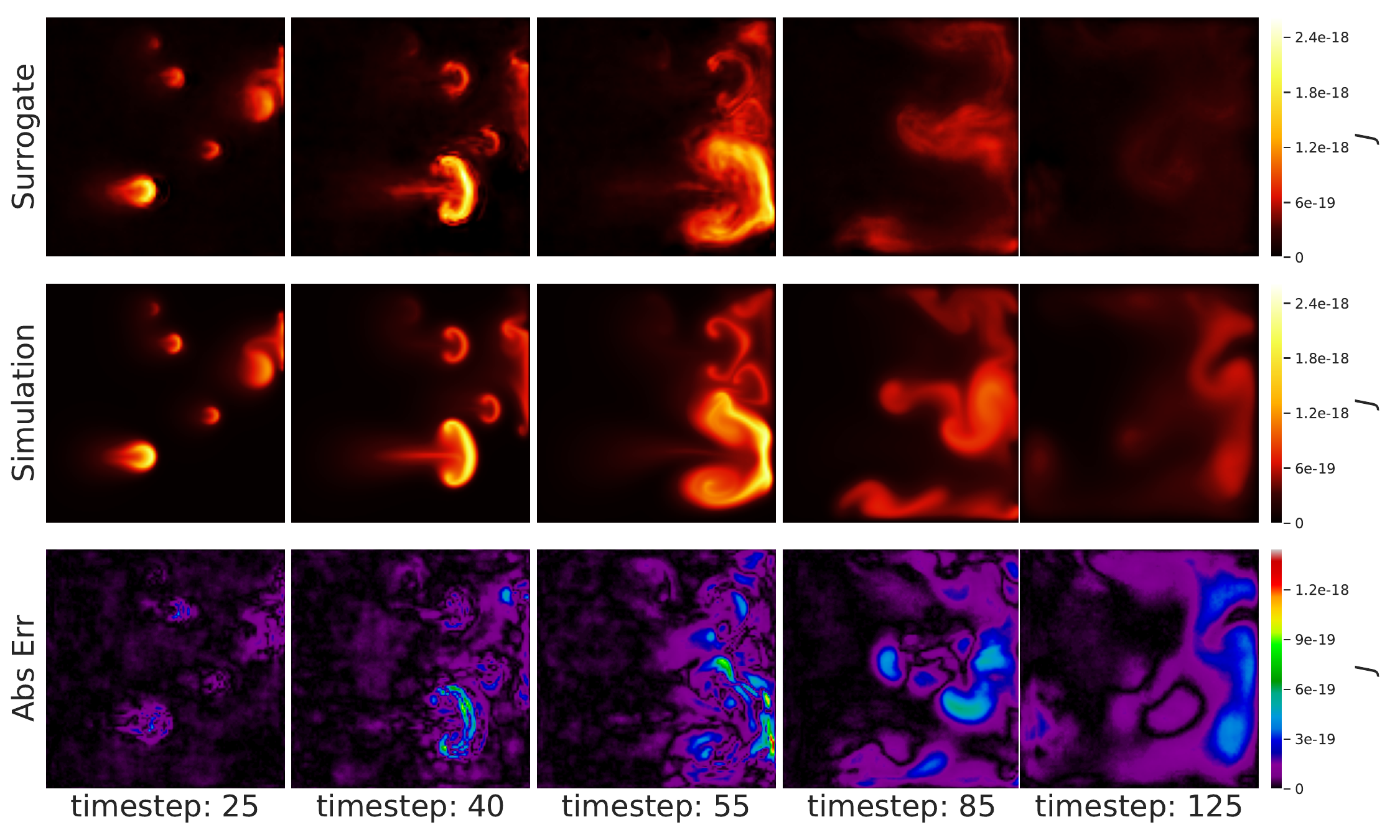}
    }
    \caption{An example electrostatic JOREK run plotted at specific rollout timesteps for fields (a) density, (b) electric potential and (c) temperature.}
    \label{fig:initial_results:example_ejorek_traj}
\end{figure}

\begin{figure}[h!]
\centering
    \subfigure[Particle density $\rho$ ($m^{-3}$)]{
    \includegraphics[width=0.48\textwidth]{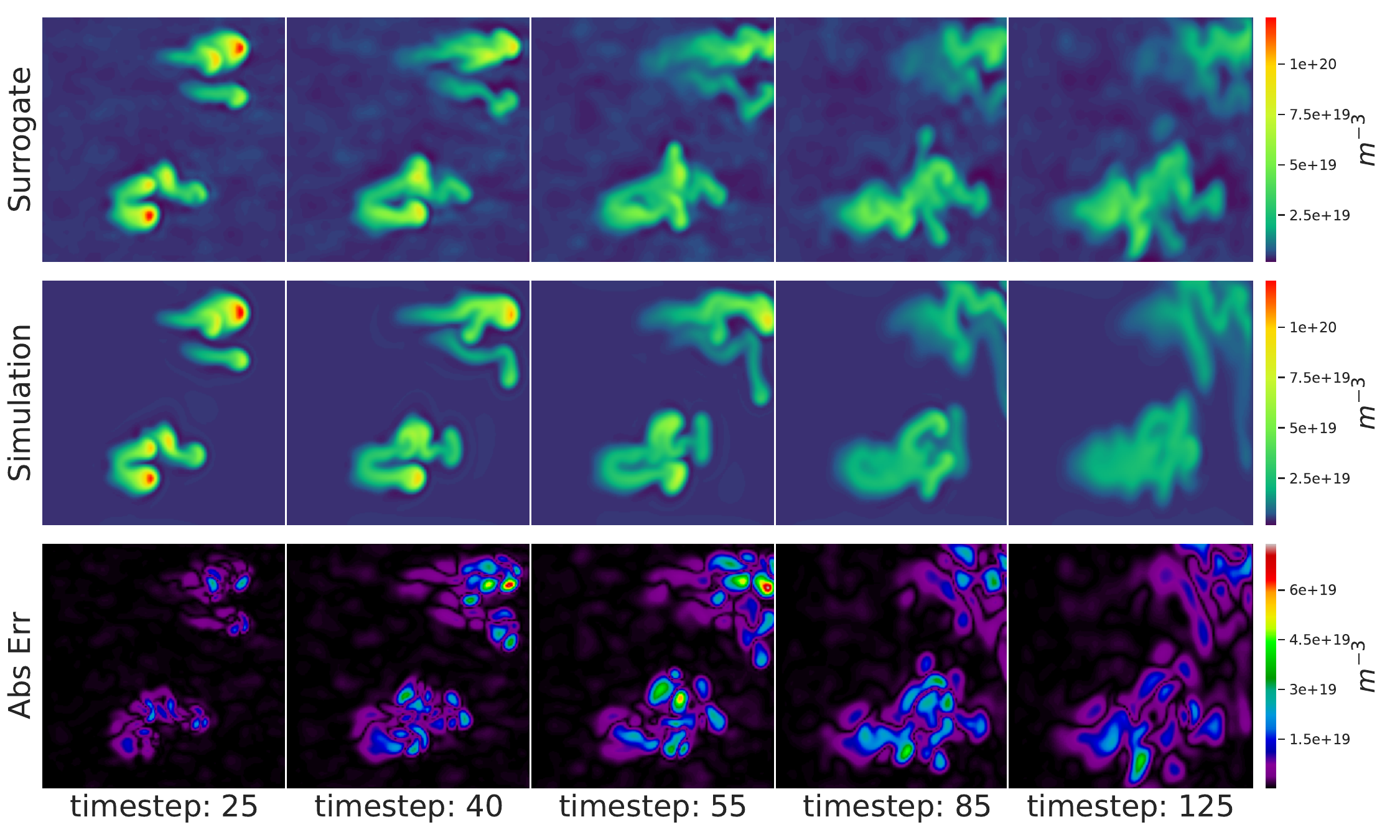}
    }
    \vfill
    \subfigure[Electric potential $\Phi$ ($V$)]{
    \includegraphics[width=0.48\textwidth]{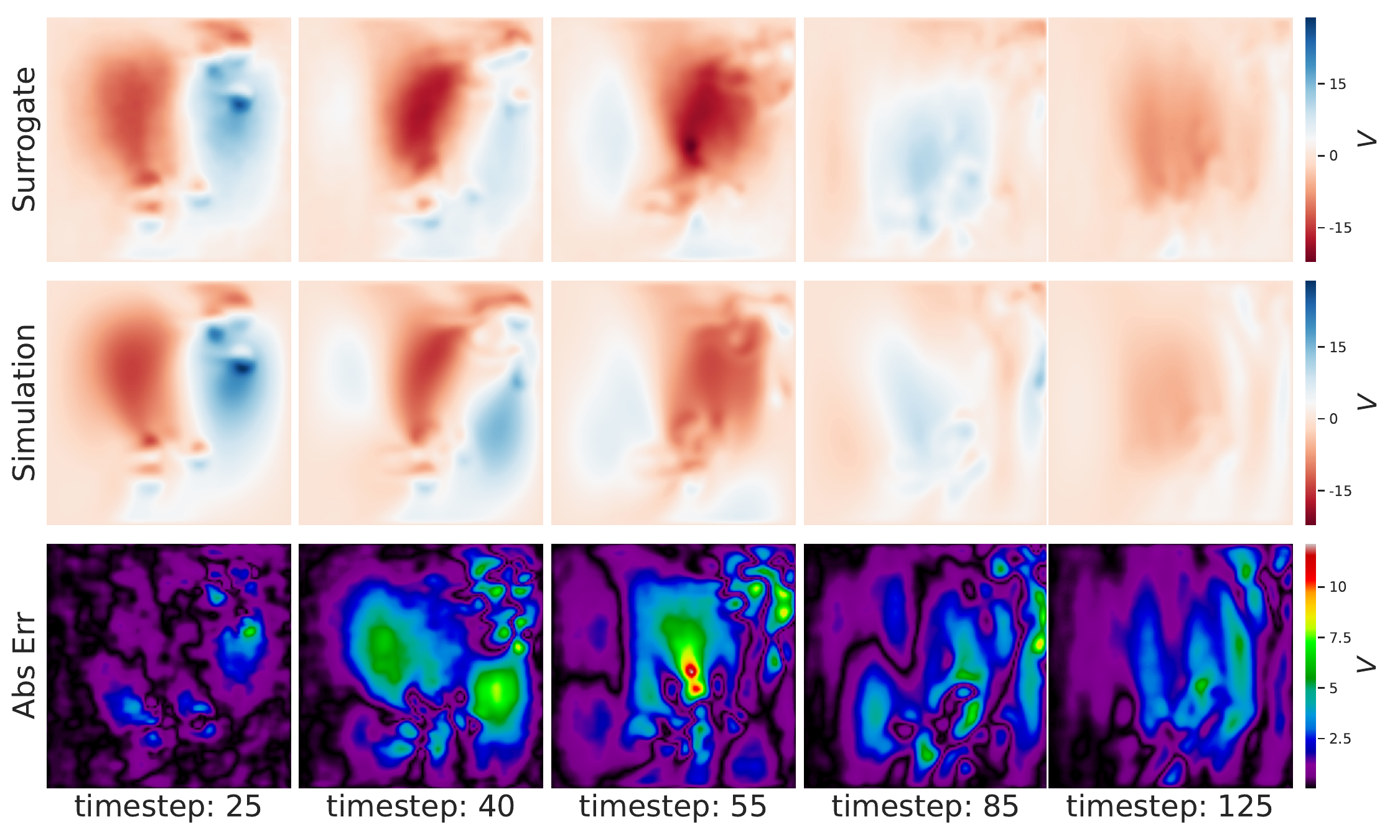}
    }
    \vfill
    \subfigure[Temperature $T$ ($eV$)]{
    \includegraphics[width=0.48\textwidth]{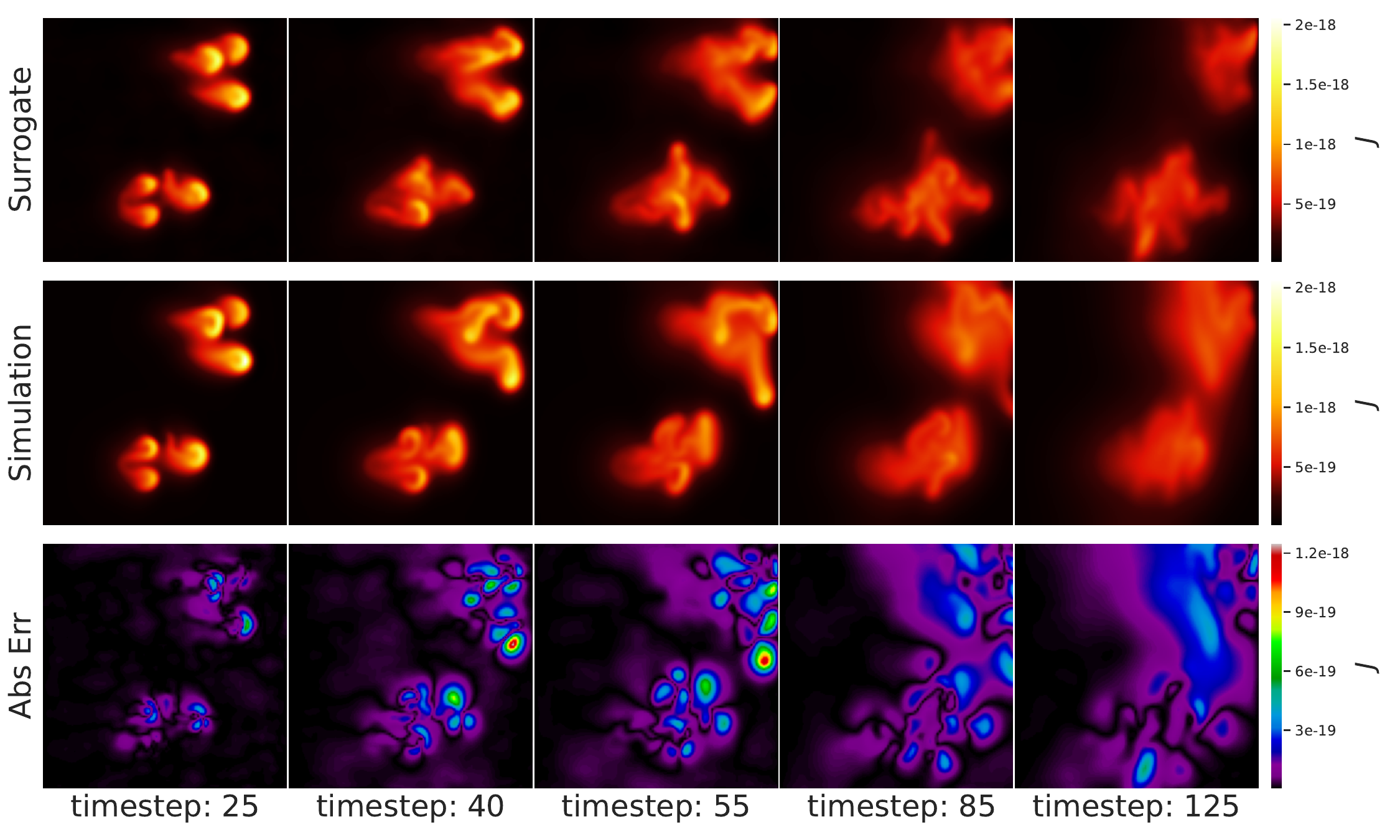}
    }
    \vfill
    \subfigure[Toroidal current $zj$ ($MA.m^{-2}$)]{\includegraphics[width=0.48\textwidth]{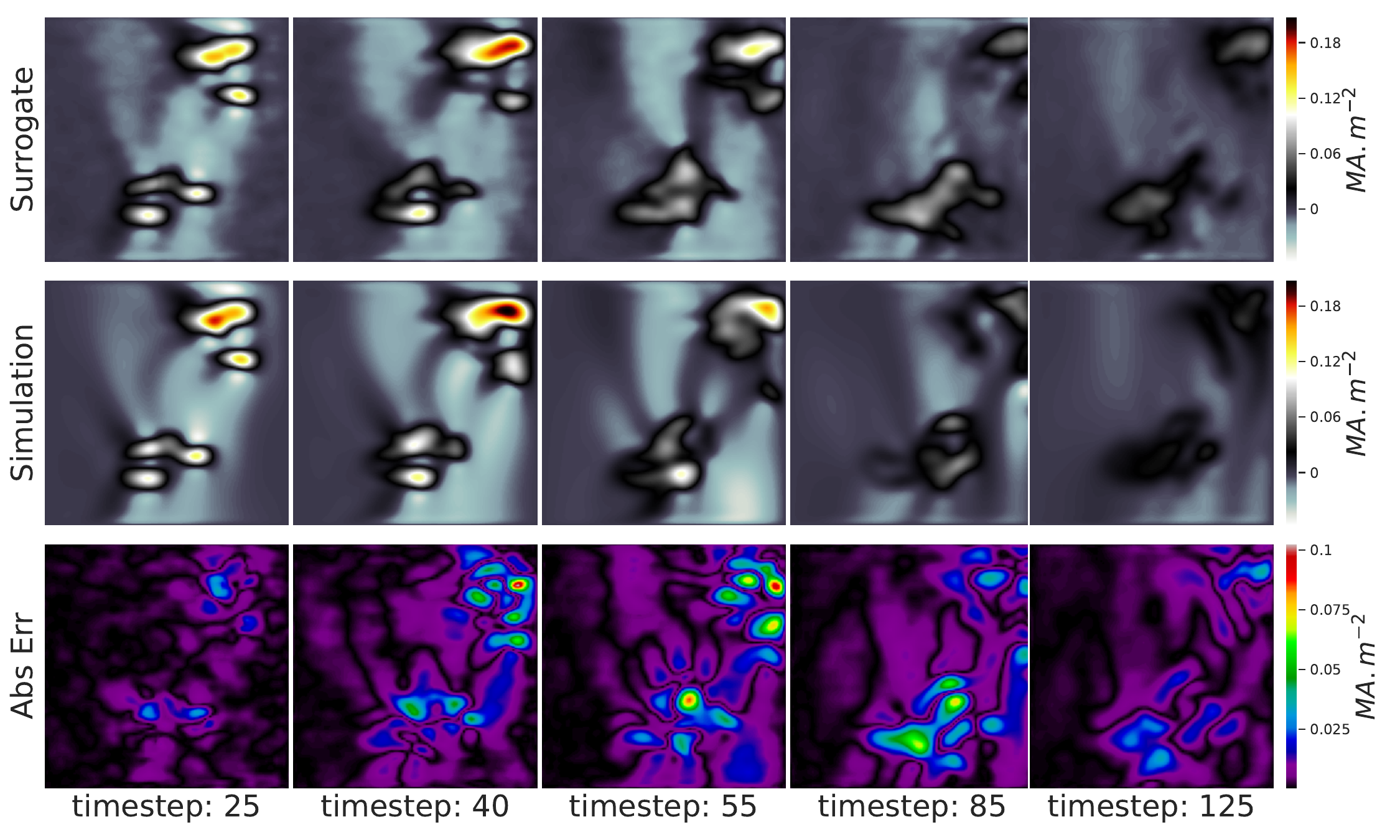}
    }
    \caption{An example reduced-MHD JOREK run for fields (a) density, (b) electric potential, (c) temperature and (d) current. The remaining fields can be found in the appendix at \ref{fig:initial_results:example_mjorek_traj_additional}}
    \label{fig:initial_results:example_mjorek_traj}
\end{figure}

\begin{figure}[h!]
    \centering
    \subfigure[Particle density $\rho$ ($m^{-3}$)]{
    \includegraphics[width=0.7\textwidth]{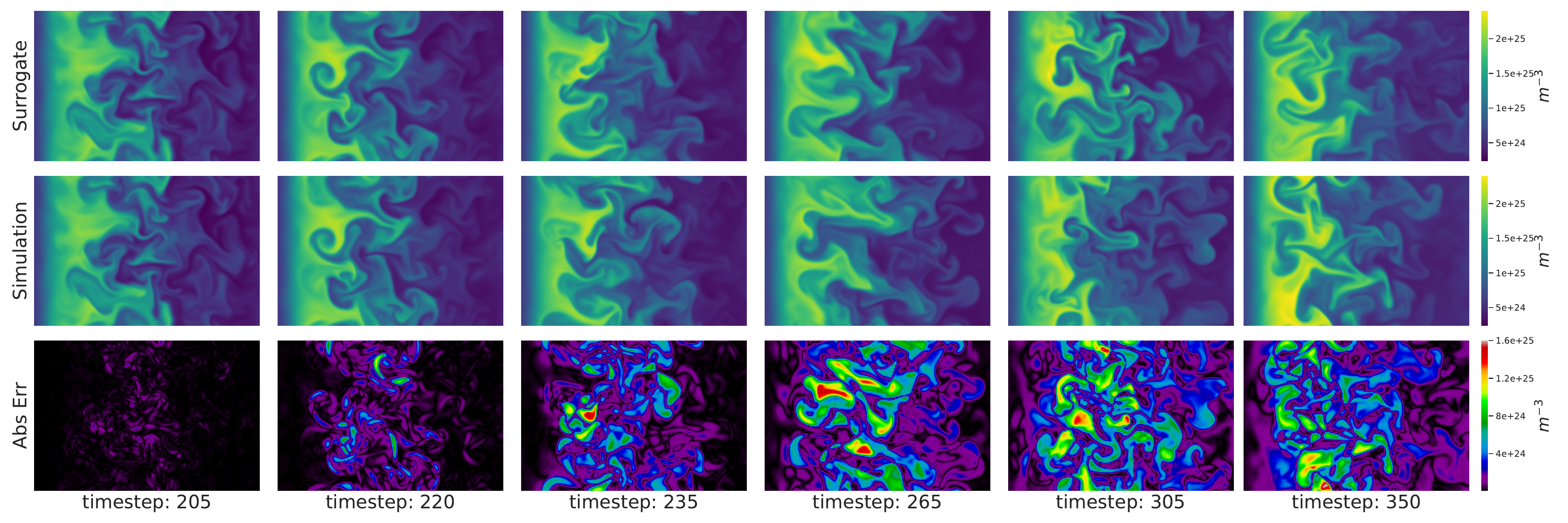}
    }
    \subfigure[Electric potential $\Phi$ ($V$)]{
    \includegraphics[width=0.7\textwidth]{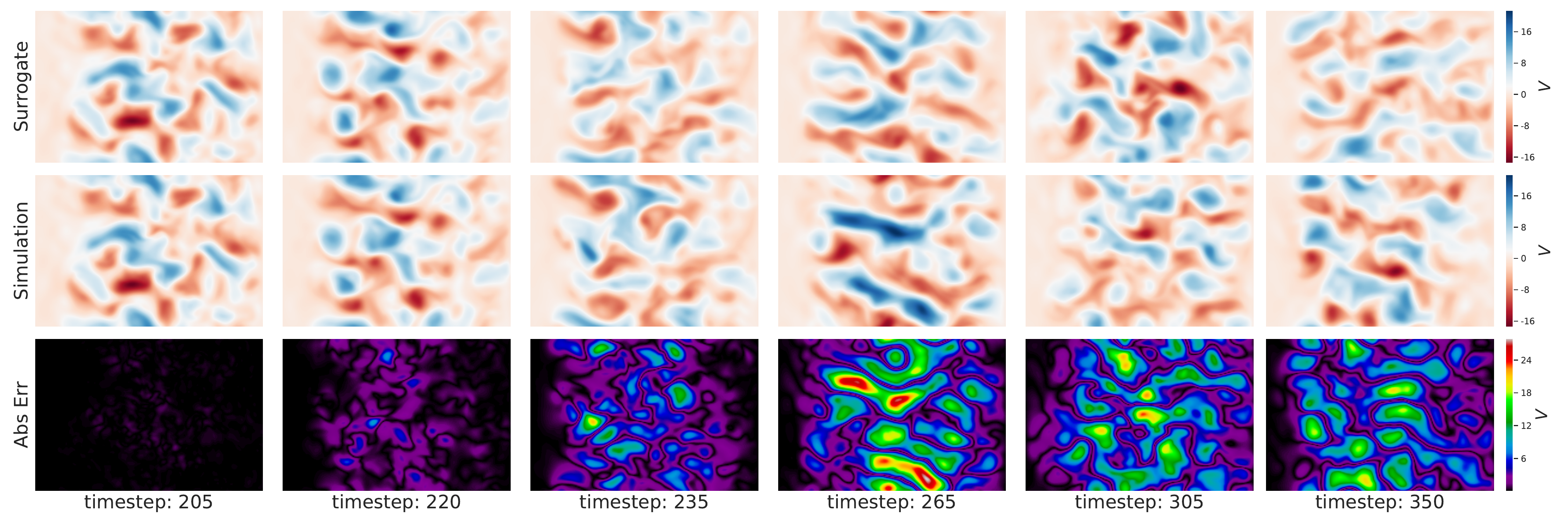}
    }
    \caption{An example STORM run for fields (a) density and (b) electric potential}    \label{fig:initial_results:example_storm_traj}
\end{figure}

Example trajectory rollouts for test samples are shown in figs \ref{fig:initial_results:example_ejorek_traj}-\ref{fig:initial_results:example_storm_traj}. These rollouts are plotted at specific time points for every field showing the target simulation, the learnt surrogate and the absolute error between them. From these figures, it can be seen that the surrogates were able to learn global features, such as the location of the density source in STORM (figure \ref{fig:initial_results:example_storm_traj}) and the location of the blobs in JOREK (figure \ref{fig:initial_results:example_ejorek_traj} and  \ref{fig:initial_results:example_mjorek_traj}). However, it is clear that for longer time rollouts the finer details of the fields are quite different from the ground truth. This is shown quantitatively in figures \ref{fig:traj_loc_impact:diff_starts_ejorek2}- \ref{fig:traj_loc_impact:diff_starts_storm}, where the rollout errors are calculated as averages across the multiple trajectories in the testing dataset.

Notably, for both JOREK models (figures \ref{fig:traj_loc_impact:diff_starts_ejorek2} and \ref{fig:traj_loc_impact:diff_starts_mjorek}), an error peak is observed for many of the variables. This error peak is not present for STORM in figure \ref{fig:traj_loc_impact:diff_starts_storm}. To investigate this further, two experimental setups were employed: (1) evaluating the rollout error initiated at various points in the trajectory and (2) assessing the error in a teacher forcing setting. These are discussed in the following subsections.

We also observe periodic error spikes in several fields, being visible in temperature for electrostatic JOREK. These occur at regular five-step intervals and are believed to be aligned with the surrogate's chunked prediction strategy, where each forward pass generates five timesteps at a time and results in small discontinuities at the boundaries of prediction blocks. While temperature exhibits lower overall test error compared to other variables, the presence of these structured spikes suggests a degree of overfitting which is reflected in these discontinuities at the boundaries being more visible.

\subsubsection{Impact of trajectory location with JOREK datasets} \label{sec:traj_loc_impact}

\begin{figure}[h!]
    \subfigure[Autoregressive rollouts starting from different  points in time.]{\includegraphics[width=\textwidth]{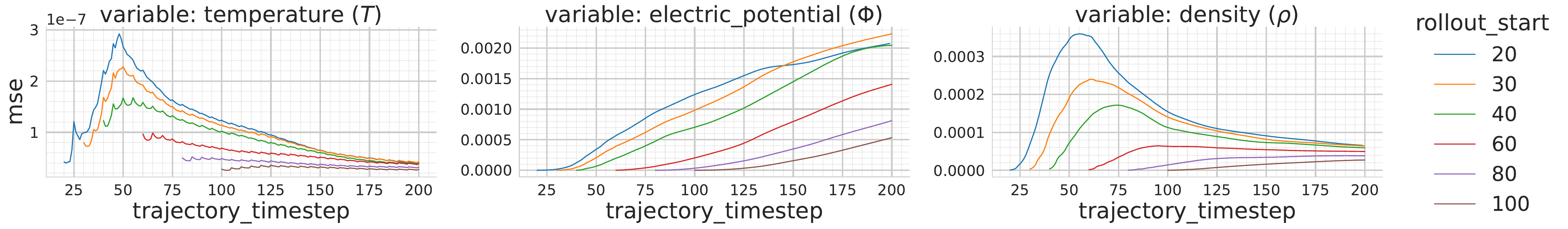}
    }
    \vfill
    \subfigure[Teacher-forcing testing.]{\includegraphics[width=0.875\textwidth]{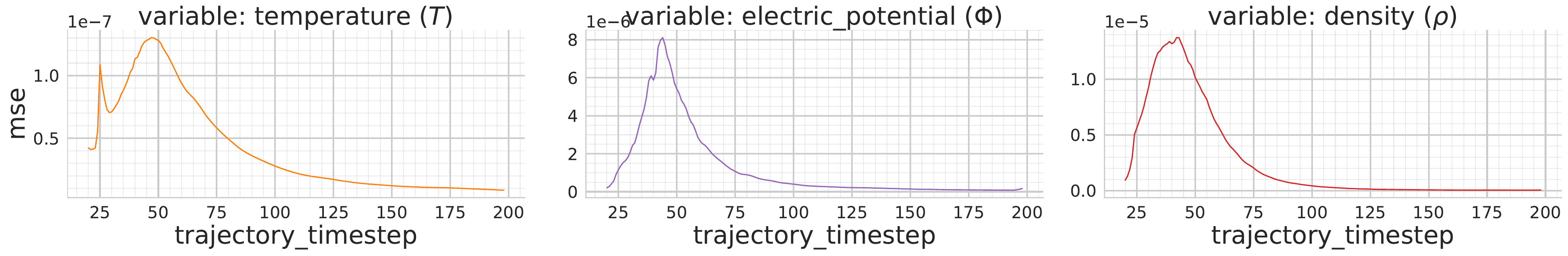}
    }
    \caption{Long rollout error for FNO model trained on electrostatic JOREK (for each field) for rollout error relative to trajectory timestep (a) with different starting points with accumulating input error and (b) with teacher-forcing testing.}    
    \label{fig:traj_loc_impact:diff_starts_ejorek2}
\end{figure}

Distinct behaviors emerged among the three fields in the Electrostatic JOREK model, as seen in fig \ref{fig:traj_loc_impact:diff_starts_ejorek2}:
\begin{enumerate}
    \item Electric Potential: Electric potential errors increased steadily over time, closely following the expected pattern of accumulating error. Unlike the other two variables, no trajectory-specific peaks were observed, suggesting that error growth here is primarily due to cumulative error effects over longer rollouts.
    \item Temperature: Unlike electric potential, rollout errors for temperature did not consistently increase. Instead, errors decreases throughout the auto-regressive rollout when these rollouts were initiated later in the trajectory. This results in a general pattern where errors are consistently highest at a specific point in the trajectory (around $t=50$) and diminishing in regions further from this point regardless of the time gap since the model last saw the ground truth. These results suggest that temperature error may be predominantly driven by specific trajectory dynamics at that point rather than accumulated input error.
    \item Density: The density field showed a combination of both behaviors. When rollouts were initiated prior to $t=50$, a distinct peak appeared in the error profile, similar to the temperature field. However a pattern observed is that this peak diminished and shifted right when rollouts were initiated later in the trajectory. A potential explanation of this can be found if both trajectory-specific dynamics (like in temperature) and accumulating error effects (like in electric potential) are occurring at the same time. When the trajectory is started sooner, the accumulating error has had less time to build and thus the errors introduced by trajectory specific dynamics around $t=50$ are a lot more dominant, causing the error peak to be shifted sooner. Whereas the reverse occurs as the trajectory is started later.
\end{enumerate}

Notably, the observed rollout behavior persisted for bth temperature and density even in a teacher forcing setting, which removes accumulating input error as a factor. This suggests that there may be a systemic time-dependent issue within the model’s handling of certain fields. This result is similar to \cite{Poels_2023} where a similar pattern was observed. Indeed, the different models used in \cite{Poels_2023} often failed in qualitatively similar regions of the trajectory, suggesting shared limitations in capturing the underlying physics. The observed behaviour is thus puzzling but not uncommon for neural surrogates. These weaknesses highlight the need for further refinement of the neural surrogate models, particularly for fields where input error accumulation is not the primary issue but systemic error remains. An additional observation is the potential correlation between the magnitude of the error in shorter rollouts and the degree to which accumulating error affects longer rollout behavior. For example, temperature, which exhibits the least error in short rollouts (see Table \ref{table:initial_results:error_table}), seems to be less impacted by accumulating error in the autoregressive process. In contrast, electric potential, which shows the most error in short rollouts, is most susceptible to the effects of accumulating error.

\begin{figure}[h!]
    \subfigure[Autoregressive rollouts starting from different  points in time.]{\includegraphics[width=\textwidth]{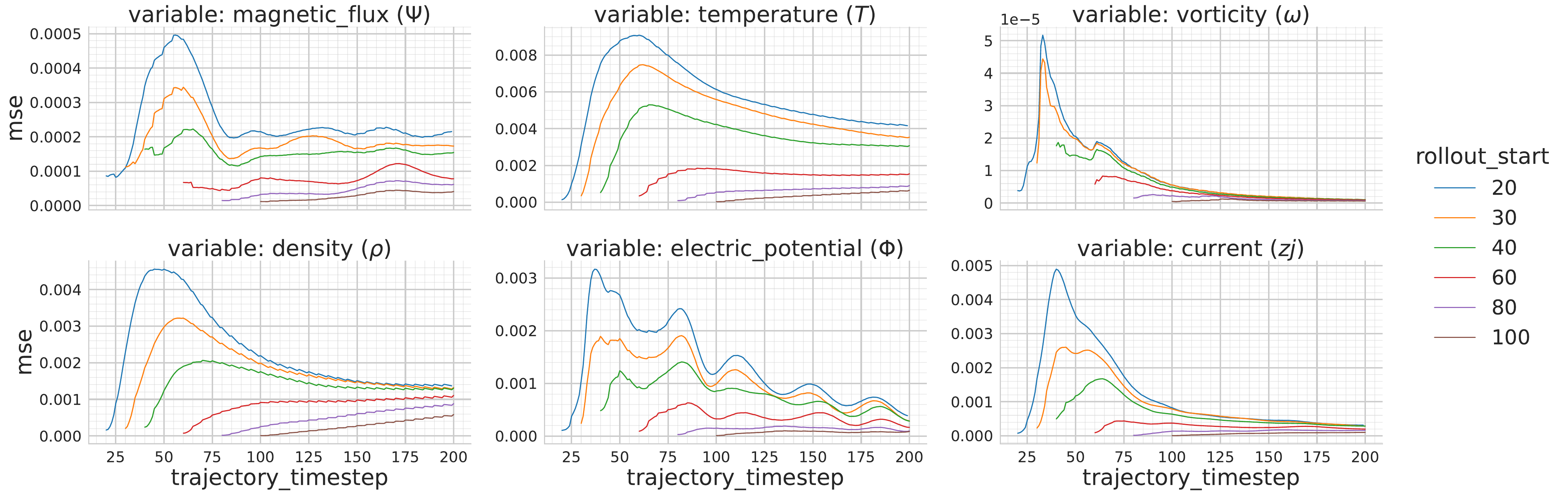}
    }
    \vfill
    \subfigure[Teacher-forcing testing.]{\includegraphics[width=0.875\textwidth]{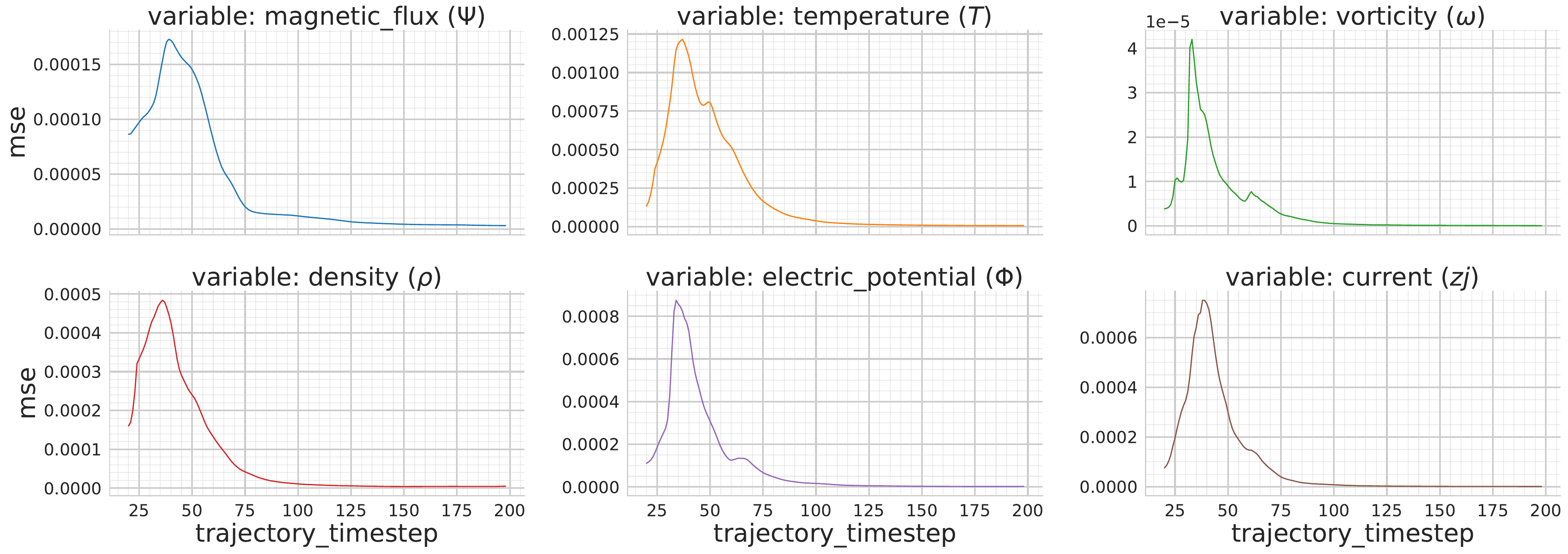}
    }
    \caption{Same as Figure \ref{fig:traj_loc_impact:diff_starts_ejorek2} but for reduced-MHD JOREK. }    \label{fig:traj_loc_impact:diff_starts_mjorek}
\end{figure}

For the reduced-MHD JOREK model, a similar pattern was observed, as in Figure \ref{fig:traj_loc_impact:diff_starts_mjorek}, with an error peak around $t=40$ for most variables, excluding temperature. In contrast to Electrostatic JOREK, error in this model consistently increased over the 10 rollout steps for all fields except magnetic flux. This suggests behavior similar to Electrostatic JOREK’s density field, where it is proposed that both simulation-specific dynamics and accumulating errors are in effect. A similar correlation between the size of the error in shorter rollouts and the degree to which accumulating error affects longer rollout behavior is also loosely observed.

\hfill \break

\subsubsection{Impact of trajectory location with STORM dataset}

\begin{figure}[h!]
    \centering
    \hspace*{1.8cm}
    \subfigure[Autoregressive rollouts starting from different  points in time.]{\includegraphics[width=0.9\textwidth]{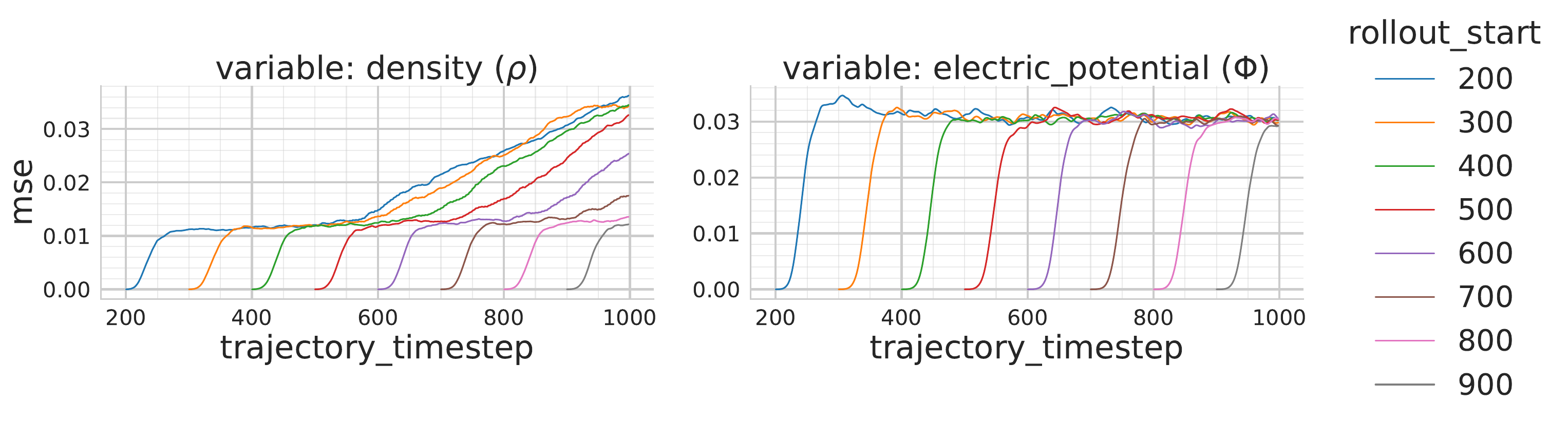}
    }
    \vfill
    \subfigure[Teacher-forcing testing.]{\includegraphics[width=0.75\textwidth]{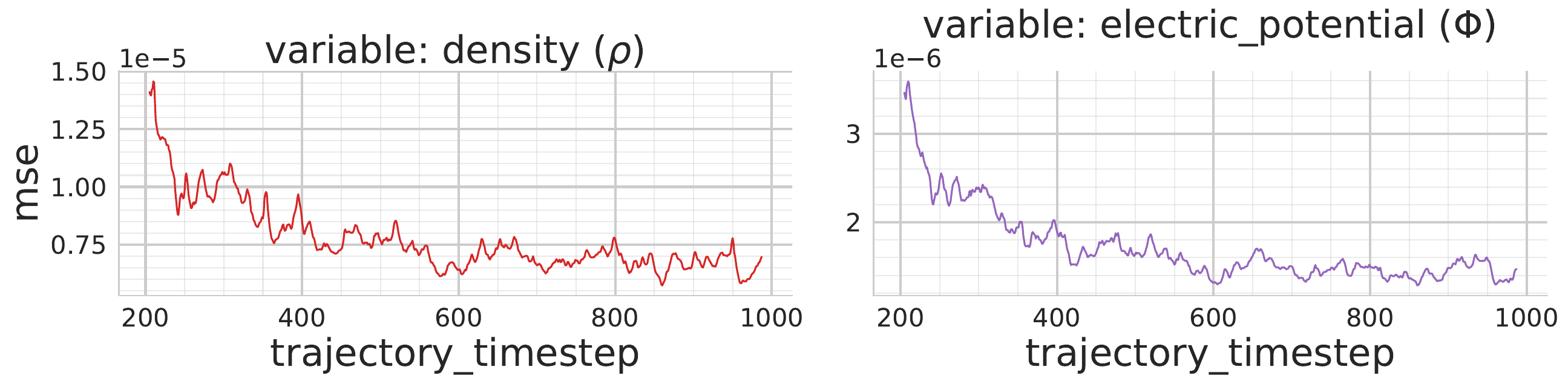}
    }
    \caption{Same as Figure \ref{fig:traj_loc_impact:diff_starts_ejorek2} but for STORM. }    
    \label{fig:traj_loc_impact:diff_starts_storm}
\end{figure}

In STORM (seen in fig \ref{fig:traj_loc_impact:diff_starts_storm}), errors exhibited more uniform patterns across fields:
\begin{enumerate}
    \item Electric Potential: This field showed an initial exponential error increase, followed by a plateau, suggesting that errors reach a saturation level in longer rollouts. This behaviour was also consistent across different starting points in the trajectory.
    \item Density: Density behaved differently, with an initial similar exponential increase followed by a sustained, gradual increase over the rollout duration, indicating no clear saturation point.
\end{enumerate}

\begin{figure}[h!]
    \centering
    \includegraphics[width=\textwidth]{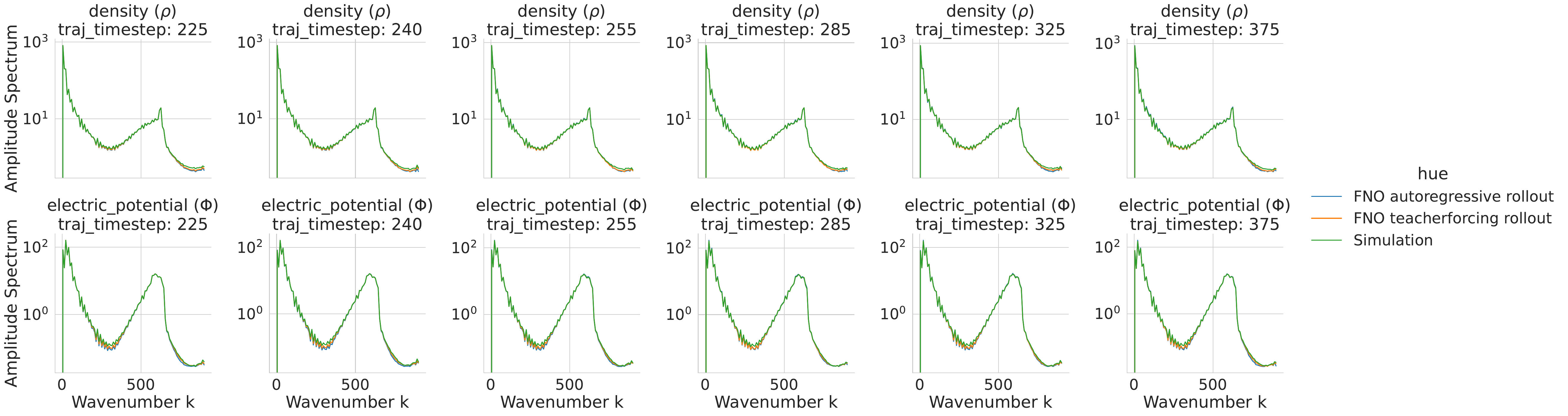}
    \caption{Radially averaged amplitude spectra for STORM at different rollout lengths. This is calculated with and without accumulating error. Accumulating error rollout is started at rollout timestep 200. }    \label{fig:freq_spectra_storm}
\end{figure}

Unlike both JOREK models, STORM fields did not exhibit trajectory-specific error peaks, pointing to more generalized error accumulation across the rollout. This is perhaps unsurprising, as STORM is used in a steady state setting, as opposed to JOREK which includes transient behaviours. In fact, although the MSE is used as a training signal for the STORM surrogate models, it may not be appropriate for predicting the long-rollout behaviour of chaotic systems. A statistical representation of the flow is instead recommended. Therefore, the radially averaged spectrum was computed for both STORM and the corresponding FNO surrogate model, as shown in figure \ref{fig:freq_spectra_storm}. The FNO captures much of the spectrum accurately at all time rollouts. This result emphasizes the need to use appropriate metrics for surrogate models of chaotic systems.

\subsection{Heatflux deposited on the wall} \label{sec:heatflux}

To evaluate the surrogate’s ability to capture key physical behaviors relevant to plasma-facing component performance, the radial heat flux at the wall was computed using the JOREK surrogates. The radial heat flux ($Q_R$) quantifies the energy transferred per unit area in the radial direction and is crucial for assessing material stress and component lifetime.

More precisely, the heat flux was computed as:
\begin{equation}
    Q_R = \rho  T V_R
\end{equation}
where the radial velocity, $V_R$, was determined using the relation:
\begin{equation}
    v_{R}=R B \frac{d\phi}{dZ}
\end{equation}
where R is the distance in the radial direction, B is the magnetic field and $\frac{d\phi}{dZ}$ is the gradient of the electrostatic potential in the toroidal direction. Predictions for the density, electric potential and temperature are produced autoregressively by the surrogate, with the rollout starting at $t=20$. The heat flux on the outmost radial slice across the poloidal direction is considered (which is equivalent to the right hand side boundary as plotted in example fig \ref{fig:initial_results:example_ejorek_traj}).

\begin{figure}
    \centering
    \subfigure[Heatflux scatter plot]{
    \includegraphics[width=0.45\textwidth]{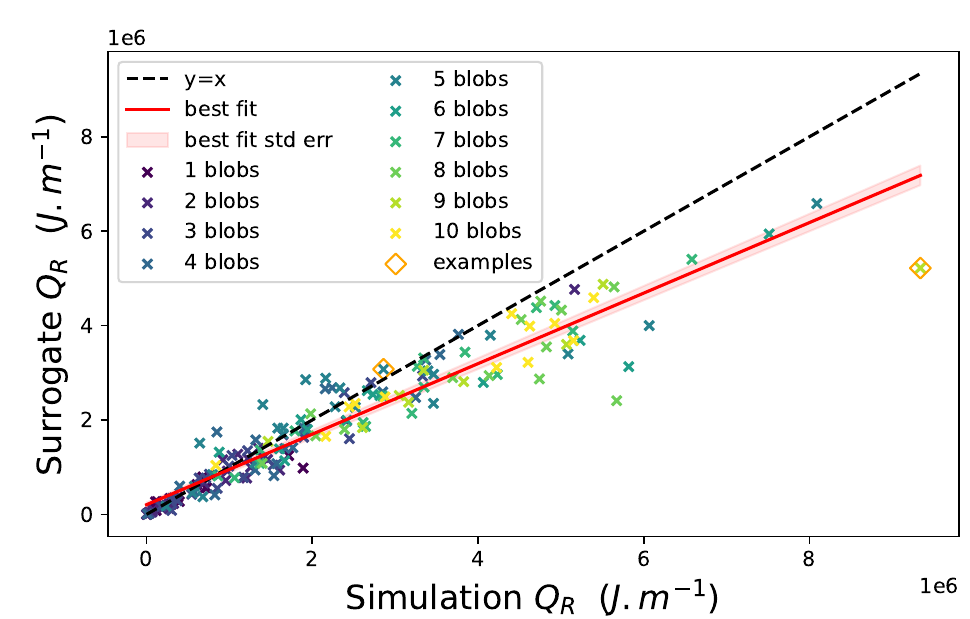}
    }
    \subfigure[Simulation heatflux histogram]{
    \includegraphics[width=0.45\textwidth]{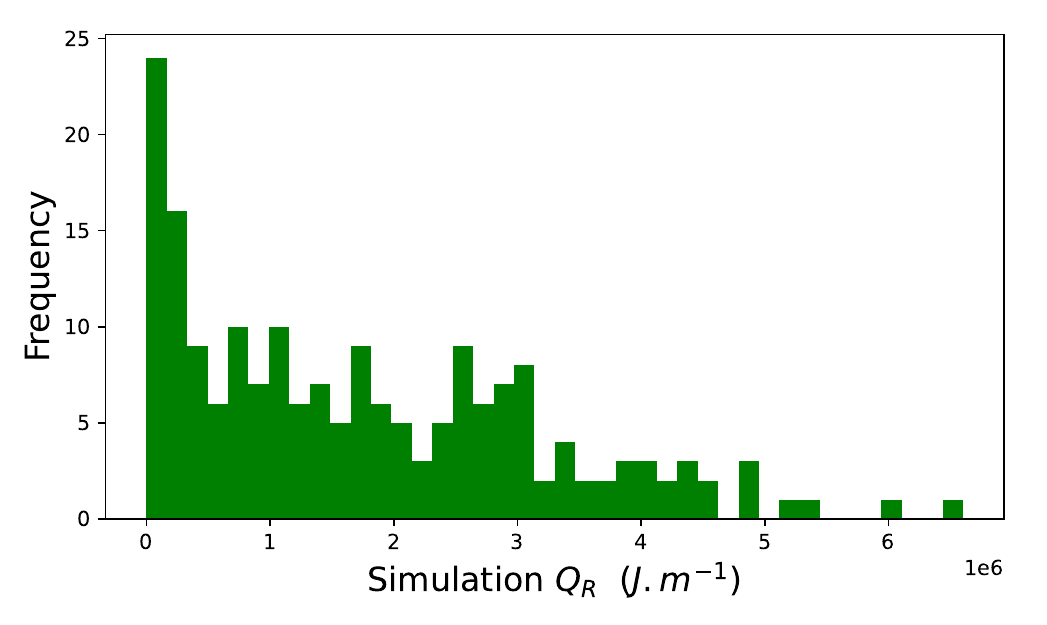}
    }
    \caption{(a) Heat flux at boundary integrated over poloidal direction (y-axis) and time comparing surrogate and simulation. Each datapoint is trajectory with the color corresponding to the number of blobs. A black line is an identity line provided as a demonstration of full agreement between surrogate and simulation. The red line is the best fit line with the coefficients ($\pm$stderror): slope $0.75\pm0.17$ and intercept $2.05\pm0.47\times10^{5}$; with a pearson coefficient of $0.95$. The orange diamonds circle the 2 example trajectories shown in the figures below. (b) is the distribution of the simulation heatflux, demonstrating that there are many more trajectories with lower heatflux values}
    \label{fig:heatflux_scatter_ejorek}
\end{figure}

The scatter plot in Figure \ref{fig:heatflux_scatter_ejorek} compares the total heat flux at this boundary, integrated over time and the poloidal direction. The surrogate model performs well for lower heat flux values ($< 0.4\times10^{6}$ J m$^{-1}$) but systematically underestimates higher fluxes, with errors increasing as flux rises. This issue is particularly pronounced for extreme cases ($>0.6\times10^{6}$) J m$^{-1}$, though such instances are rare, occurring in only $\sim$5\% of the trajectories. Similar trends are observed in the reduced-MHD JOREK dataset (see Appendix), though meaningful conclusions are limited due to the small dataset size. A strong trend with the number of blobs simulated is observed, where a higher number of blobs and high heat fluxes are correlated with a general deterioration of the surrogate performance.

\begin{figure}
    \centering
    \subfigure[Initial plasma condition]{
        \includegraphics[width=0.35\textwidth]{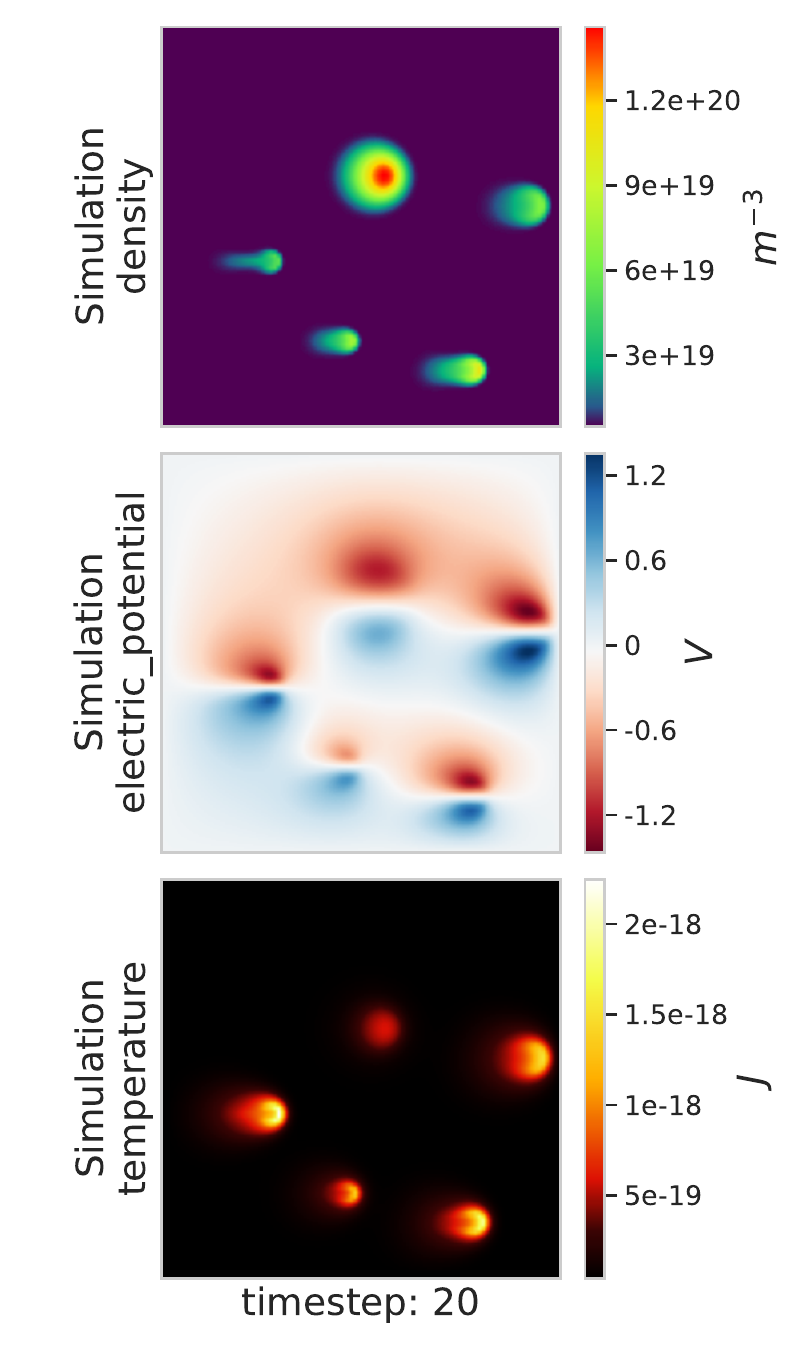}
    }
    \subfigure[Temporal evolution of plasma at RHS boundary]{
        \includegraphics[width=0.56\textwidth]{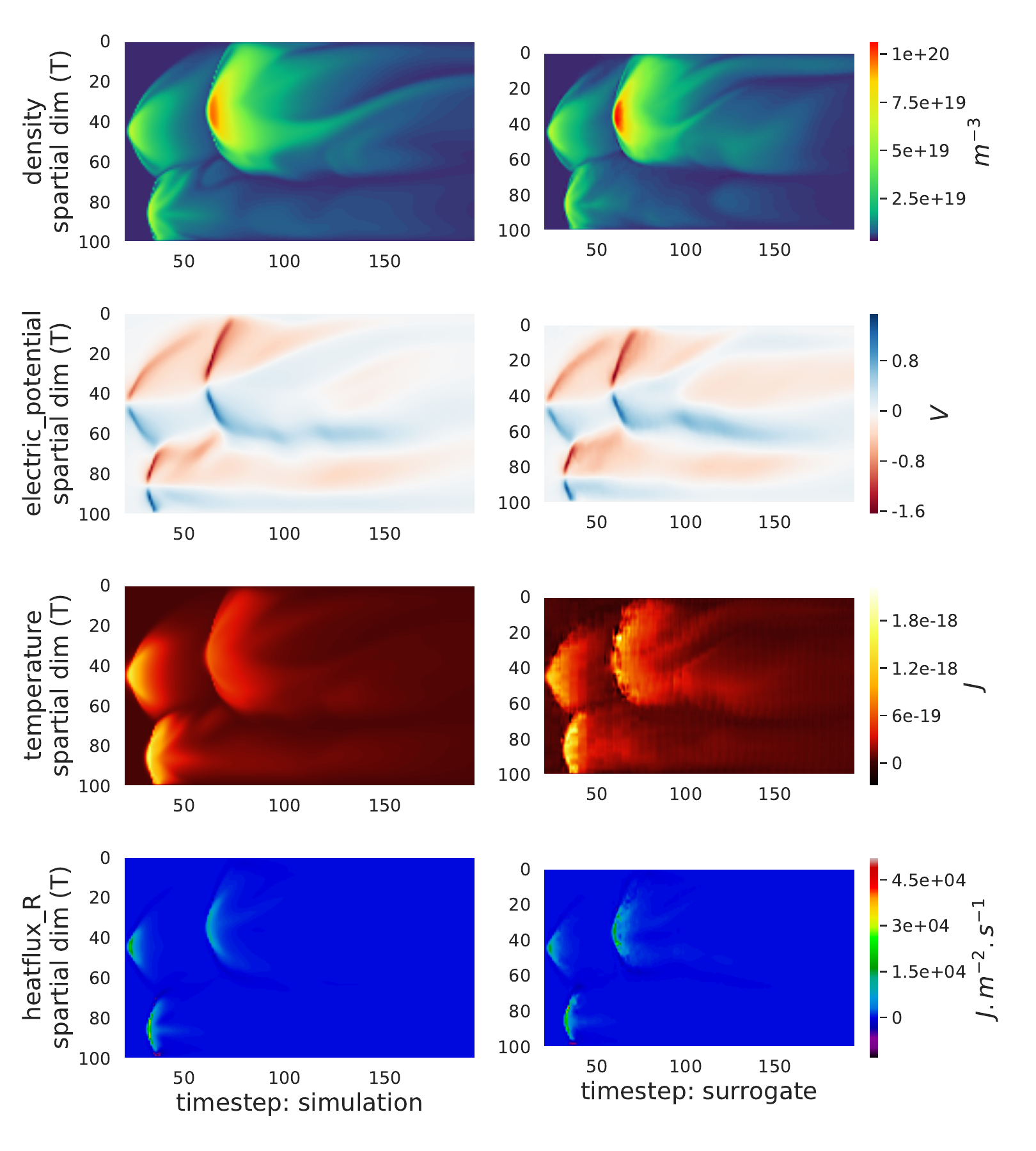}
    }
    \begin{flushright}
    \subfigure[Resulting summed heatflux at boundary over time]{
        \includegraphics[width=0.54\textwidth]{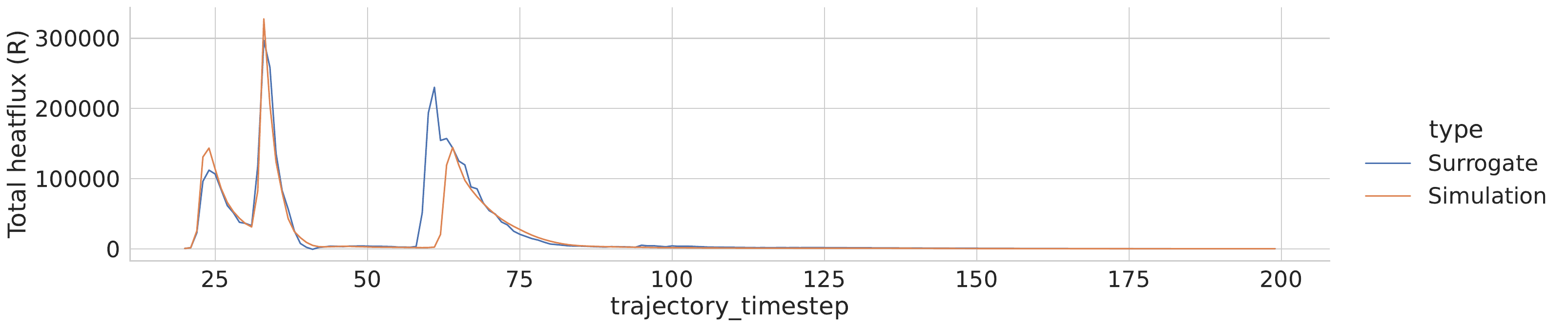}
    }
    \end{flushright}
    \caption{Comparison of heat flux at the right-hand side boundary for electrostatic JOREK during a rollout with a low total heat flux and blob count of 5. (a) shows the initial plasma condition, (b) depicts the temporal evolution of density, temperature, electric potential, and heat flux along the boundary, and (c) presents the summed heat flux over time, highlighting strong agreement in the timing of blob impacts that cause heat flux spikes.}
    \label{fig:heatflux_examples_ejorek_1}
\end{figure}

\begin{figure}
    \centering
    \subfigure[Initial plasma condition]{
        \includegraphics[width=0.35\textwidth]{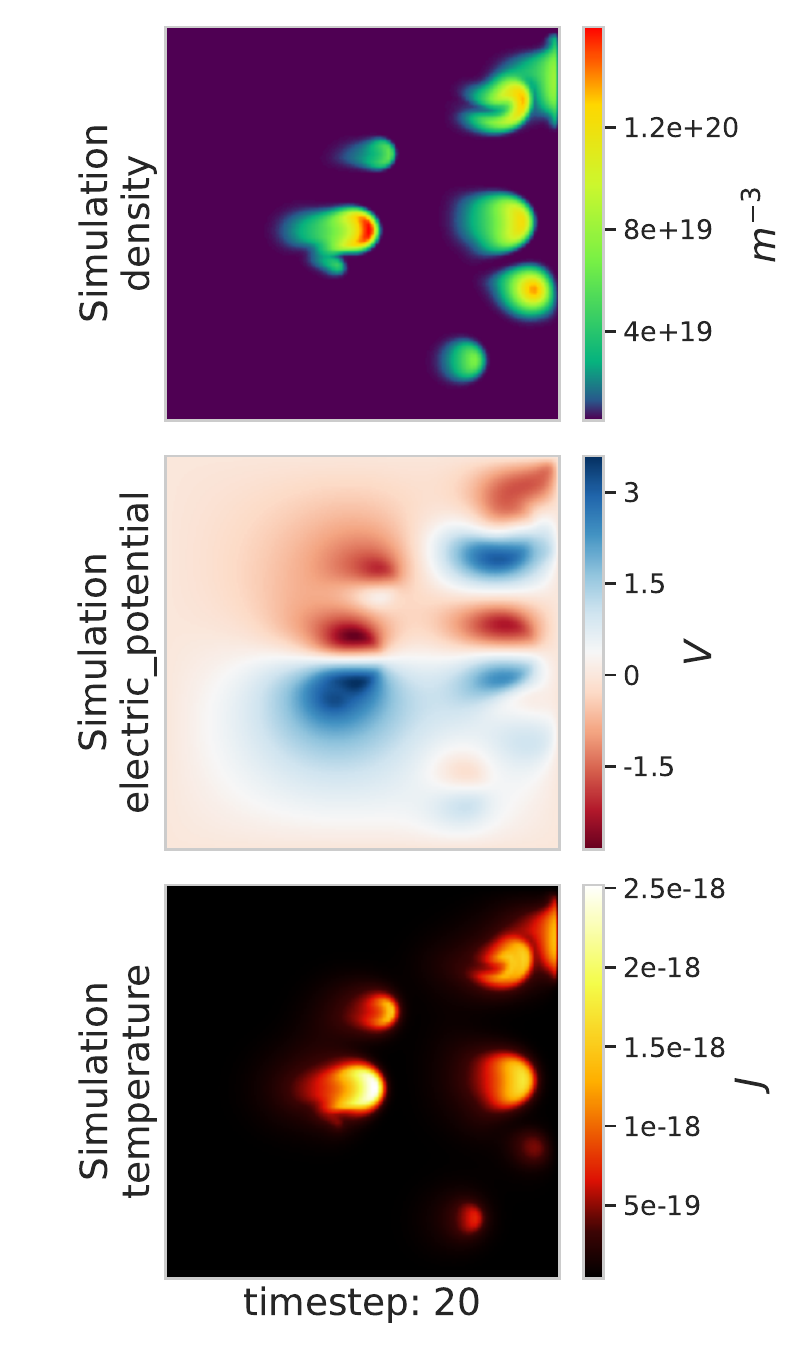}
    }
    \subfigure[Temporal evolution of plasma at RHS boundary]{
    \includegraphics[width=0.54\textwidth]{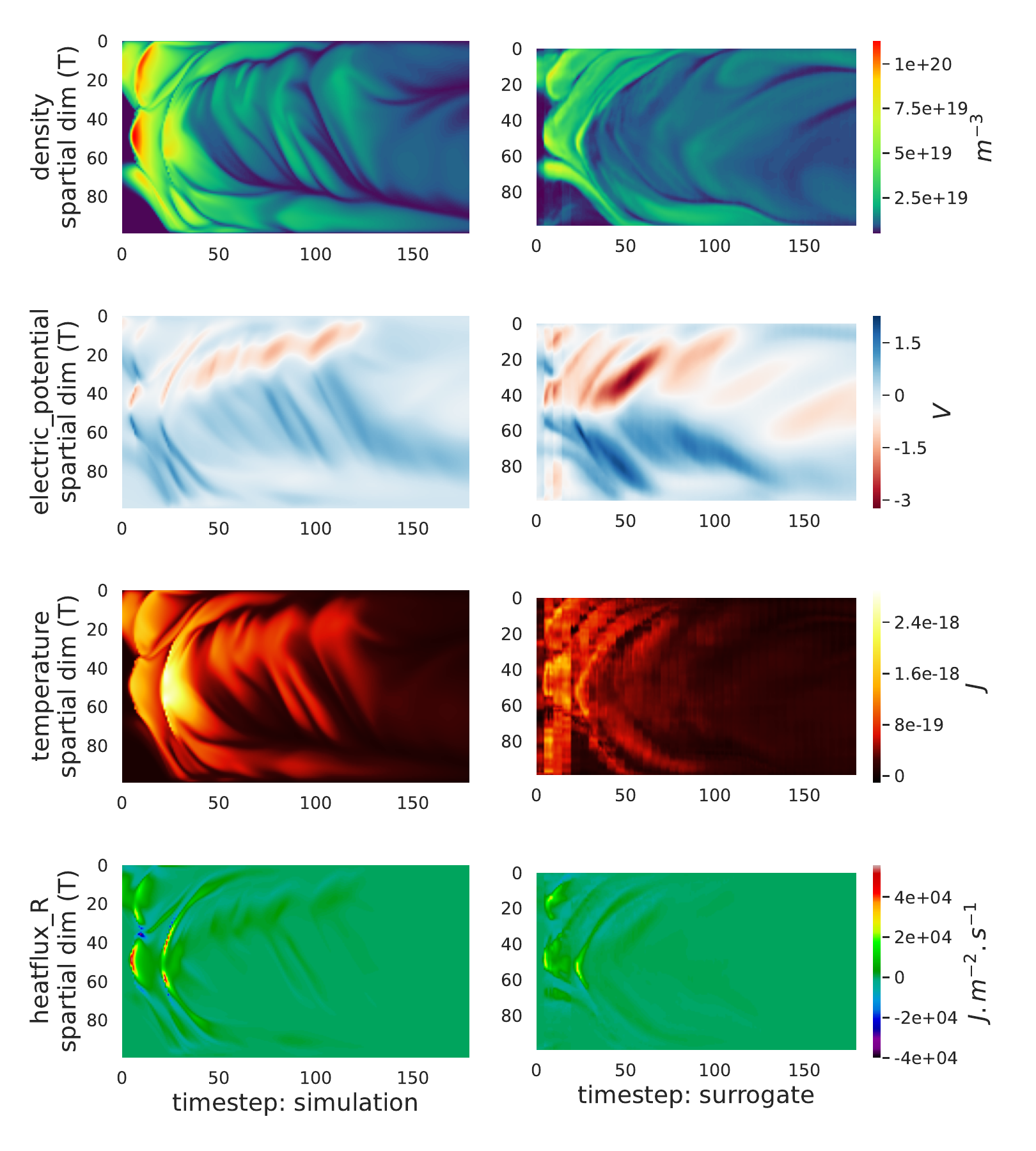}
    }
    \begin{flushright}
    \subfigure[Resulting summed heatflux at boundary over time]{
        \includegraphics[width=0.56\textwidth]{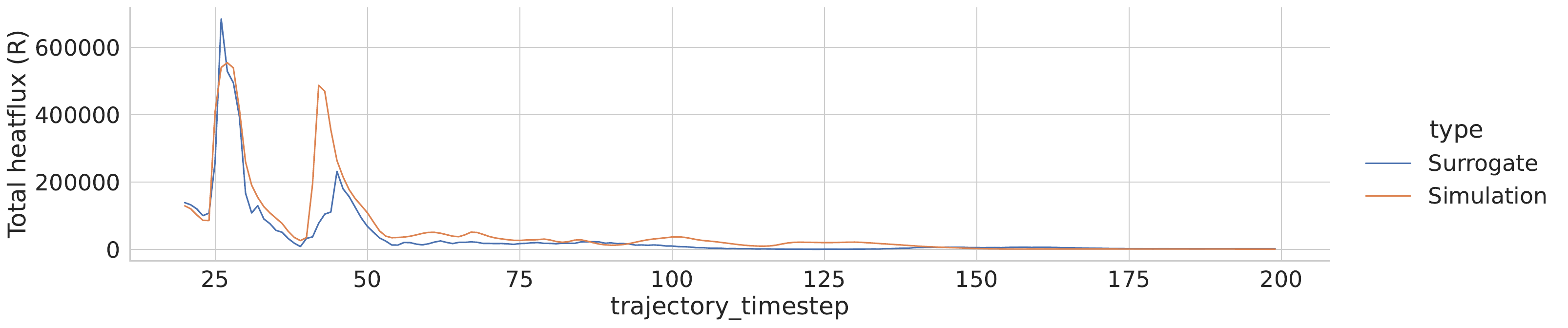}
    }
    \end{flushright}
    \caption{Comparison of heat flux at the right-hand side boundary for electrostatic JOREK during a rollout with high total heat flux and a blob count of 8. This trajectory corresponds to one of the largest heat flux prediction errors in \ref{fig:heatflux_scatter_ejorek}. (a) shows the initial plasma condition, (b) depicts the temporal evolution of density, temperature, electric potential, and heat flux along the boundary, and (c) presents the summed heat flux over time, highlighting strong agreement in the timing of blob impacts that cause heat flux spikes.}
    \label{fig:heatflux_examples_ejorek_2}
\end{figure}

To illustrate these findings, two example trajectories are shown in \ref{fig:heatflux_examples_ejorek_1} and \ref{fig:heatflux_examples_ejorek_2}: one where the integrated heat flux over the wall is low (figure \ref{fig:heatflux_examples_ejorek_1}) and another where it is high (figure \ref{fig:heatflux_examples_ejorek_2}). The integration over the poloidal boundary provides a physically meaningful measure of the total heat load on the wall, helping to assess the model’s predictive accuracy. In the first case, the model correctly captures both the timing and magnitude of heat flux spikes as blobs impact the wall. In contrast, the second example highlights a case where, while the model correctly predicts the timing of two spikes, it consistently underestimates their magnitude.

Example trajectories against time and space are provided for each of the relevant physical quantities trained on. A visual artefact is also apparent in these examples - most notably in the temperature field for figure \ref{fig:heatflux_examples_ejorek_1} and \ref{fig:heatflux_examples_ejorek_2} - where banding or pixelation can be seen along the time axis. This is not due to spatial down-sampling or this specific experiment, but rather results from the model's chunked rollout strategy. Predictions are generated in blocks of five timesteps, each using the previous 20 steps as input. Discontinuities at the boundaries between these blocks lead to stepwise patterns that become visible in time-resolved plots. This behaviour is consistent with the periodic error spikes seen in Figure \ref{fig:traj_loc_impact:diff_starts_ejorek2}(a).

These results reinforce that while the surrogate effectively models typical heat flux behavior, it struggles with high-flux events — likely due to an inadequate learned representation of extreme plasma dynamics. Importantly, the model was not trained to predict heat flux directly; rather, it was assessed on whether such physically meaningful quantities could be reconstructed from its field predictions. This design choice allows for evaluating the surrogate's ability to generalize to downstream physical quantities without explicitly including them in the training objective.

Although incorporating additional terms into the loss function to directly penalize discrepancies in quantities like heat flux may improve fidelity, doing so for all potentially relevant diagnostics - such as particle flux - would result in a highly task-specific and potentially over-constrained training process. Instead, this analysis (motivated from the successes of foundation models for text and image) serves as a test of whether the surrogate captures the underlying physics sufficiently well to enable accurate post hoc computation of derived quantities. However, the results presented here highlight the limitations of such an approach and future work should explore hybrid loss formulations to better preserve physically coupled behaviors.

\subsection{Investigation into error spikes in JOREK datasets} 

In both the Electrostatic and reduced-MHD JOREK datasets, significant error spikes were observed at specific points in the simulation trajectory, independent of accumulated input error. These peaks occurred at around $t=50$ for the electrostatic JOREK model and $t=30$ for the reduced-MHD JOREK model. Notably, trajectories across both JOREK datasets exhibited consistent physical behaviors: blobs were initialized at random positions, traveled to the right, interacted with the boundary, and eventually dissipated. These consistent dynamics suggest that specific physical events occurring at certain points in the trajectory, rather than a gradual accumulation of inaccuracies over time, are likely to be the dominant contributors to the observed rollout error.

\begin{figure}[h!]
    \centering
    \includegraphics[width=0.85\textwidth]{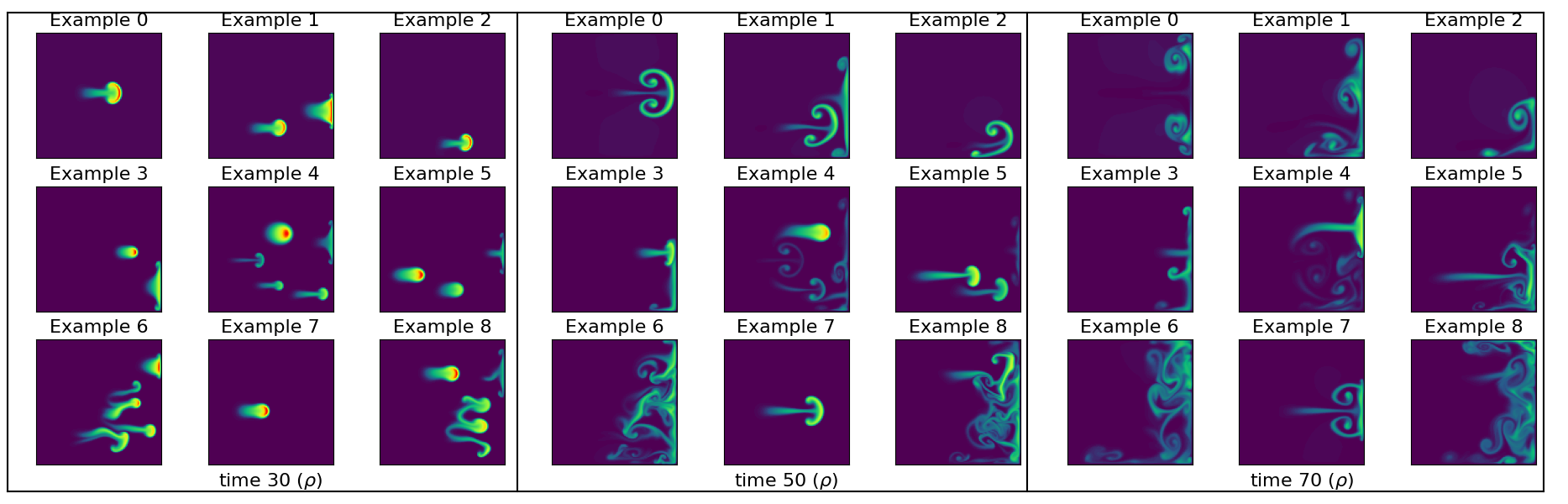} 
    \caption{FNO model trained electrostatic JOREK rollout examples at different points in time for the (rescaled) density field; demonstrating that the blobs tended to hit the wall around $t=50$.}
    \label{fig:boundary_cond:snapshots_ejorek}
\end{figure}

\begin{figure}[h!]
    \centering
    \includegraphics[width=0.9\textwidth]{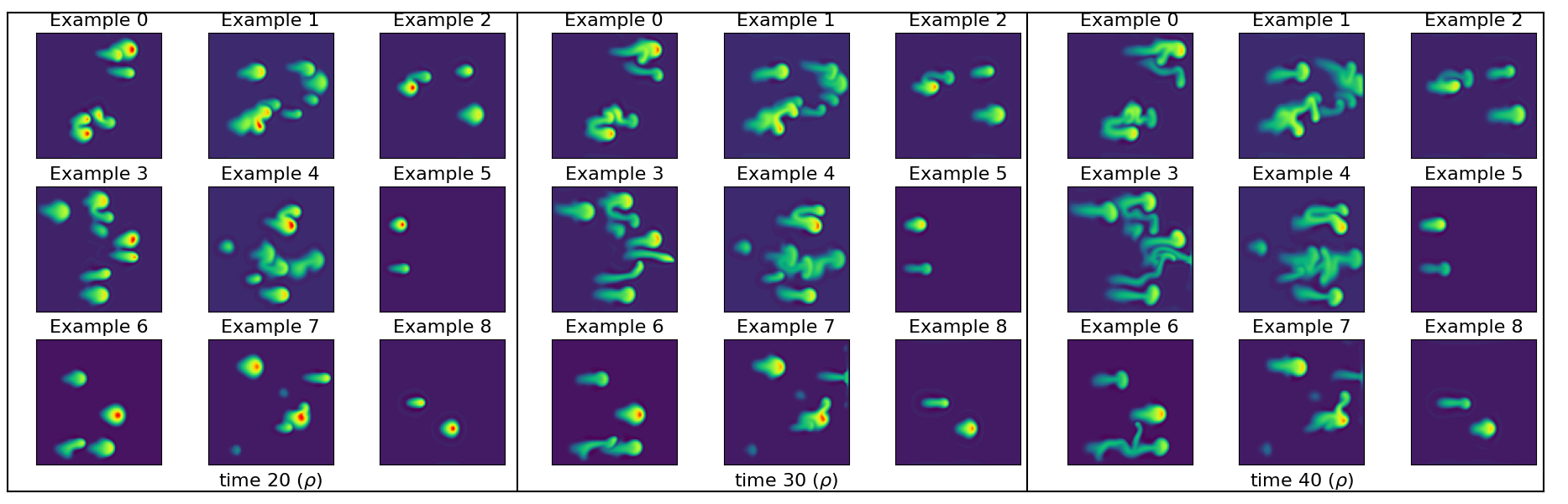}    
    \caption{FNO model trained reduced-MHD JOREK rollout examples at different points in time for the density field; demonstrating that the blobs did not hit the around $t=30$.}
    \label{fig:boundary_cond:snapshots_mjorek}
\end{figure}

This finding is notable because it suggests, somewhat counterintuitively, that input error accumulation is not always the primary driver of performance issues in these datasets for long rollouts. Instead, the errors can also be driven by specific transitions in physical behavior. Many studies on neural operators focus on understanding and improving long-time rollouts accuracy, but this finding highlights that the usual behavior — where errors accumulate steadily over time — is not observed uniformly across all variables. In this case, error accumulation appears significant only for specific variables, such as the electric potential $\phi$ for elecrostatic JOREK, underscoring the importance of analyzing errors in the context of the underlying physical dynamics.

When looking at specific examples, it can be noticed that the error peaked at $t=50$, aligning with blob-wall collisions (fig \ref{fig:boundary_cond:snapshots_ejorek}). However, this pattern was absent in reduced-MHD JOREK, where blobs took longer to reach the boundary and dissipated beforehand (fig \ref{fig:initial_results:example_mjorek_traj}).

Various factors were systematically ruled out, with all being explored for electrostatic JOREK and, where possible, for reduced-MHD JOREK. Analysis of pointwise error for electrostatic JOREK (see \ref{sec:pointwise_error_jorek}) revealed distinct spatial patterns, with errors predominantly concentrated near boundaries, particularly the right-hand side wall during early timesteps.), before spreading to other boundaries over time. This suggests that boundary interactions present a challenge for the neural operator, which lacks explicit boundary condition enforcement beyond observed edge behavior in the dataset.  However for reduced-MHD JOREK, the error was not concentrated near the boundary, suggesting additional contributing factors.

Resampling around the error peaks (see \ref{sec:sampling_ejorek}, not possible for reduced MHD-JOREK due to its fragmented nature) to improve representation of these dynamics did not enhance performance, suggesting that the issue is not due to under-representation in the training dataset. Furthermore, while the magnitude of error increased with the number of blobs in a trajectory (see \ref{sec:split_blob_num_ejorek} for figures, reduced-MHD JOREK validation dataset was too small with only 2 samples per blob number on average), the timing and frequency of error spikes were not correlated with blob count. 

Furthermore, the lower MSE observed later in the rollout might initially appear to be a consequence of decreasing overall solution magnitudes. This interpretation was further examined using normalized error metrics. Specifically, both pointwise and timewise normalized errors were computed to account for changes in scale and smoothness of the solution over time (see \ref{sec:nmse_jorek}). These metrics normalize the prediction error by scaling up the error relative to either the pointwise target value or the spatially averaged target at each time step. Importantly, the observed peaks and trends in error remained consistent under these alternative metrics, confirming that the surrogate performance is influenced not solely by solution magnitude.

Another possible explanation is that the neural operator struggles to capture the rapid acceleration of blobs. This process involves a quick increase in velocity, peaking before the blobs collide with the wall, followed by a fast deceleration during collision and eventual dissipation. Such abrupt transitions in dynamics may be difficult for the FNO to resolve, especially in localized regions. This is consistent with findings in \cite{lippe2023pderefiner}, which suggest that these models do not effectively capture fine-scale spatial features. Whilst increasing the input buffer size (which is already reasonably large at $t_{in}=20$) for rollout predictions could potentially improve accuracy by capturing more information about acceleration, it would also introduce inefficiencies in downstream applications, where many timesteps need to be simulated before passing the data to the neural operator. The objective employed may also contribute to the observed error patterns. Mean squared error (MSE), while effective for tracking precise blob positions and localized features, may inadequately capture the more turbulent or diffusive phases where the properties that are more interesting are statistical properties like average heat flux. 

Overall, the dynamics behind the observed error spikes remain unclear but these findings highlight the importance of considering both the physical dynamics of the system and the limitations of the model when analyzing long-term predictions.

\subsection{Transfer learning}

In this section the results on transfer learning are presented. The methodology is explained in sections \ref{sec:transfer_learning} and \ref{sec:transfer_learning2}. Table \ref{table:transfer_code_cases} provides an overview of the transfer learning case studies, detailing the source and target datasets and variables for each of the cases.

\begin{table}[h!]
\centering
\begin{adjustbox}{width=\textwidth}
\footnotesize
\begin{tabular}{@{}l|ll|ll@{}}
\toprule
 & \multicolumn{2}{c|}{Source} & \multicolumn{2}{c}{Target} \\
Case & Dataset & Variables & Dataset & Variables \\ \midrule
1    & Electrostatic JOREK & density, electric potential & STORM & density, electric potential               \\
2    & STORM & density, electric potential               & Electrostatic JOREK & density, electric potential \\
3    & Electrostatic JOREK & density, electric potential & Reduced-MHD JOREK & density, electric potential  \\
4    & Electrostatic JOREK & density, electric potential & Reduced-MHD JOREK & temperature, current  \\ \bottomrule
\end{tabular}
\end{adjustbox}
\caption{The different transfer learning cases investigated for transfer learning. EJOREK is electrostatic JOREK and MJOREK is Reduced-MHD JOREK}
\label{table:transfer_code_cases}
\end{table}

\subsubsection{Cross-Simulation-Code Transfer learning} \label{sec:cross_code_transfer}

For transfer learning between independently developed simulation codes, the adaptability of neural operator surrogates was tested using data from the STORM and JOREK models. These codes simulate distinct physical phenomena and operate under different frameworks, providing a test for cross-simulation-code transfer. The goal was to assess whether models trained on one dataset could effectively transfer knowledge to another, highlighting the potential for cross-simulation-code surrogate modelling in fusion research.

\begin{figure}[h!]
    \centering
    \subfigure[Case 1 Transfer learning from STORM to electrostatic JOREK]{
        \includegraphics[width=0.9\textwidth]{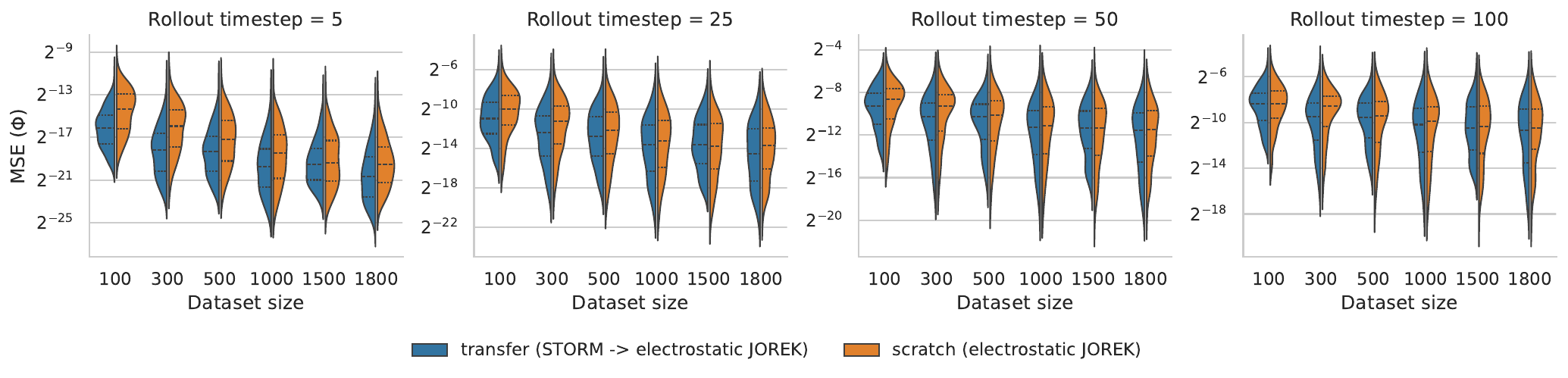}
    }
    \subfigure[Case 2 Transfer learning from electrostatic JOREK to STORM]{
        \includegraphics[width=0.9\textwidth]{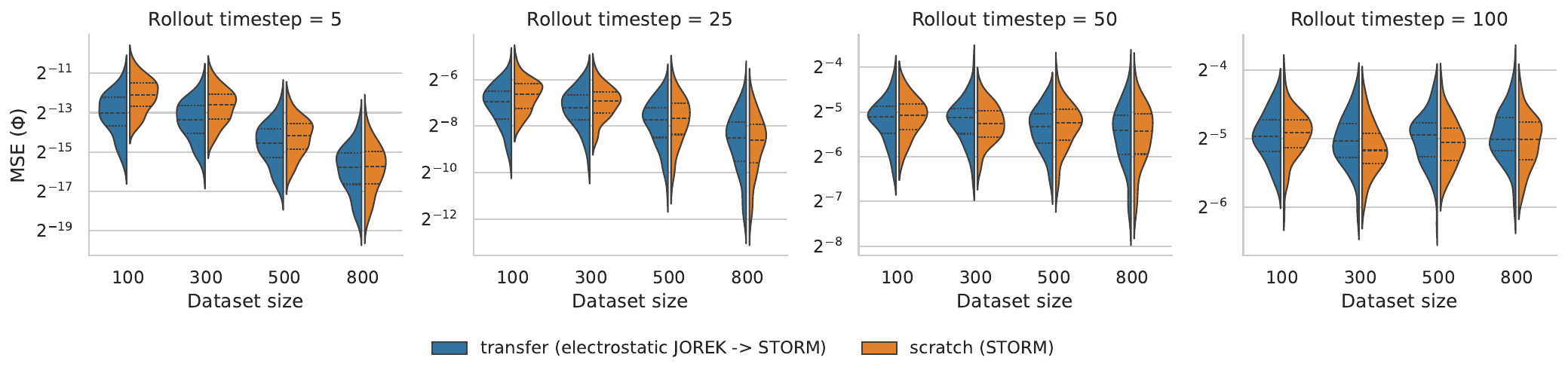}
    }
    \caption{Scratch and transfer model error for electric potential at different timesteps and different dataset sizes. The line is the medium error and the error bars correspond to the 16th and 84th percentile.}
    \label{fig:transfer_dataset_size_electric_potential}
\end{figure}

The details of the two experimental cases (cases 1 and 2) are summarized in table \ref{table:transfer_code_cases}. In both scenarios, the MSE was analyzed as a function of dataset size and rollout length. Errors were measured at specific timesteps (e.g., 5, 25, 50, and 100) to assess performance trends over time.  As illustrated in figure \ref{fig:transfer_dataset_size_electric_potential} for electric potential, transfer learning models demonstrated improved performance compared to scratch models for shorter rollouts, particularly when smaller training datasets were used.

However, for longer rollouts, the performance of the transfer learning models progressively converged to similar results with that of the scratch models. Increasing the dataset size improved the performance of both approaches, though this trend became less stable for longer rollouts. A similar pattern was observed across varying dataset sizes: as the dataset size increased, the performance gap between transfer learning and scratch models narrowed. This indicates that transfer learning’s advantage diminishes for extended rollouts.

These findings underscore the need for more robust transfer learning strategies capable of improving long-term rollout accuracy.

\subsubsection{Transfer Across Simulation Physics-Fidelity Levels in JOREK}

The third scenario explored transfer learning between different fidelity levels within the JOREK simulation code, moving from low-fidelity-physics to high-fidelity-physics datasets. The objective was to assess whether transfer learning could enable the FNO to leverage lower-fidelity data to enhance performance on higher-fidelity datasets. This approach aimed to reduce the need for generating large high-fidelity datasets, thereby mitigating the computational burden associated with high-fidelity modeling.

The sole case in Table \ref{table:transfer_code_cases} represents this fidelity transfer experiment. Similar to the cross-simulation-code analysis, the MSE error was evaluated against dataset size and rollout length. Compared to the cross-simulation-code results, transfer learning between fidelities yielded more substantial improvements in accuracy for short rollouts, with the improvements diminishing more gradually. For example, in the density field at the smallest dataset size of 3797 slices, transfer learning achieved an error reduction of an order of magnitude for a short rollout (timestep 5), decreasing the MSE from $5.74\times10^{-4}$ to $7.62\times10^{-5}$. For a longer rollout (timestep 100), the error reduced by a factor of 2 with the MSE decreasing from $1.13\times10^{-3}$ to $5.80\times10^{-4}$. However, this advantage still disappeared for longer rollouts.

The findings demonstrate that transfer learning is most effective when the source and target datasets share considerable similarities. This suggests the potential for a practical pathway for accelerating model training by leveraging cheaper, low-fidelity simulations as a foundation for high-fidelity model development. With this strategy, the computational costs associated with generating large high-fidelity simulation datasets could be drastically reduced.

\begin{figure}
    \centering
    \subfigure[Density]{
    \includegraphics[width=1\textwidth]{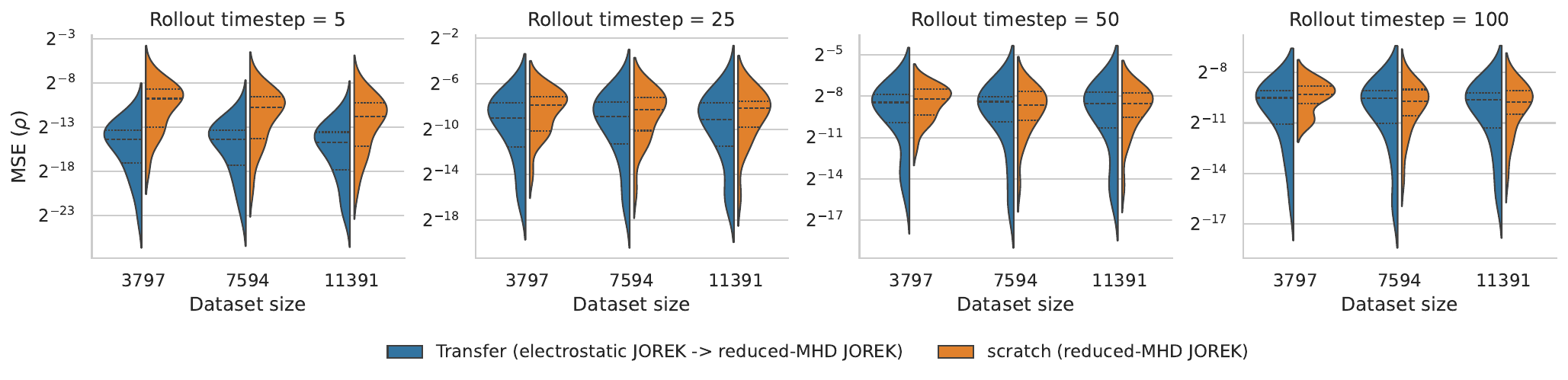}
    }
    \subfigure[Electric potential]{
    \includegraphics[width=1\textwidth]{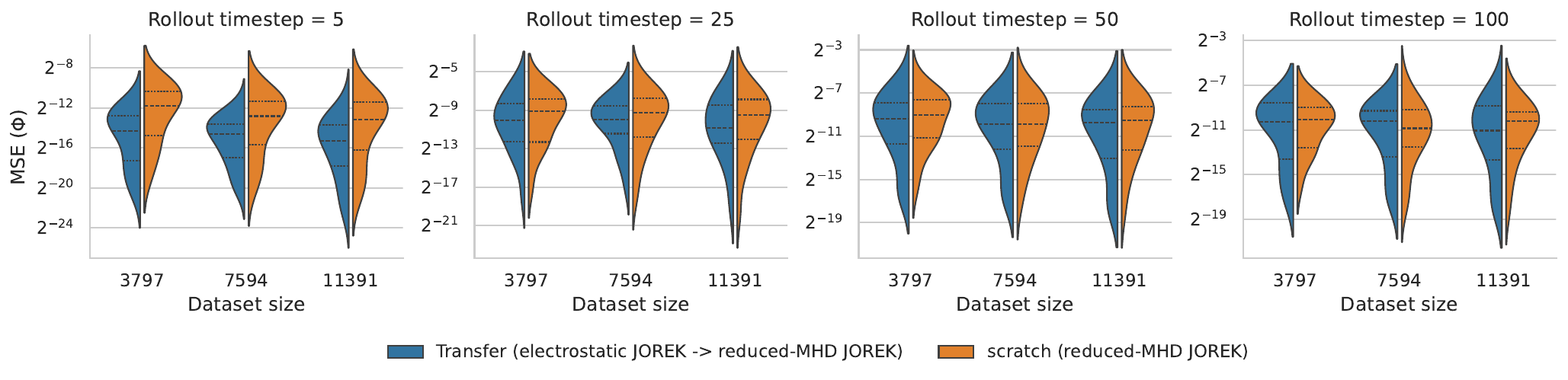}
    }
    \caption{Scratch and transfer model error at different timesteps and different dataset sizes for case 3. Scratch was trained from scratch on the reduced-MHD JOREK dataset whilst transfer model was first trained on electrostatic JOREK and then finetuned on reduced-MHD JOREK. The line is the medium error and the error bars correspond to the 16th and 84th percentile.}
    \label{fig:transfer_dataset_size_density}
\end{figure}

\subsubsection{Transfering to unseen variables for higher fidelity}

A challenge encountered when transferring from low- to high-fidelity JOREK simulations was that the target dataset (reduced-MHD JOREK) contained variables absent in the source dataset (electrostatic JOREK) such as current. Neural operators like the Fourier Neural Operator (FNO) lack the flexibility to seamlessly incorporate prior knowledge from learned variables into datasets with new variables. To address this, a two-step transfer learning approach was devised, as detailed in section \ref{sec:additional_details}. This method aims to allow the model to transfer learnt information to previously unseen variables in the target datasets previously unseen variables to evaluate applicability as demonstrated in this study using the reduced-MHD JOREK dataset.

\begin{figure}
    \centering
    \subfigure[Temperature]{
        \includegraphics[width=1\textwidth]{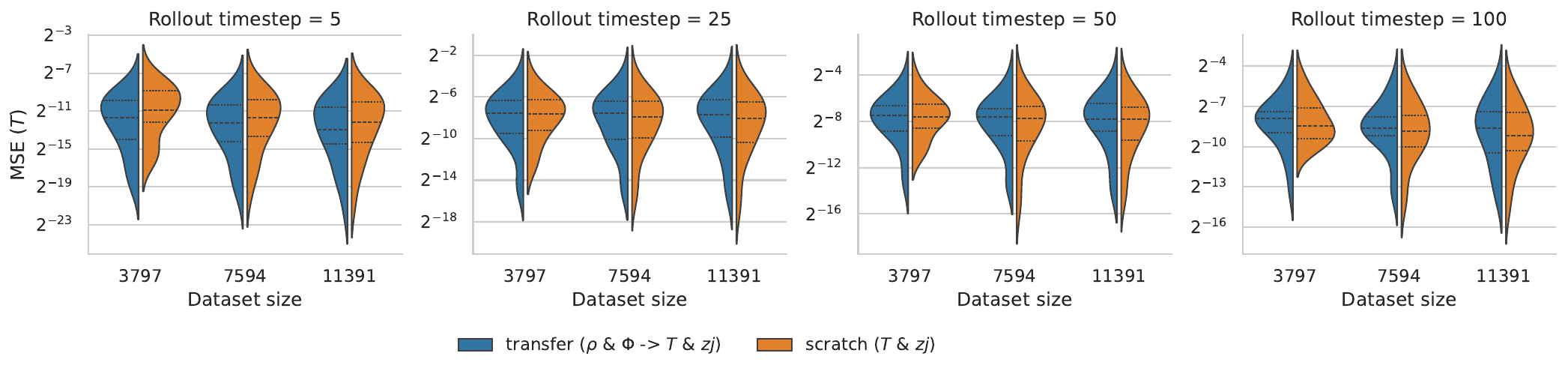} 
    }
    \subfigure[Current]{
        \includegraphics[width=1\textwidth]{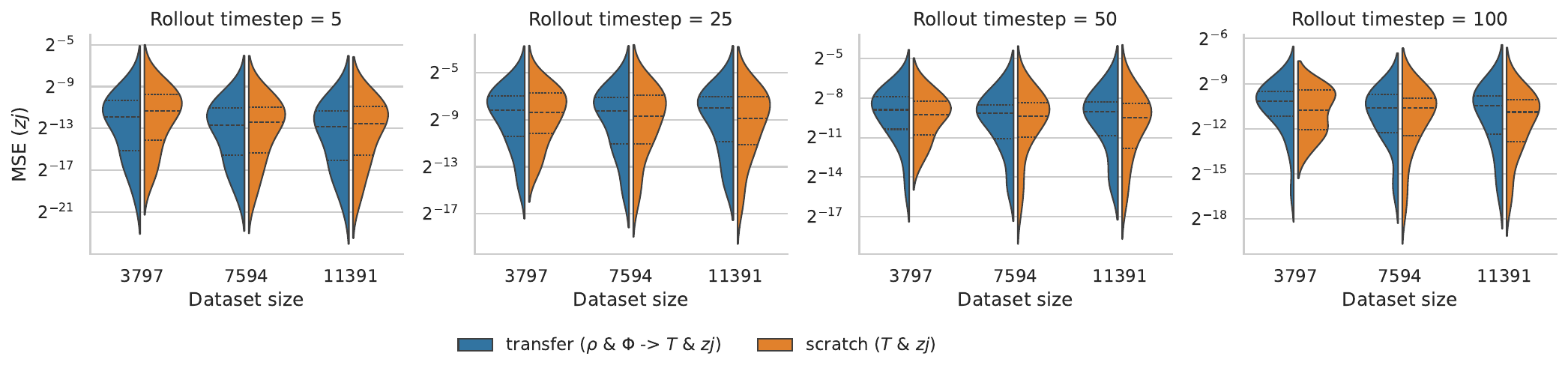} 
    }
    \caption{Scratch and transfer model error at different timesteps and dataset sizes on reduced-MHD JOREK. Scratch model was trained directly on the corresponding variables on the dataset. Transfer x2 model was first trained on electrostatic JOREK density and electric potential, finetuned on reduced-MHD JOREK density and electric potential (similar to prior section) and the transferred to reduced-MHD JOREK temperature and current. The line is the medium error and the error bars correspond to the 16th and 84th percentile.}
    \label{fig:transfer_dataset_size_diff_var}
\end{figure}

The same error metrics were produced, as illustrated in figure \ref{fig:transfer_dataset_size_diff_var}. Consistent with previous findings, the transfer models generally outperformed the scratch models during shorter rollouts. However, relative to the previous findings, the performance improvement diminished more rapidly for longer rollouts. In addition, the scratch model consistently demonstrated better average performance than the transfer model in longer rollouts for the temperature field. This pattern suggests that transferring knowledge between variables might improve short-term performance but could also lead to negative transfer effects \cite{Zhang_2023}. This result highlights the need for careful adaptation of transfer learning when working with datasets that involve previously unseen variables.

\section{Conclusion}

\subsection{Key Observations}

This work demonstrates the potential of neural operators as surrogate models for simulating the evolution of variables in tokamak plasma systems. Fourier Neural Operators were trained using the STORM and JOREK simulation datasets and the surrogate model's ability to replicate plasma behaviour autoregressively was evaluated for both short and long temporal rollouts.

Key observations and contributions include the following:
\begin{enumerate}
    \item \textbf{Impact of Starting Points, Boundary Interactions and long-term degradation}: the FNO model's error patterns are influenced by both accumulating input errors and specific trajectory events, with boundary interactions playing a significant role in the errors observed in the Electrostatic JOREK model, highlighting the importance of considering underlying physical dynamics when evaluating model performance.   
    \item \textbf{Transfer Learning for Dataset Efficiency}: Given the high computational cost of generating simulation datasets, transfer learning techniques were explored to leverage information across similar and more complex datasets. Results indicate that transfer learning offers a notable improvement when the source and target datasets share physical similarities, such as in cases involving  electrostatic and reduced-MHD JOREK datasets. Findings suggest that transfer learning from lower- to higher-fidelity JOREK simulations can be very effective, even in longer rollouts, helping to reduce the computational resources and expert human effort required for generating large databases of high-fidelity simulations. However, the benefits of transfer learning were observed to diminish in longer rollouts, especially when the datasets differed in their underlying physics. This suggests that while transfer learning can be a powerful tool for dataset efficiency, its advantages are more pronounced in shorter rollouts and datasets with similar physical properties.
    \item \textbf{Variable-to-Variable Transfer Challenges}: The FNO architecture lacks flexibility to accommodate a varying number of variables, and transferring knowledge to datasets with new variables is not straightforward. To address this, a two-step approach was explored. While this method outperformed models trained from scratch in short rollouts, its performance deteriorated in longer rollouts relatively quickly. This suggests that transferring knowledge between distinct variables governed by different physical processes (e.g., the reduced-MHD JOREK dataset) can suffer from negative transfer effects. These challenges highlight the need for more flexible transfer learning techniques when dealing with datasets that involve varying or previously unseen variables.
\end{enumerate}

\subsection{Limitations and Future work}

While this study represents a step forward in leveraging neural operators for plasma physics simulations, there are several limitations that warrant further investigation.

The FNO and U-Net are powerful baselines, however they are now routinely outperformed by attention-based methods   \cite{alkin2024universalphysicstransformersframework}. Future work should focus on more performing architectures, which could improve long-term predictive accuracy and better handle non-overlapping variable sets. Additionally, physical constraints, such as boundary conditions, were not explicitly considered in this work. Including them in future work may help preventing errors from escalating over time. Moreover, some of the observed performance degradation could be attributed to poor representations of certain physical processes, such as blob acceleration, which bespoke surrogate model architectures could better account for.

Although transfer learning provided promising results, particularly for datasets with similar physical properties, the performance gains disappeared for long rollouts. Future research could explore more sophisticated methods such as LoRA \cite{hu2021loralowrankadaptationlarge} and domain adaptation \cite{setinek2025simshiftbenchmarkadaptingneural} to improve target domain performance. Attention mechanisms could also be incorporated to prioritize relevant features during knowledge transfer and strategies like varying learning rates or selectively freezing layers could help identify and keep the parts of the network most responsible for effective transfer.

The neural operators presented in this work were trained using mean-squared error (MSE) between normalised predicted and ground truth fields, a common choice in neural operator literature due to its numerical stability and compatibility with variables spanning different physical units. However, this loss tends to prioritise large-amplitude, low-frequency structures and may underemphasise fine-scale, low-amplitude features. Recent advancements, such as the Pushforward Trick \cite{brandstetter2023message}, PDE-Refiner \cite{lippe2023pderefiner}, and autoregressive diffusion models \cite{kohl2024benchmarkingautoregressiveconditionaldiffusion}, have demonstrated potential in enhancing the robustness of neural operators for long-term predictions. These methods use techniques such as refining predictions iteratively, correcting for low-amplitude spatial frequencies, and incorporating input errors directly into the training process as a kind of data augmentation. Such techniques could substantially improve the stability and accuracy of long rollouts. Incorporating these approaches into future work could help overcome some of the limitations observed, in particular for transfer learning. 

Applications of neural operators to more complex physics models, such as higher-dimensional, multiphysics simulations in complex geometry, is needed for AI-based methods to deliver on the promise of faster modelling, design tasks and control studies. A recent step in this direction was taken by \cite{mitsuru_multimodal_2023,galletti20255dneuralsurrogatesnonlinear} to emulate nonlinear gyrokinetic simulations in a flux tube, demonstrating that neural surrogates can, in principle, scale to higher-dimensional problems given appropriate inductive biases and architectural design. The primary bottleneck in surrogate models of simulators that are more complex than the ones presented in this work
is the generation of sufficient high-quality simulation data in complex geometry and dimension higher or equal than 3, which requires substantial domain expertise and extensive computational resources. With the advent of Exascale computing, the large scale volume of simulations needed to train surrogate models will be costly but not out of reach \cite{Jenko2025}. Realistic plasma profiles must be parametrised and several magnetic equilibria and other initial and boundary conditions specified. Due to the curse of dimensionality, it may be possible to modify only a reduced set of those. Transfer learning, which has been explored in this work, and active learning \cite{Zanisi2024}, which can be used to query only the most informative simulations, will be crucial to ensure that the cost of simulations is manageable and that suitable training sets can be gathered in a timely manner. Hybrid numerical-AI strategies such as Neural Parareal \cite{pamela2024neuralparareal} may also be adopted to this end.

In summary, while this work highlights the potential of FNO-based surrogates for plasma physics, there are clear avenues for improvement. Significant efforts should be devoted to (1) surrogate model architectural innovations and (2) data efficiency considerations. By addressing the limitations discussed and integrating recent advancements in neural network architectures and techniques, future research can enhance the reliability, scalability, and versatility of neural operators in fusion research.
% \lzcomm{AWESOME work :)}

\section{Acknowledgments}

This work has been part-funded by the EPSRC Energy Programme [grant number EP/W006839/1].  To obtain further information on the data and models underlying this paper please contact PublicationsManager@ukaea.uk. This work has also been part-funded by the Fusion Futures Programme. As announced by the UK Government in October 2023, Fusion Futures aims to provide holistic support for the development of the fusion sector. The authors thank the reviewers for their very constructive comments that substantially helped improve the the paper.

\printbibliography

\newpage

\appendix

\section{Simulation parameter space}

The range of input parameters are as follows for each of the simulation datasets in table \ref{table:jorek_dataset_input_param_range} and \ref{table:storm_dataset_input_param_range}.

\begin{table}[]
\centering
\begin{tabular}{@{}llll@{}}
\toprule
Input parameter & Min/max & Type \& unit \\ \midrule
number of blobs & [1 : 10] & discrete \\
R-position of blob & [9.6 : 10.4] & continuous [m] \\
Z-position of blobs & [-0.4 : 0.4] & continuous [m] \\
width of blobs & [0.02 : 0.1] & continuous [m] \\
density amplitude of blobs & [0.1 : 0.4] & continuous [$10^{20}m^{-3}$] \\
temperature amplitude of blobs & [12 : 72] & continuous [eV] \\ \bottomrule
\end{tabular}
\caption{The shared input parameter range used for both JOREK datasets.}
\label{table:jorek_dataset_input_param_range}
\end{table}

\begin{table}[]
\centering
\begin{tabular}{@{}lll@{}}
\toprule
Input parameter             & Min/max          & Type \& unit \\ \midrule
amplitude of density source & [0.05 : 0.5] & continuous [$10^{18}m^{-3}$]   \\
width of density source     & [1 : 10]     & continuous [1.414213562mm]  \\ \bottomrule
\end{tabular}
\caption{The input parameter range used for the STORM dataset.}
\label{table:storm_dataset_input_param_range}
\end{table}

\section{Model hyparpameters and additional training details} \label{sec:model_hyperparam}

The models were trained on all fields (unless otherwise noted) using a mean squared error (MSE) loss function, optimized over a maximum of 72 hours. The Adam optimizer, with a cosine annealing learning rate scheduler starting at $5\times10^{-4}$, $2\times10^{-4}$ and $2\times10^{-4}$ for electrostatic JOREK, reduced-MHD JOREK and STORM respectively and a minimum of 1.e-7, was used throughout. To determine an optimal learning rate, a small grid search was run for the model trained on each full dataset, halving the learning rate until performance stopped improving. Once selected, this learning rate was consistently applied across all models trained on smaller data subsets.

To enhance training stability, gradient clipping was applied with a standard threshold of $0.125$, apart from electrostatic JOREK which used a value of $0.25$.

All these hyperparameter are described in the tables below (table \ref{table:data_consthyperparam} for hyperparameters constant for all models and table \ref{table:data_diffhyperparam} for hyperparameters that are modified for each dataset) below which directly correspond to the config files used in pdearena.

\begin{table}[h!]
\centering 
\resizebox{0.5\textwidth}{!}{
\begin{tabular}{@{}ll@{}}
\toprule
Name                    & Value             \\ \midrule
Model                   & FNO               \\
FNO modes               & 32                \\
Num FNO blocks          & 3 x 1             \\
Hidden channels         & 128               \\
Activation layer        & GeLu              \\
Loss                    & MSE               \\
Batch size              & 64                \\
Optimizer               & Adam              \\
Learning rate schedular & CosineAnnealingLR \\
Weight decay            & 1e-5              \\
Eta min lr              & 1e-7              \\ \bottomrule
\end{tabular}}
\caption{Hyperparameters consistent between datasets}
\label{table:data_consthyperparam}
\end{table}

\newpage

\begin{table}[h!]
\centering 
\resizebox{0.8\textwidth}{!}{
\begin{tabular}{@{}llll@{}}
\toprule
Name                       & Electrostatic JOREK & Reduced-MHD JOREK & STORM \\ \midrule
Early stopping patience    & 10                  & 50                & 10    \\
Gradient clipping val      & 0.25                & 0.125             & 0.125 \\
Input timesteps ($t_in$)   & 20                  & 5                 & 20    \\
Output timesteps ($t_out$) & 5                   & 5                 & 5     \\
Starting Learning rate     & 5e-4                & 5e-4              & 2e-4  \\
Max epochs                 & 2000                & 2000              & 500   \\ \bottomrule
\end{tabular}}
\caption{Hyperparameters different for each datasets}
\label{table:data_diffhyperparam}
\end{table}

\section{U‑Net Experiments} \label{sec:model_unet}

\begin{figure}[h!]
    \centering
    \subfigure[Particle density $\rho$ ($m^{-3}$)]{
    \includegraphics[width=0.45\textwidth]{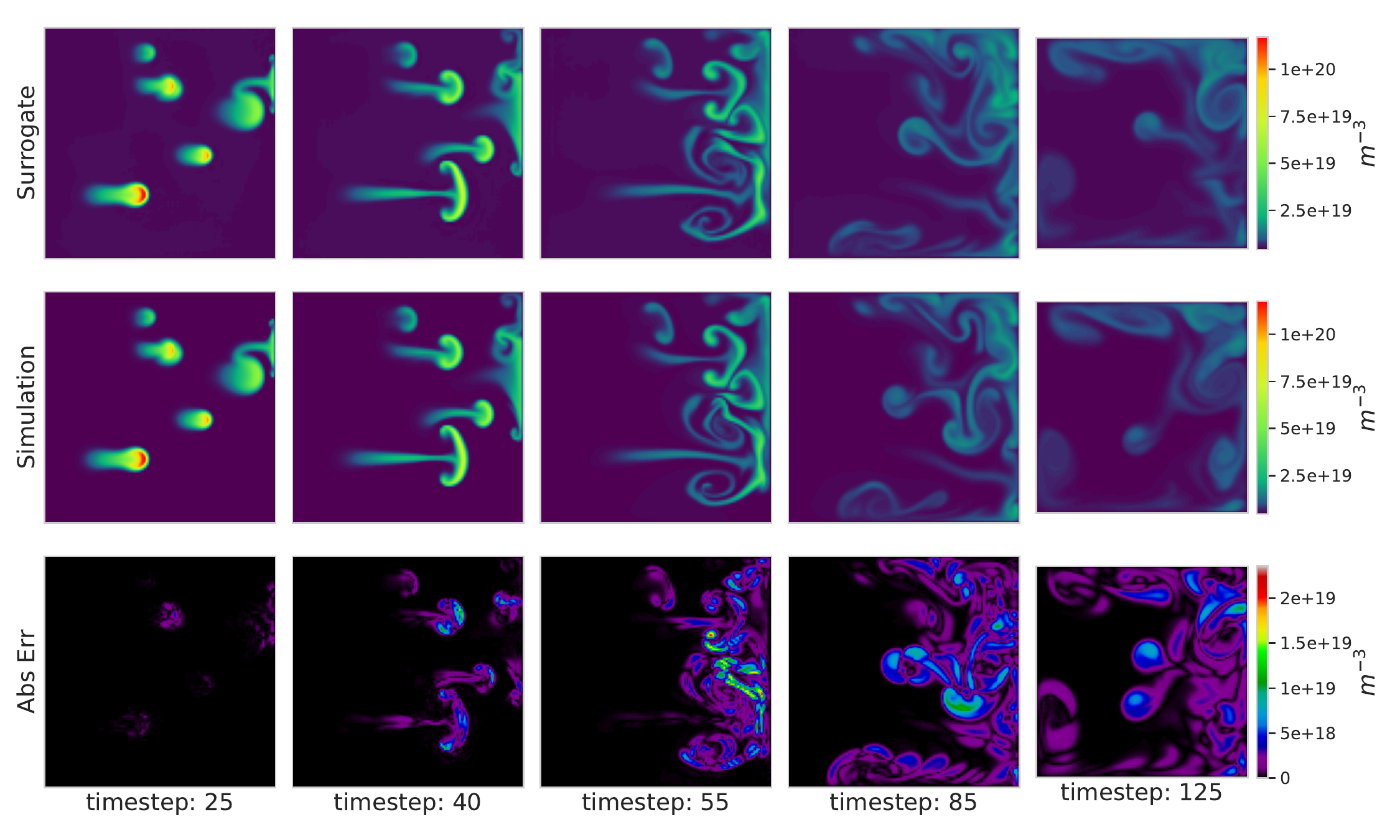}
    }
    \subfigure[Electric potential $\Phi$ ($V$)]{
    \includegraphics[width=0.45\textwidth]{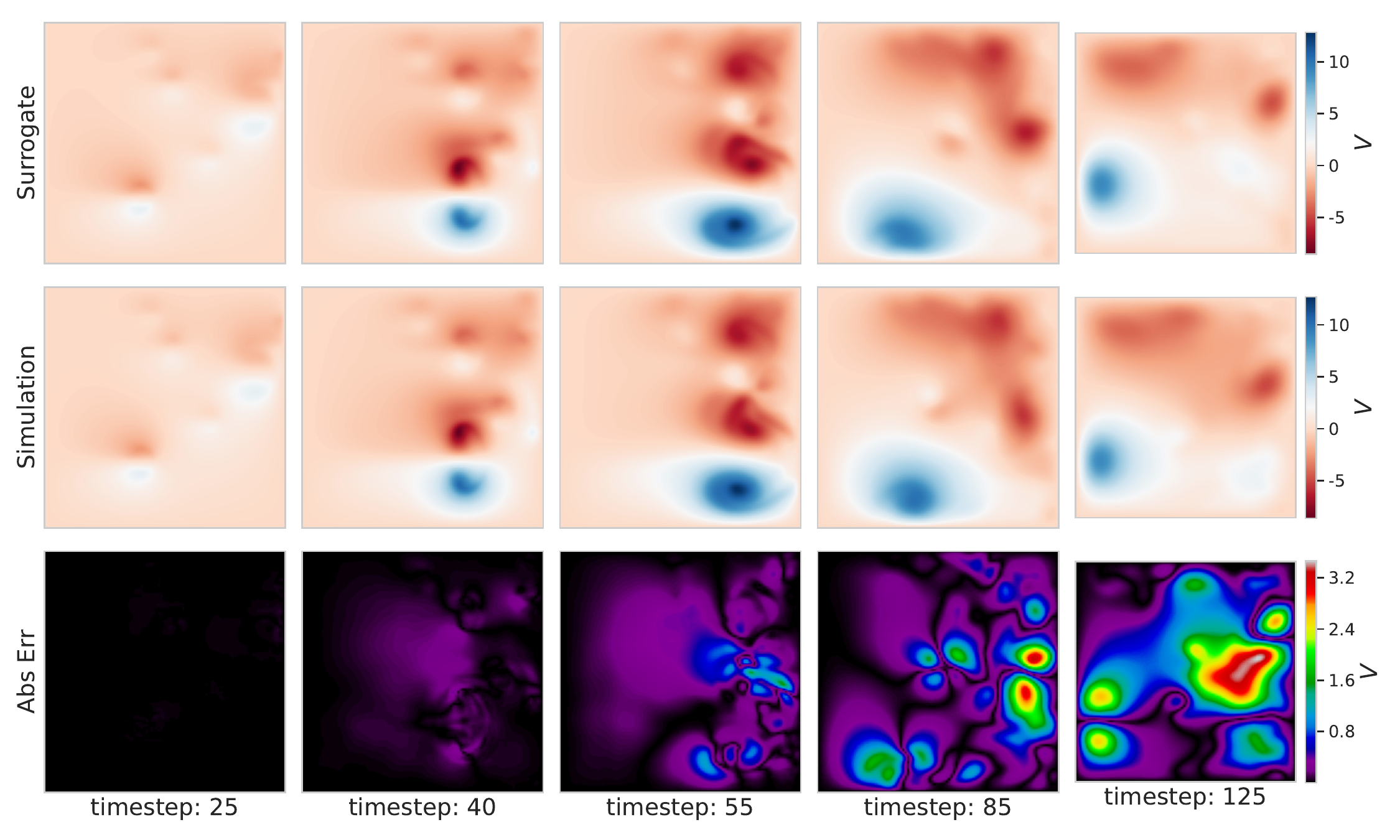}
    }
    \subfigure[Temperature $T$ ($eV$)]{
    \includegraphics[width=0.45\textwidth]{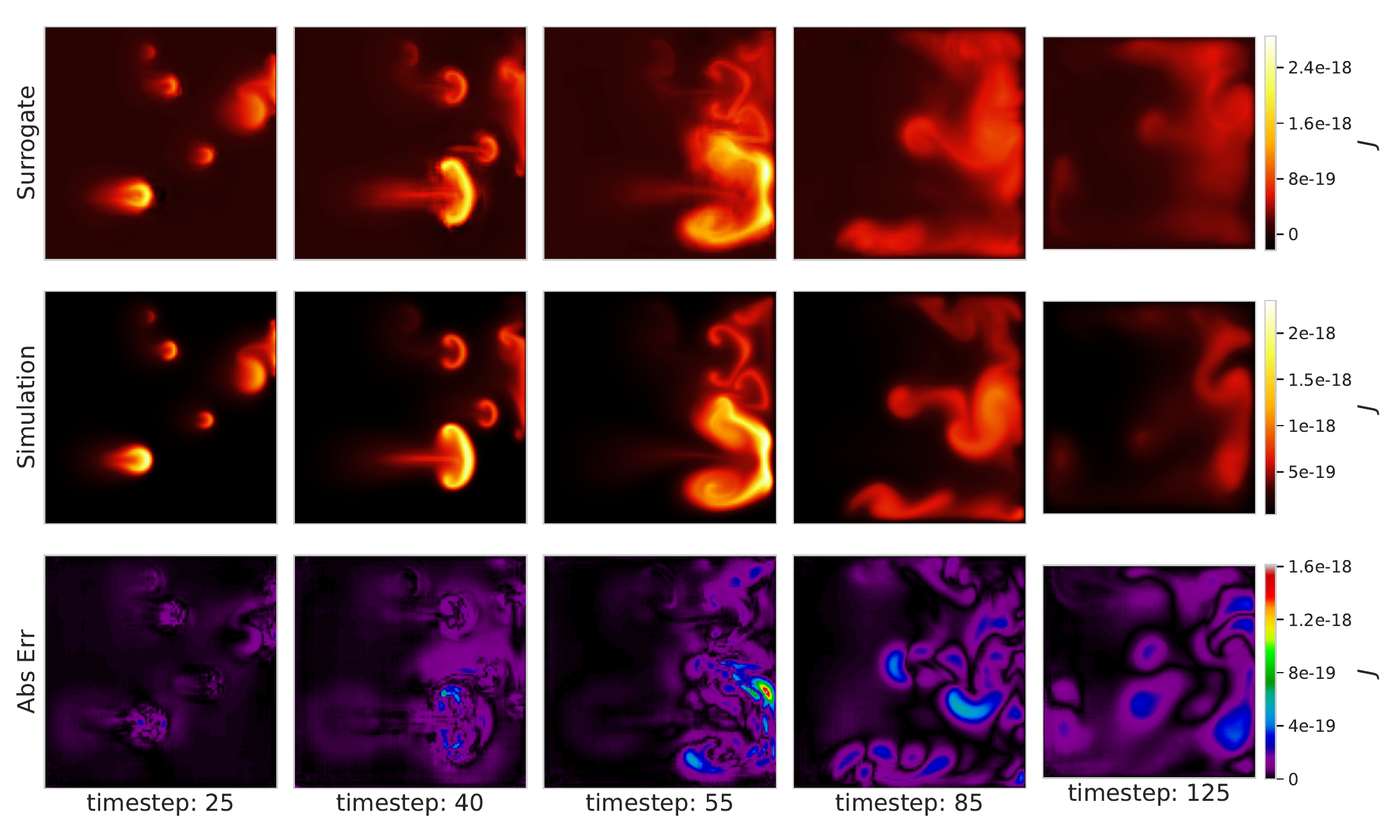}
    }
    \caption{An example electrostatic JOREK run plotted at specific rollout timesteps for fields (a) density, (b) electric potential and (c) temperature for the U-net.}
    \label{fig:initial_results:example_ejorek_traj_unet}
\end{figure}

\begin{figure}[h!]
\centering
    \subfigure[Particle density $\rho$ ($m^{-3}$)]{
    \includegraphics[width=0.45\textwidth]{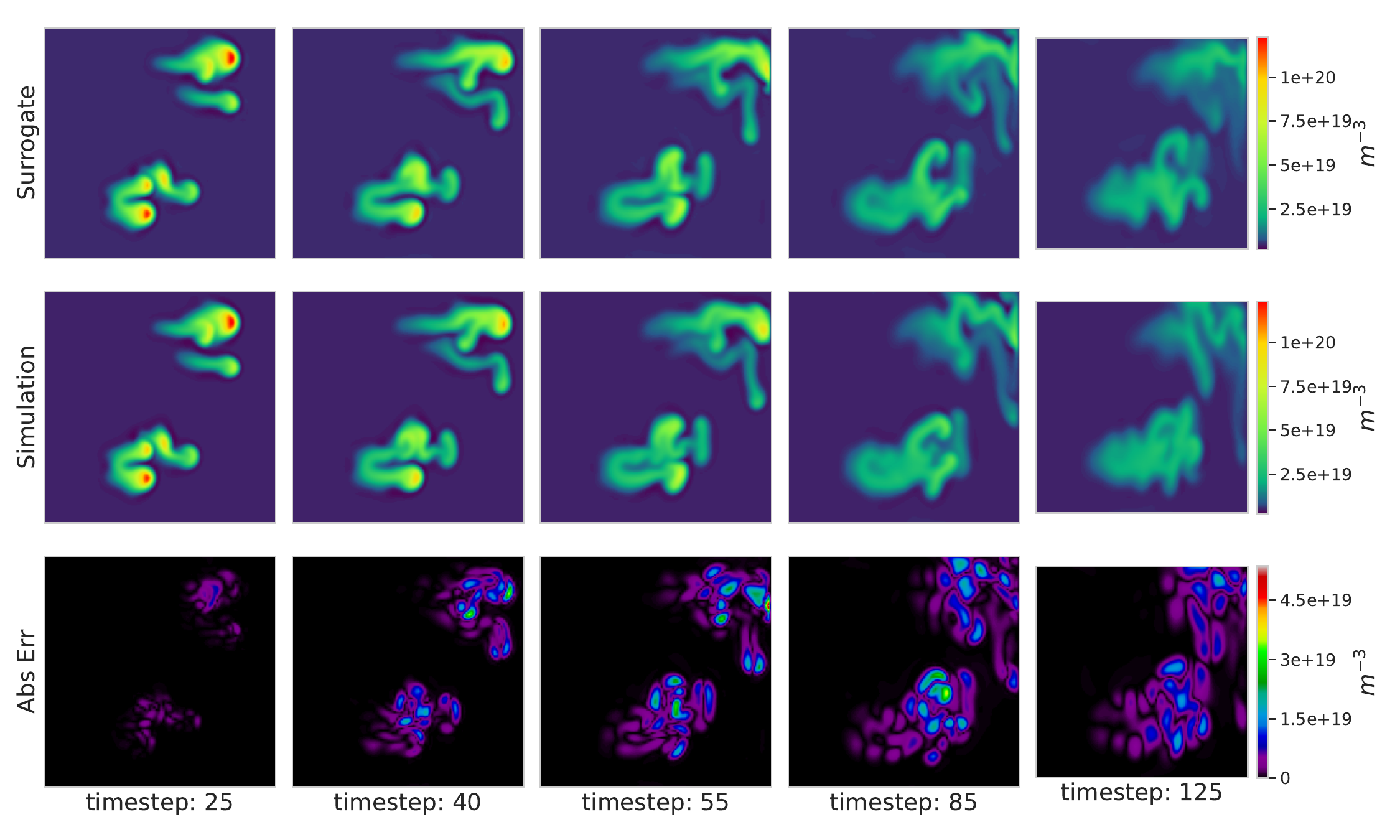}
    }
    \subfigure[\newline Electric potential $\Phi$ ($V$)]{
    \includegraphics[width=0.45\textwidth]{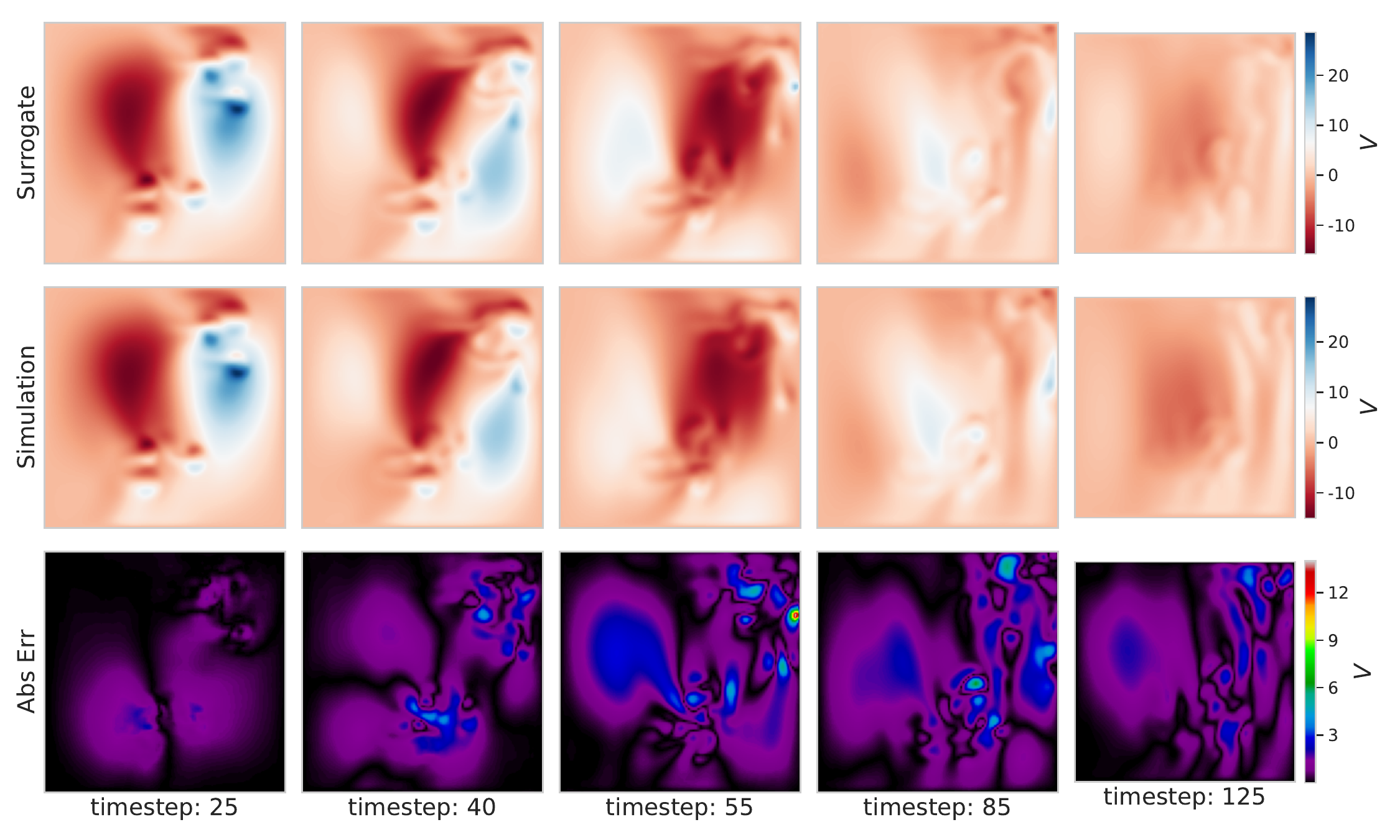}
    }
    \subfigure[Temperature $T$ ($eV$)]{
    \includegraphics[width=0.45\textwidth]{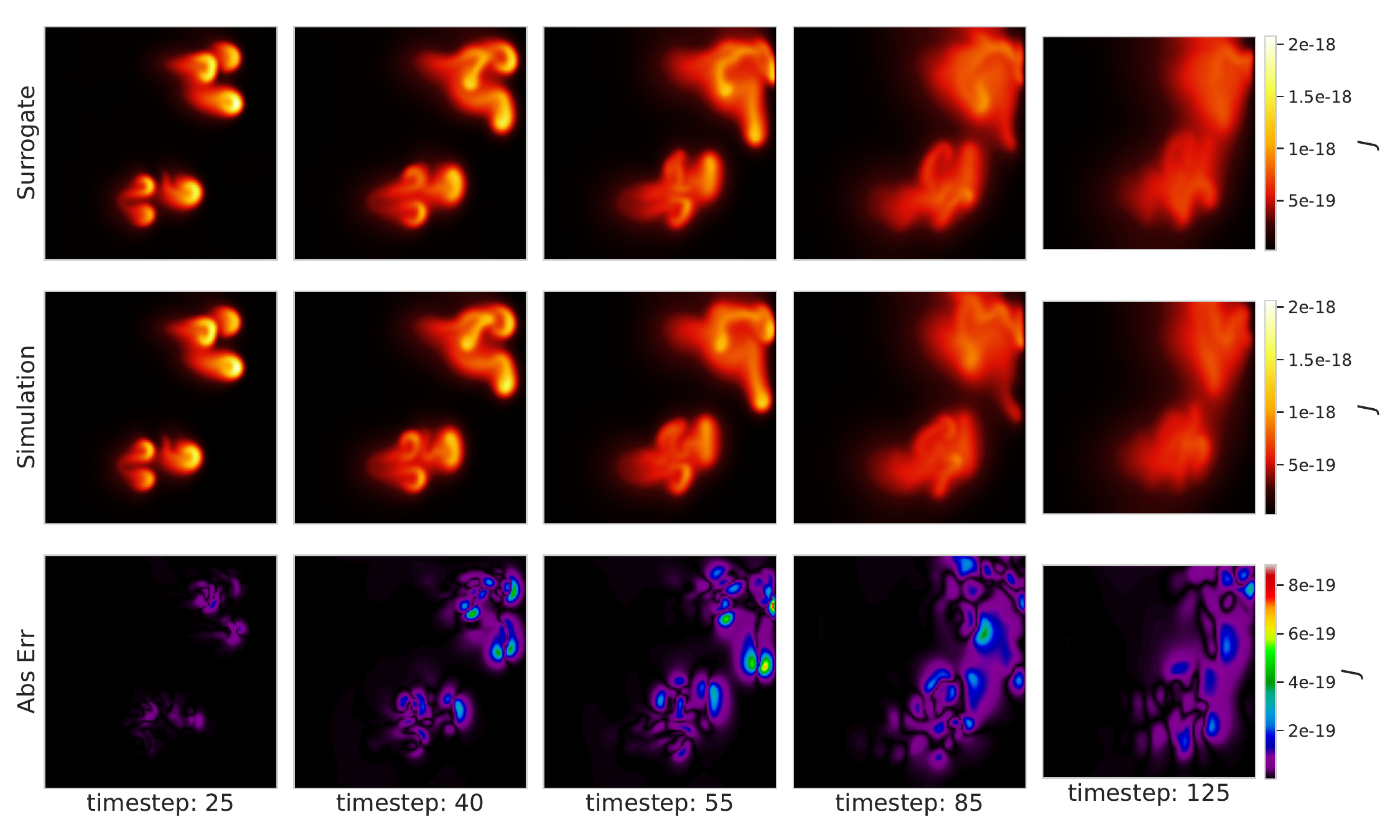}
    }
    \subfigure[Toroidal current $zj$ ($MA.m^{-2}$)]{\includegraphics[width=0.45\textwidth]{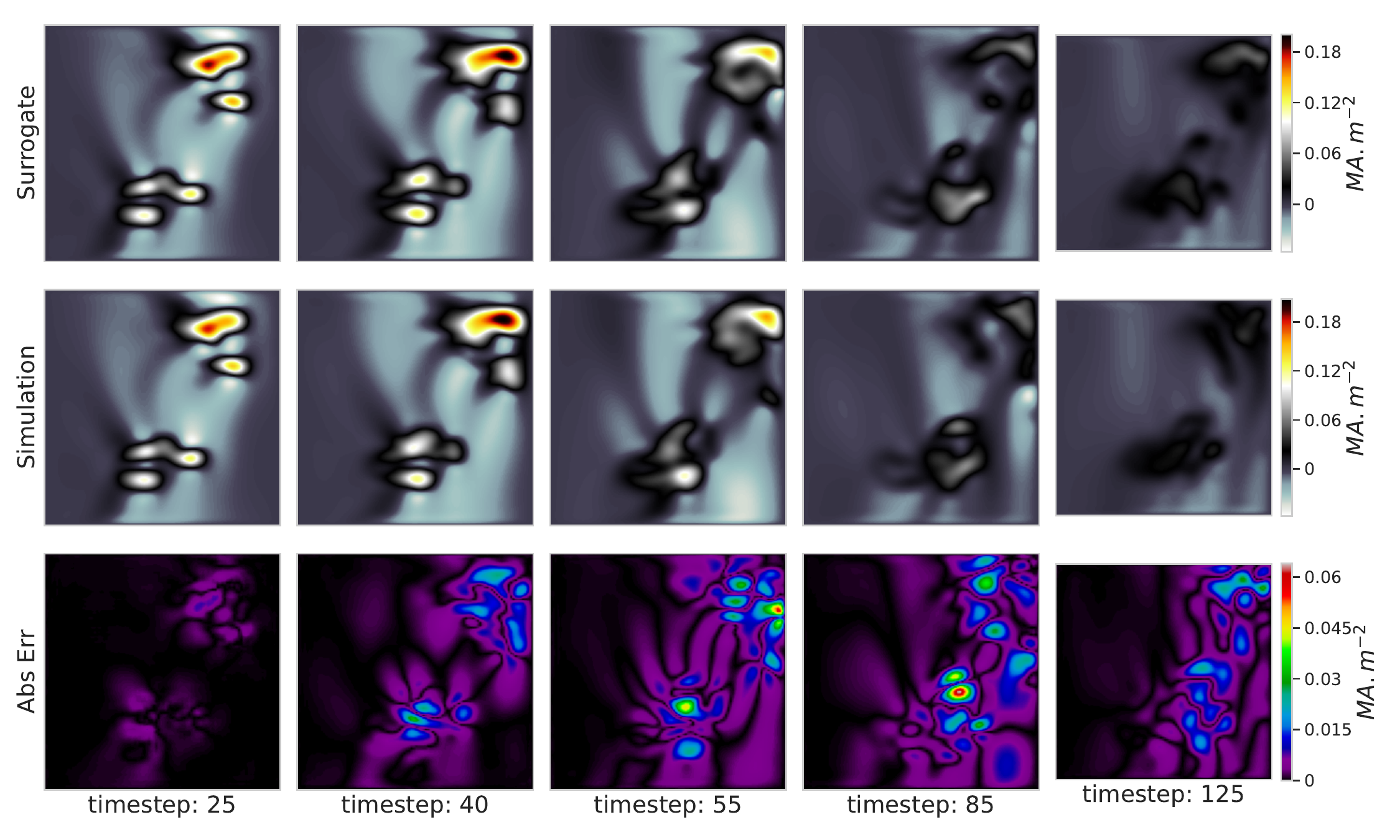}
    }
    \subfigure[Poloidal magnetic flux $\Psi$ ($T.m^{2}$)]{
    \includegraphics[width=0.45\textwidth]{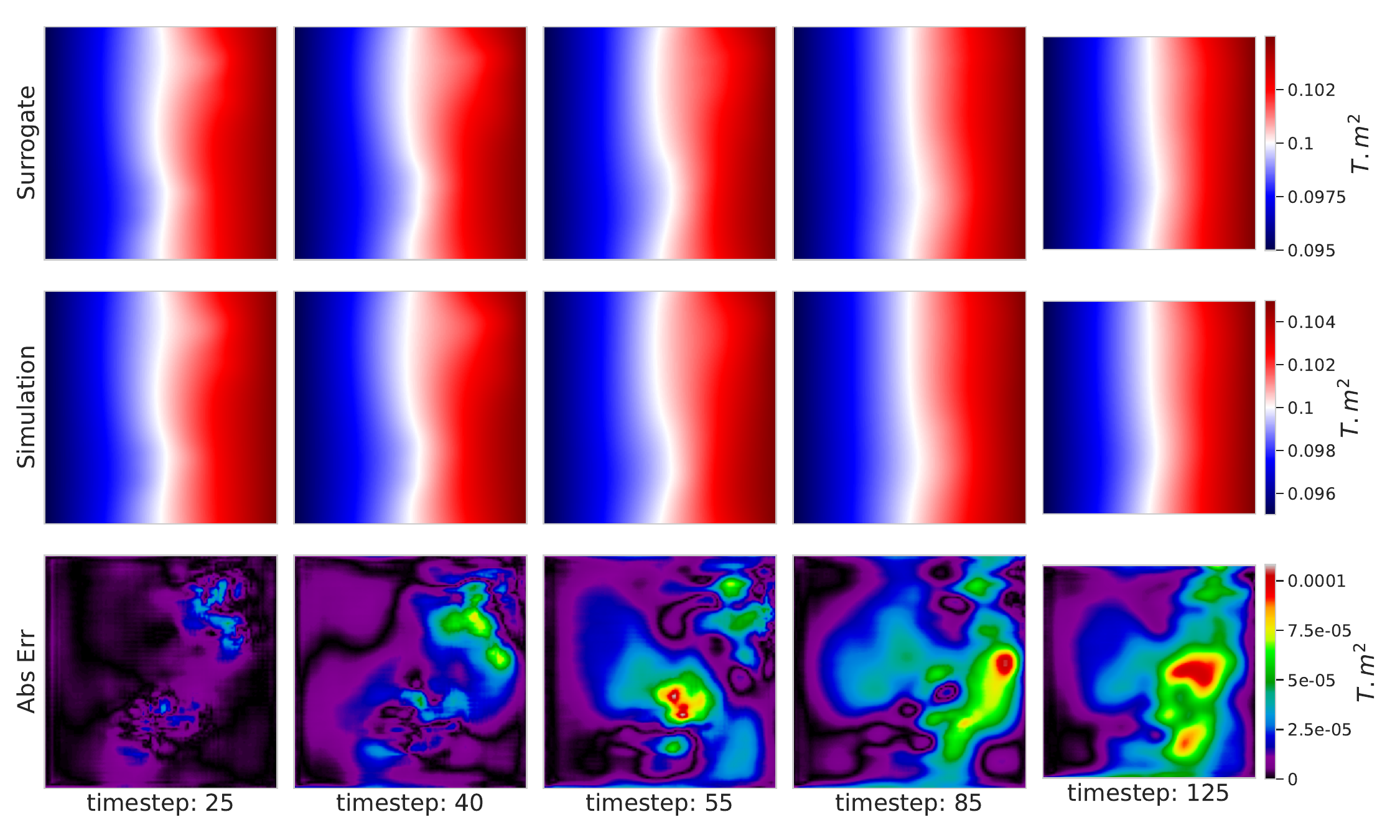}
    }
    \subfigure[Toroidal vorticity $\omega$]{
    \includegraphics[width=0.45\textwidth]{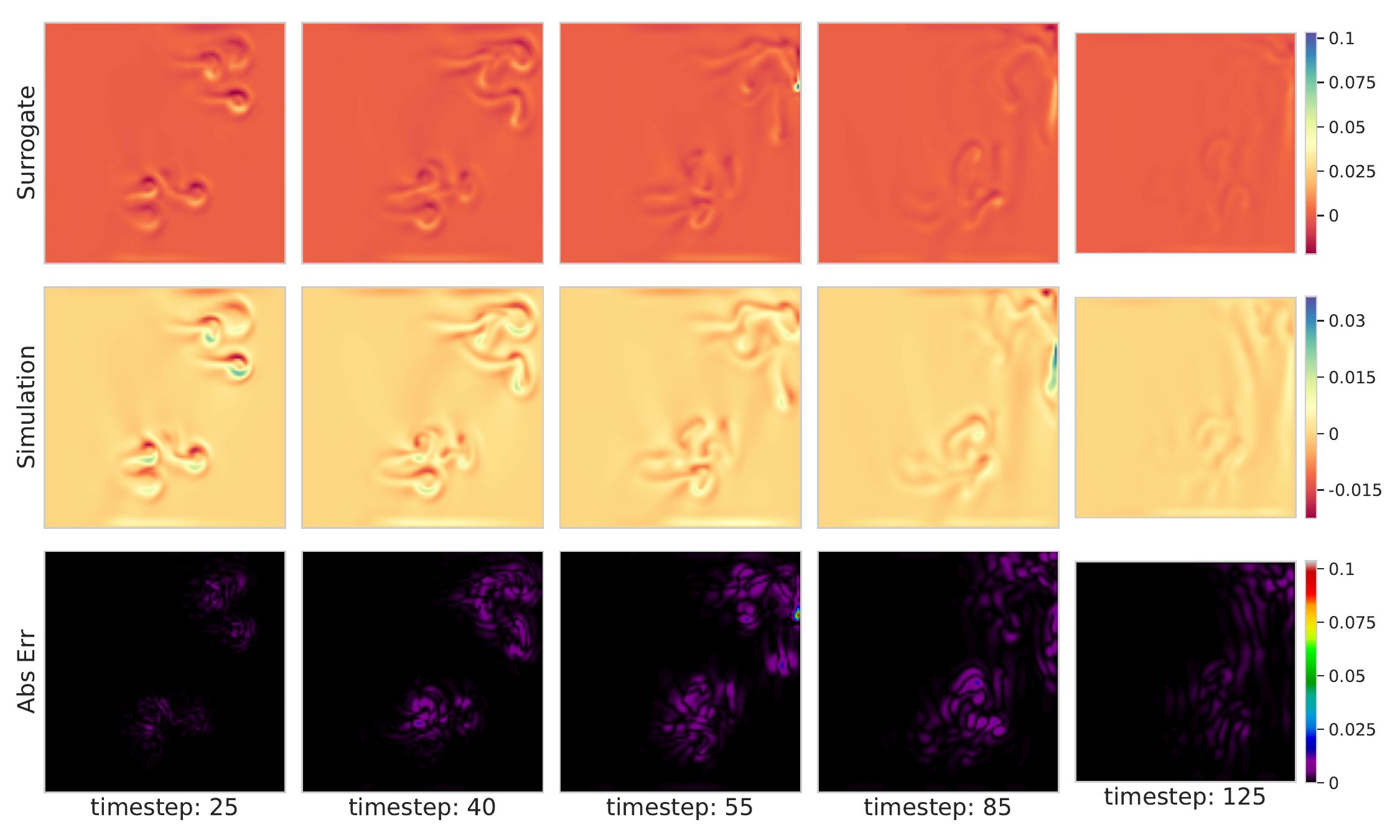}
    }
    \caption{An example reduced-MHD JOREK run for fields (a) density, (b) electric potential, (c) temperature, (d) current, (e) magnetic flux and (f) vorticity for the U-net.}
    \label{fig:initial_results:example_mjorek_traj_unet}
\end{figure}

\begin{figure}[h!]
    \subfigure[Different starting points with accumulating input error.]{\includegraphics[width=\textwidth]{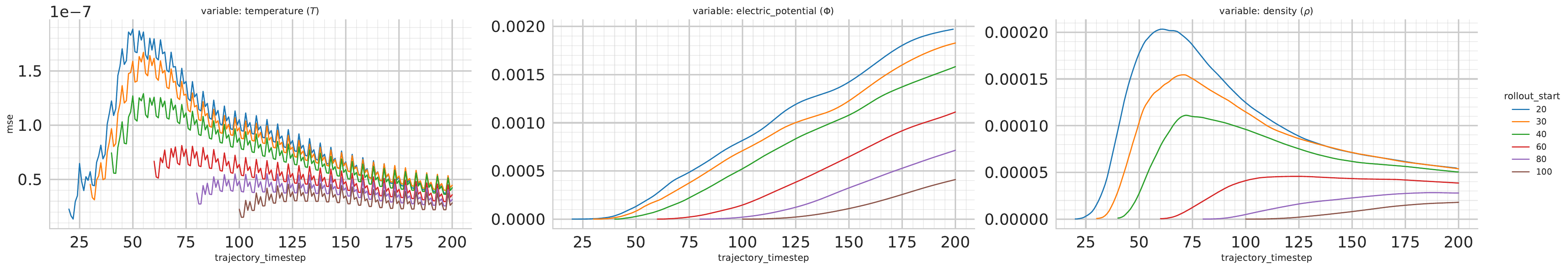}
    }
    \vfill
    \subfigure[Teacher forcing testing.]{\includegraphics[width=0.95\textwidth]{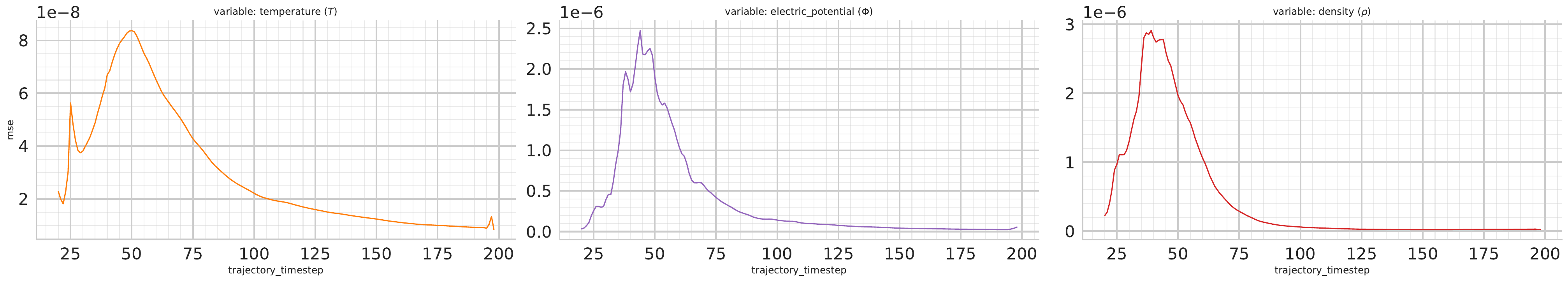}
    }
    \caption{Long rollout error for UNET model trained on electrostatic JOREK (for each field) for rollout error relative to trajectory timestep (a) with different starting points with accumulating input error and (b) with only ground truth simulation data as input (in other words, not running the surrogate in rollout set-up, but just looking at single, short-term predictions).}    
    \label{fig:traj_loc_impact:diff_starts_ejorek_unet}
\end{figure}

\begin{figure}[h!]
    \subfigure[different starting points and accumulating error]{\includegraphics[width=\textwidth]{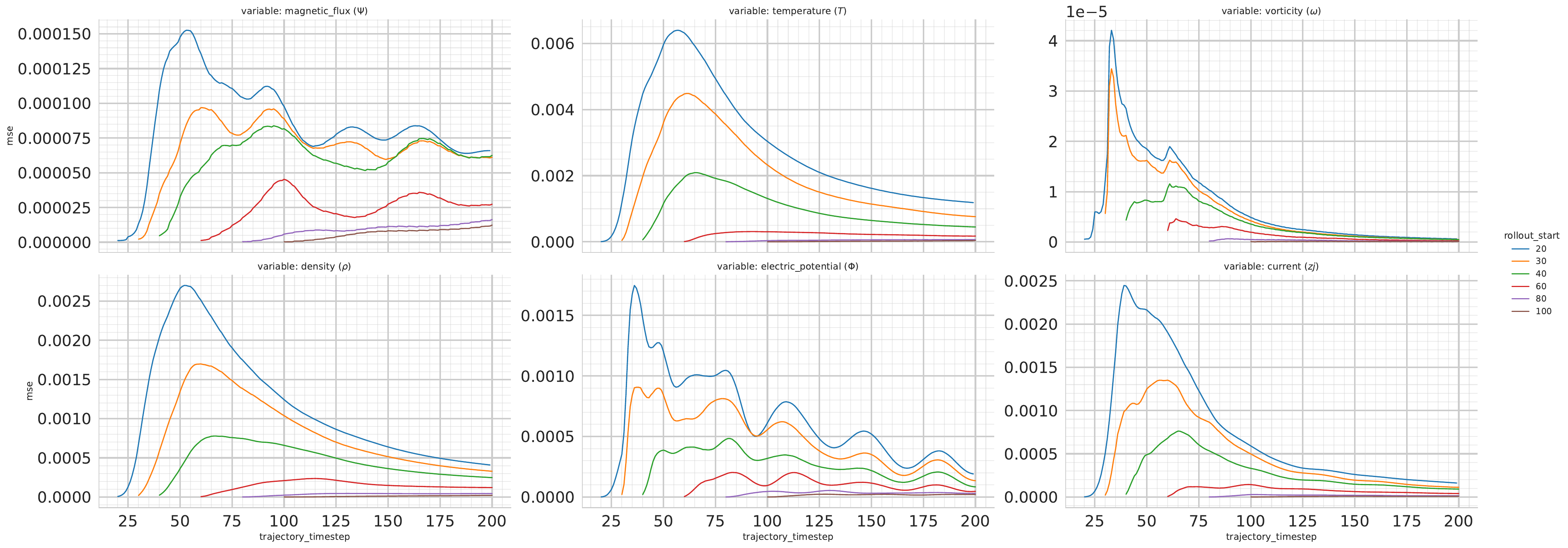}
    }
    \vfill
    \subfigure[only ground truth simulation data as input]{\includegraphics[width=0.95\textwidth]{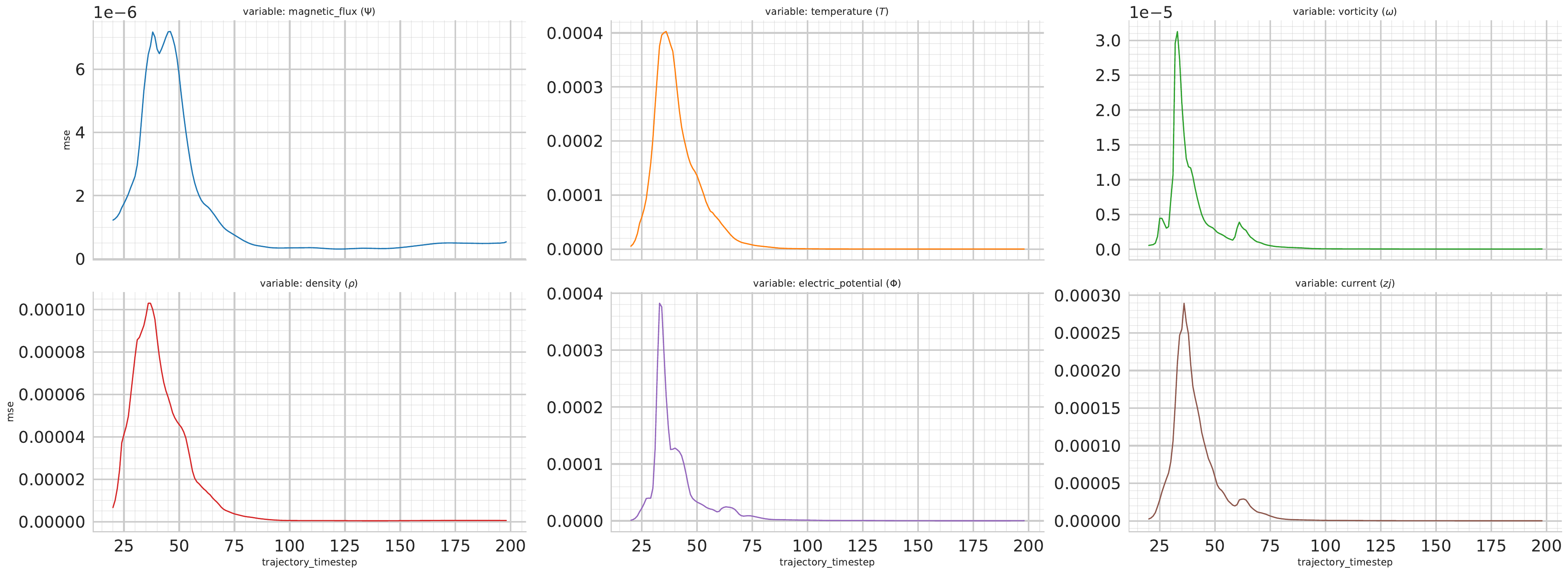}
    }
    \caption{Long rollout error for UNET model trained on reduced-MHD JOREK (for each field) for rollout error relative to trajectory timestep (a) with different starting points with accumulating input error and (b) with only ground truth simulation data as input (in other words, not running the surrogate in rollout set-up, but just looking at single, short-term predictions). }
    \label{fig:traj_loc_impact:diff_starts_mjorek_unet}
\end{figure}

To contextualize the task difficulty and evaluate FNO performance, we conducted baseline experiments using a U‑Net architecture from the PDEArena suite called "Unetmod-64". The U‑Net was selected to have a parameter count similar to that of our FNO model to ensure fair comparison. We repeated all experiments under identical settings with the same training hyperparameters (apart from gradient clipping which was increased due to training instability to 0.25, 0.0125 for electrostatic/reduced-MHD JOREK), and performance metrics for both models are presented here. The U-net outperforms for both JOREK datasets under short rollouts (see table \ref{table:initial_results:error_table_with_unet}) with the same peaked behaviour at the corresponding points in the trajectory (see fig \ref{fig:traj_loc_impact:diff_starts_ejorek_unet}, \ref{fig:traj_loc_impact:diff_starts_mjorek_unet}) and similar performance and behaviour for long rollouts. An example trajectory is included in figures \ref{fig:initial_results:example_ejorek_traj_unet} and \ref{fig:initial_results:example_mjorek_traj_unet} for the U-Net.

\begin{table}[h!]
\centering
\begin{adjustbox}{width=\textwidth}
\begin{tabular}{@{}llllllll@{}}
\toprule & \multicolumn{2}{c}{\textbf{Electrostatic JOREK}} & \multicolumn{2}{c}{\textbf{Reduced-MHD JOREK}} \\
& \textbf{FNO} & \textbf{UNET} & \textbf{FNO} & \textbf{UNET}\\
\textbf{Variable} & \textbf{(MSE$\pm$STD)} & \textbf{(MSE$\pm$STD)}& \textbf{(MSE$\pm$STD)} & \textbf{(MSE$\pm$STD)} \\\midrule
Temperature ($T$) & $4.34\pm 11.1\times10^{-8}$ & $3.25\pm 7.74\times10^{-8}$ & $3.21 \pm 13.14\times10^{-4}$ & $8.61\pm 54.2 \times10^{-5}$ \\
Electric potential ($\Phi$) & $2.95\pm 13.0 \times 10^{-6}$ & $9.91\pm 58.2 \times 10^{-7}$  & $1.48 \pm 6.39\times 10^{-4}$ & $4.44\pm 28.3 \times10^{-5}$ \\
Density ($\rho$) & $5.80 \pm 22.7\times10^{-6}$ & $9.41 \pm 65.0\times10^{-7}$ & $1.36 \pm 4.52\times 10^{-4}$ & $2.68\pm 12.8 \times10^{-5}$ \\
Vorticity ($\omega$) & & & $5.00 \pm 20.0\times10^{-6}$ & $2.46\pm 15.4 \times10^{-6}$ \\
Magnetic flux ($\Psi$) & & & $3.90 \pm 13.5\times10^{-5}$ & $1.84\pm 8.34 \times10^{-6}$ \\
Current ($zj$) & & & $1.41 \pm 5.62\times 10^{-4}$ & $4.68\pm 24.5 \times10^{-5}$ \\ \bottomrule
\end{tabular}
\end{adjustbox}
\caption{MSE on mininmum output length (meaning 5 timesteps) for each rescaled dataset variable averaged across different starting points for all dataset trajectories for the FNO and UNET model.}
\label{table:initial_results:error_table_with_unet}
\end{table}

\section{Example trajectory for additional fields for reduced-MHD JOREK} \label{sec:additional_fig_mjorek}

\begin{figure}[h!]
    \centering
    \subfigure[Poloidal magnetic flux $\Psi$ ($T.m^{2}$)]{
    \includegraphics[width=0.45\textwidth]{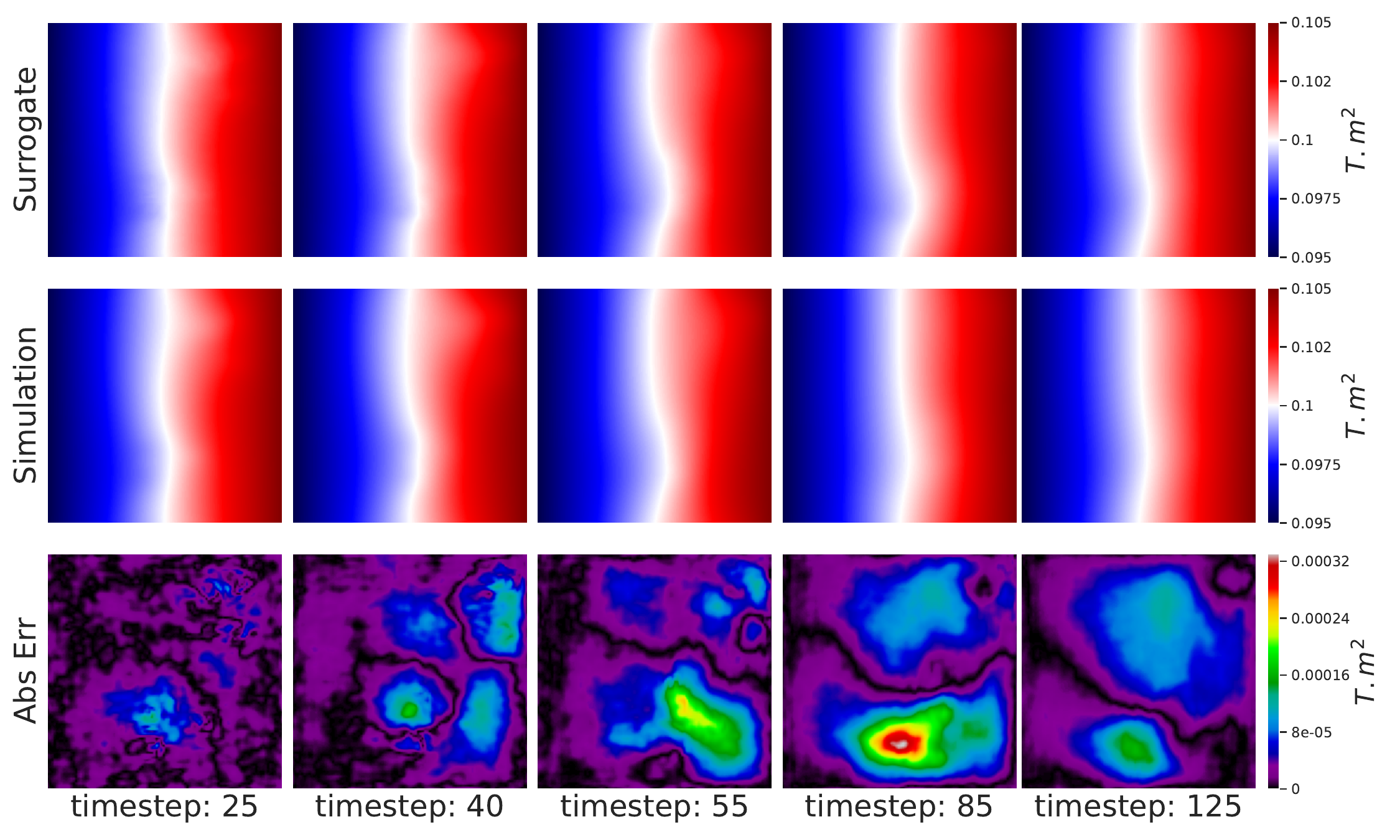}
    }
    \subfigure[Toroidal vorticity $\omega$]{
    \includegraphics[width=0.45\textwidth]{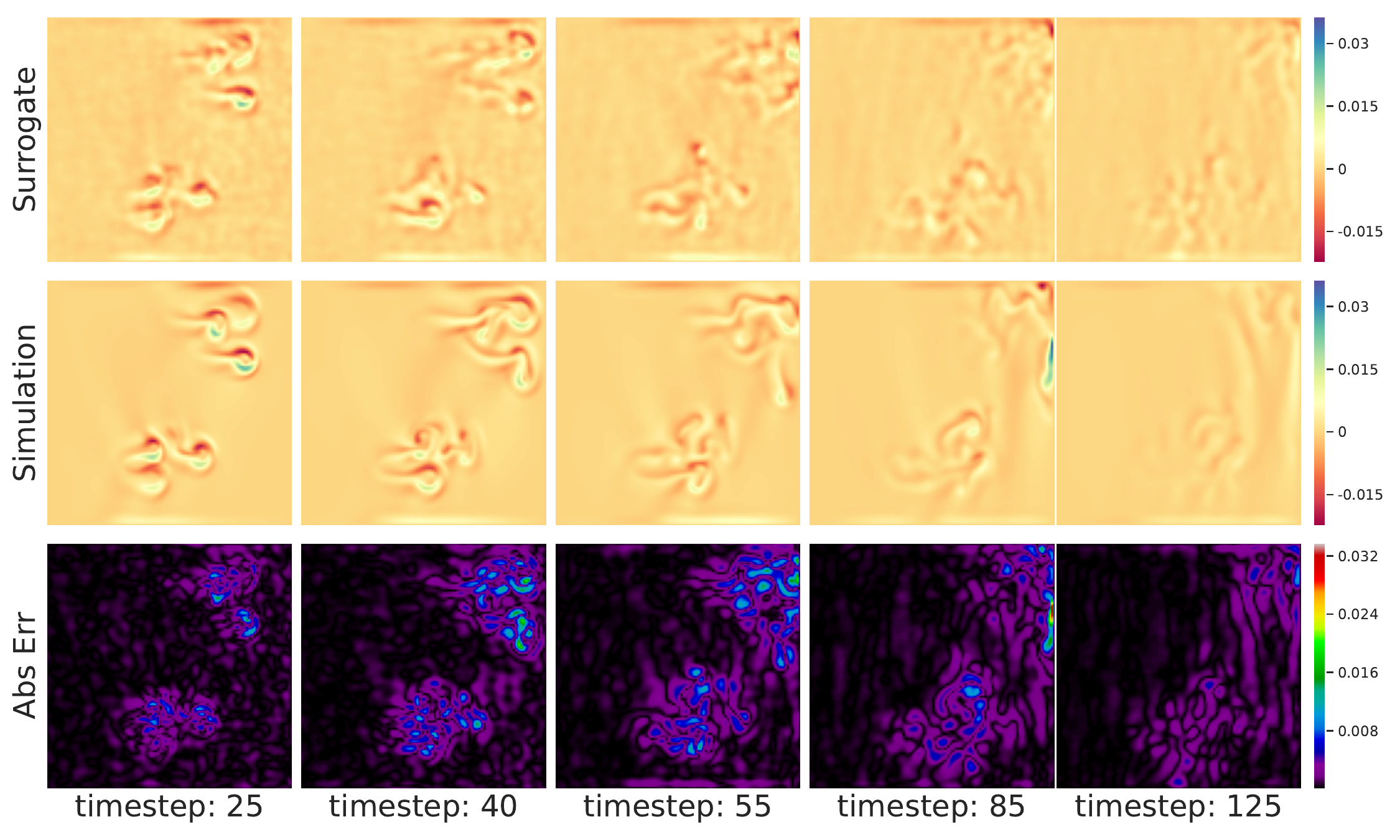}
    }
    \caption{An example reduced-MHD JOREK run for additional fields (a) magnetic flux and (b) vorticity}    \label{fig:initial_results:example_mjorek_traj_additional}
\end{figure}

Example trajectory for additional fields not included in \ref{sec:initial_results} are included in figure \ref{fig:initial_results:example_mjorek_traj_additional}.

\section{STORM different resolution Experiments}

\begin{figure}[h!]
    \centering
    \subfigure[Full resolution (256 x 192)]{
    \includegraphics[width=0.9\textwidth]{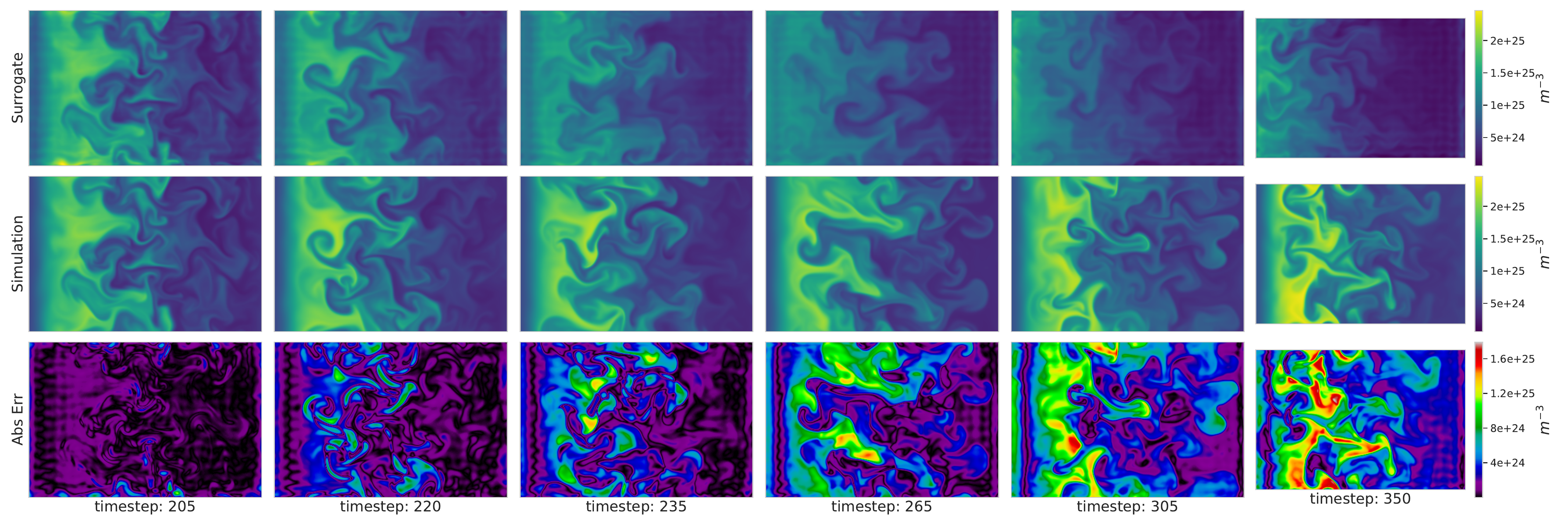}
    }
    \subfigure[Half resolution (model trained on) (128 x 96)]{
    \includegraphics[width=0.9\textwidth]{figures/4_1_storm_fno_example_traj_density.pdf}
    }
    \subfigure[Quarter resolution (64 x 48)]{
    \includegraphics[width=0.9\textwidth]{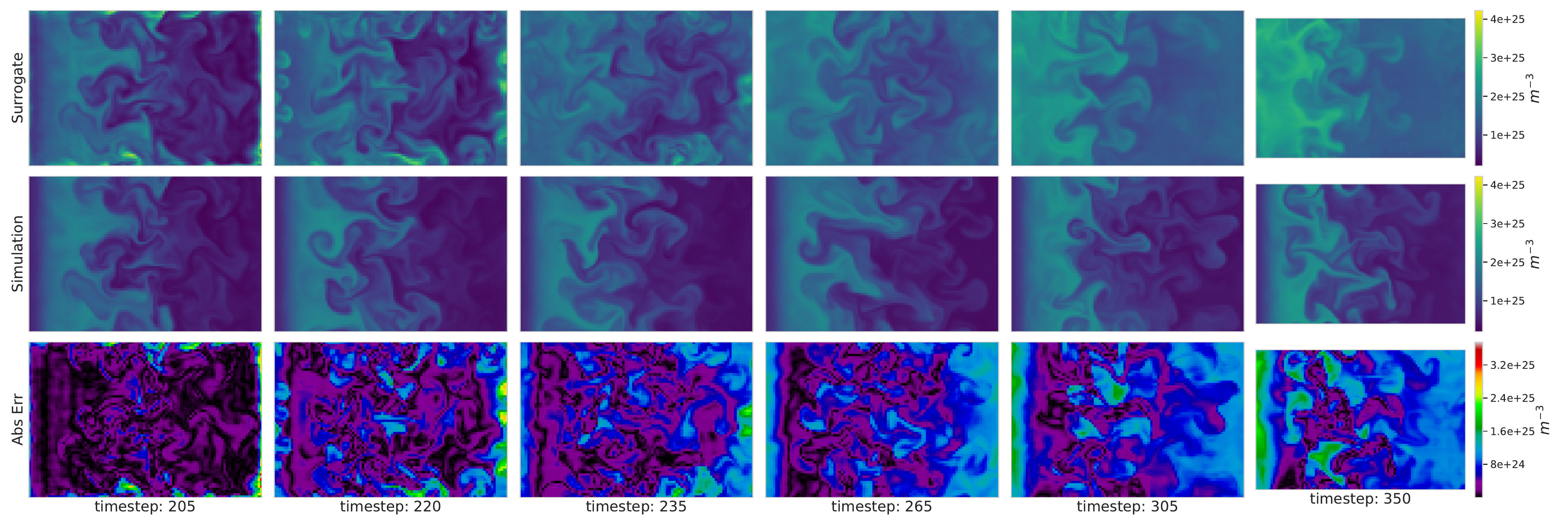}
    }
    \caption{An example STORM run plotted at specific rollout timesteps for field particle density $\rho$ ($m^{-3}$)}
    \label{fig:initial_results:example_storm_traj_diff_res_density}
\end{figure}

\begin{figure}[h!]
    \centering
    \subfigure[Full resolution (256 x 192)]{
    \includegraphics[width=0.9\textwidth]{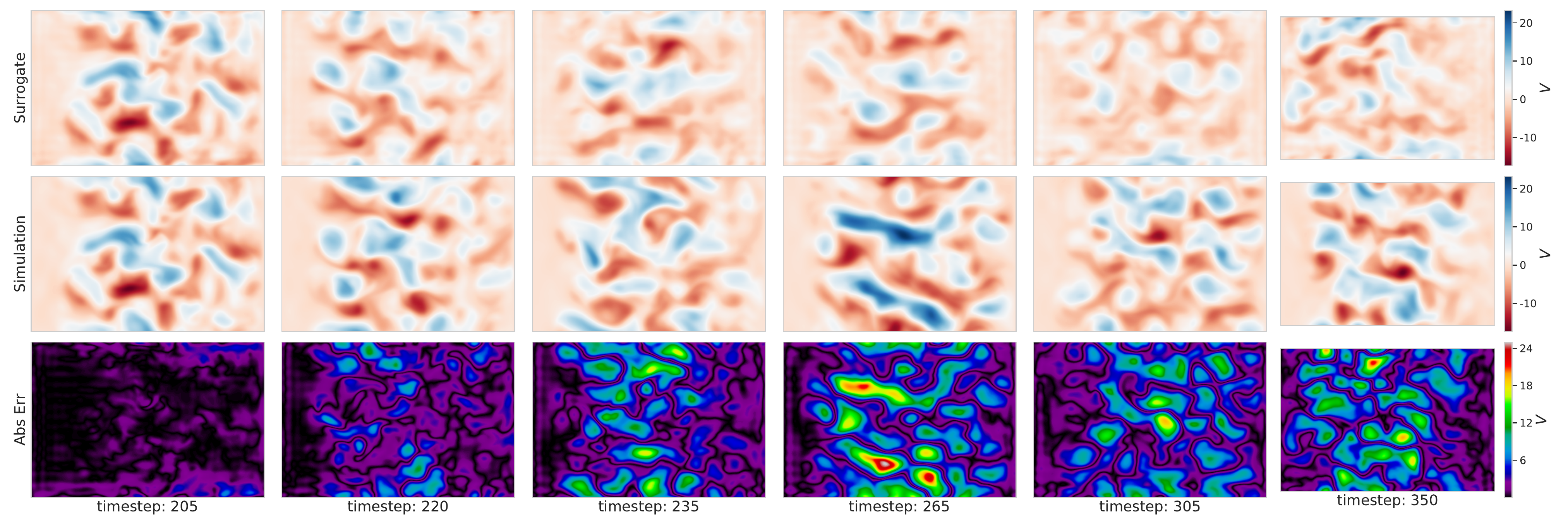}
    }
    \subfigure[Half resolution (model trained on) (128 x 96)]{
    \includegraphics[width=0.9\textwidth]{figures/4_1_storm_fno_example_traj_e_potential.pdf}
    }
    \subfigure[Quarter resolution (64 x 48)]{
    \includegraphics[width=0.9\textwidth]{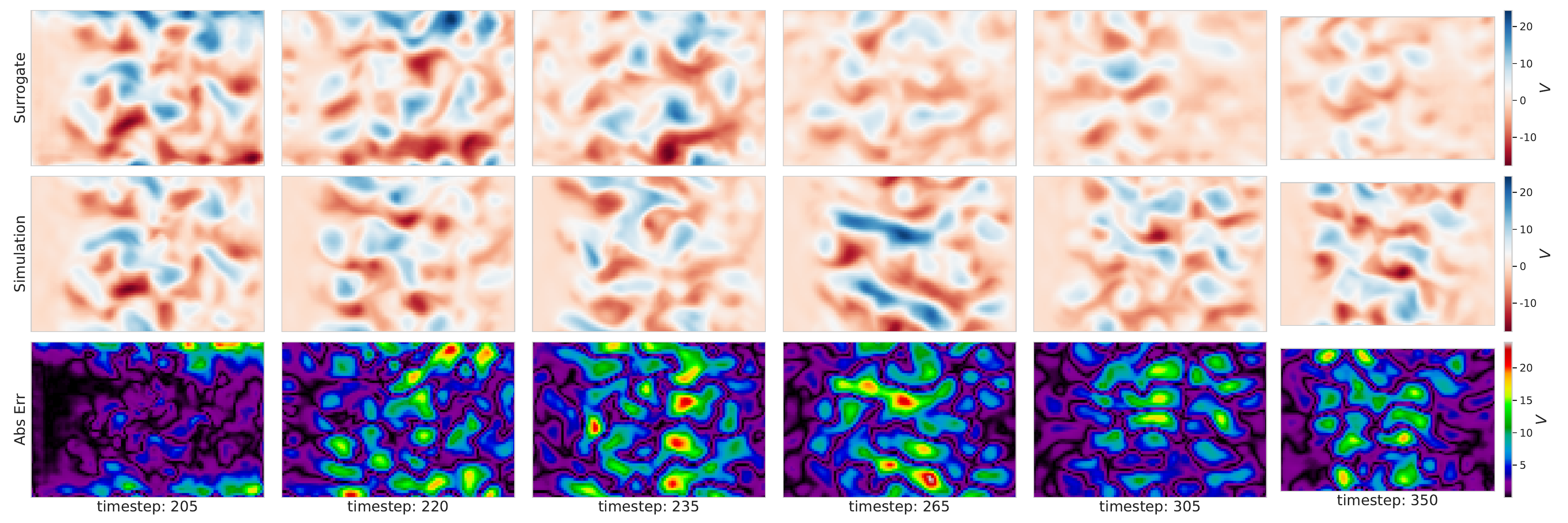}
    }
    \caption{An example STORM run plotted at specific rollout timesteps for field electric potential $\Phi$ ($V$)}
    \label{fig:initial_results:example_storm_traj_diff_res_electric_potential}
\end{figure}

\begin{figure}[h!]
    \subfigure[Different starting points with accumulating input error]{\includegraphics[width=\textwidth]{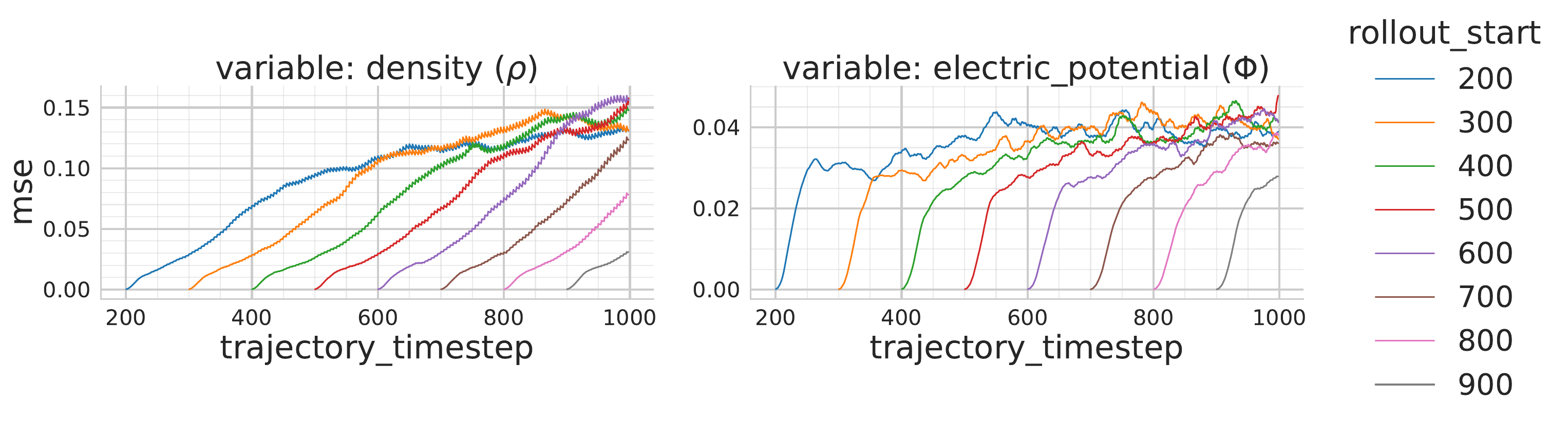}
    }
    \vfill
    \subfigure[Only ground truth simulation data as input]{\includegraphics[width=0.875\textwidth]{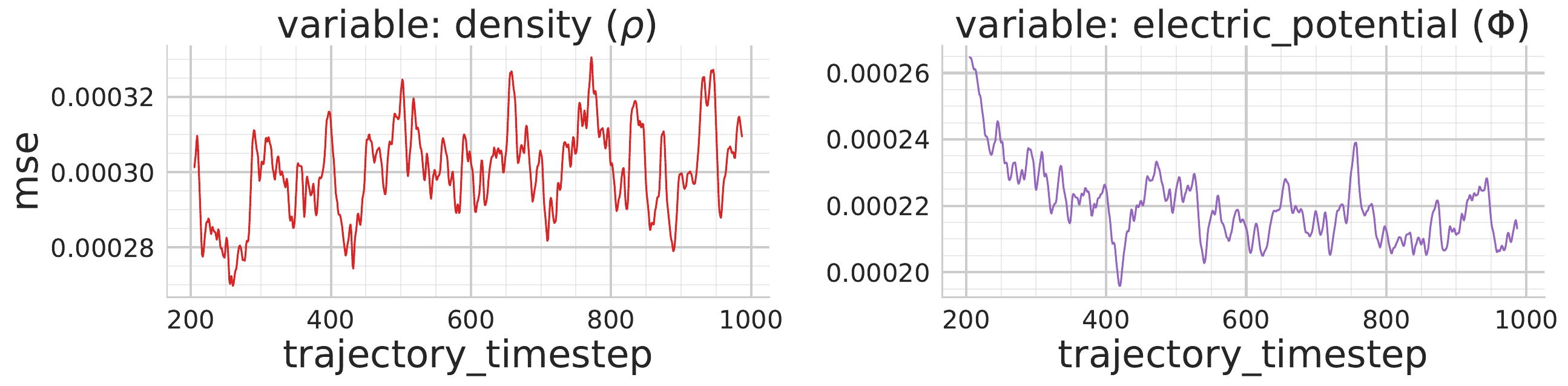}
    }
    \caption{Long rollout error for FNO model trained on STORM half resolution and evaluated on full resolution (for each field) for rollout error relative to trajectory timestep (a) with different starting points with accumulating input error and (b) with only ground truth simulation data as input (in other words, not running the surrogate in rollout set-up, but just looking at single, short-term predictions). For (a), 2 out of the 100 trajectories resulted in the model diverging for atleast 1 accumulated rollout; many for multiple. Those 2 trajectories are not included in plot (a).}    
    \label{fig:traj_loc_impact:diff_starts_storm_full_res}
\end{figure}

\begin{figure}[h!]
    \subfigure[Different starting points with accumulating input error]{\includegraphics[width=\textwidth]{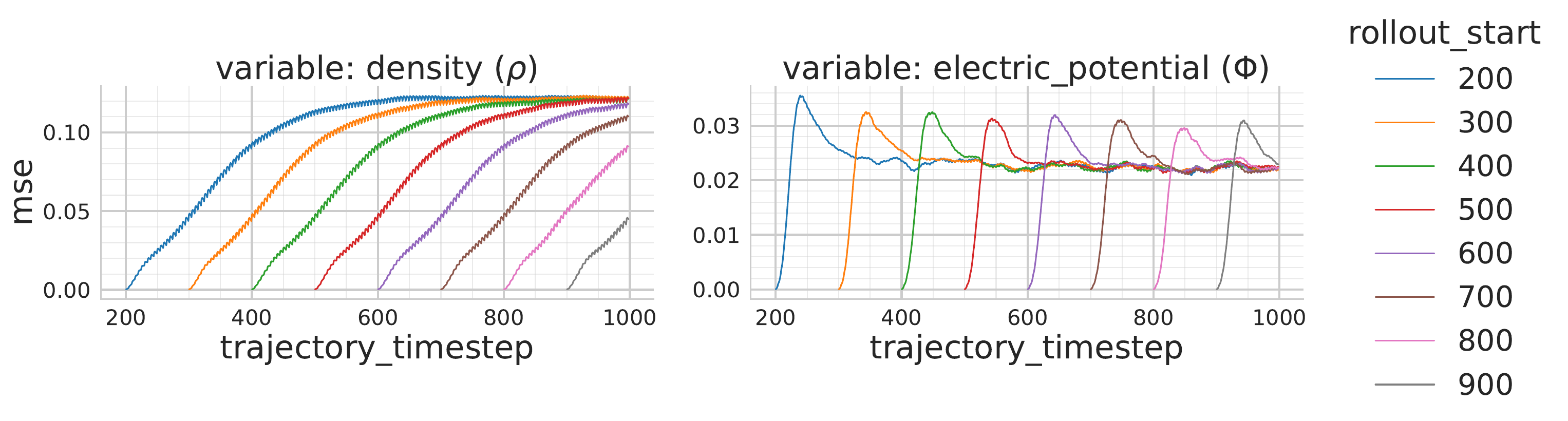}
    }
    \vfill
    \subfigure[Only ground truth simulation data as input]{\includegraphics[width=0.875\textwidth]{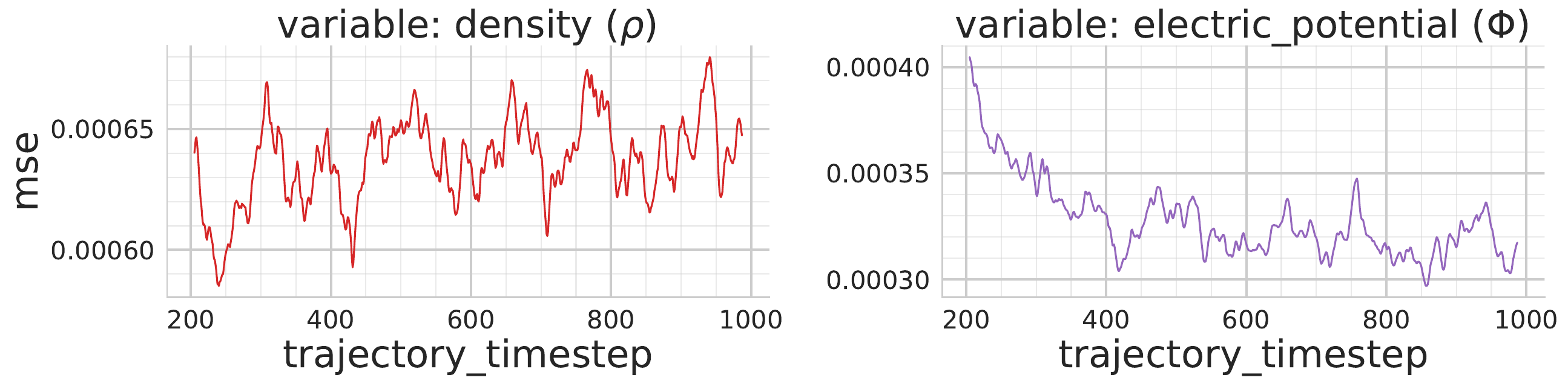}
    }
    \caption{Long rollout error for FNO model trained on STORM half resolution and evaluated on quarter resolution (for each field) for rollout error relative to trajectory timestep (a) with different starting points with accumulating input error and (b) with only ground truth simulation data as input (in other words, not running the surrogate in rollout set-up, but just looking at single, short-term predictions). For (a), 73 out of the 100 trajectories resulted in the model diverging for atleast 1 accumulated rollout; many for multiple. Those 73 trajectories are not included in plot (a).}    
    \label{fig:traj_loc_impact:diff_starts_storm_quarter_res}
\end{figure}

To assess the impact of spatial downsampling on model performance, we conducted experiments where the FNO was trained on half-resolution (2× downsampled) data and evaluated on full, half, and quarter resolutions. While the model performed best at the resolution it was trained on, accuracy degraded similarly when tested on both higher and lower resolutions (see table \ref{table:initial_results:error_table_diff_res_storm} for numerical figures). This suggests that the decline is not simply due to resolution mismatch but reflects a limited ability of the model to generalise across spatial scales. An example trajectory run is included in figure \ref{fig:initial_results:example_storm_traj_diff_res_density} and \ref{fig:initial_results:example_storm_traj_diff_res_electric_potential}. In figures \ref{fig:traj_loc_impact:diff_starts_storm_full_res}-\ref{fig:traj_loc_impact:diff_starts_storm_quarter_res}, a significant drop in long-term rollout stability was observed at mismatched resolutions even only considering stable runs. Additionally, 2 out of 100 trajectories diverged at quarter resolution, and 73 diverged at full resolution, with rapid error jumps exceeding $10^{23}$ beyond 200 rollout steps. These results challenge the notion of FNOs resolution invariance and highlight the need for further investigation - such as fine-tuning on higher resolutions - to better understand and potentially improve cross-resolution generalization.

\begin{table}[h!]
\centering
\begin{adjustbox}{width=\textwidth}
\begin{tabular}{@{}lllllll@{}}
\toprule & \textbf{Full resolution (256 x 192)} & \textbf{Half resolution (128 x 96)} & \textbf{Quarter resolution (64 x 48)} \\
\textbf{Variable} & \textbf{(MSE$\pm$STD)} & \textbf{(MSE$\pm$STD)} \\\midrule
Electric potential ($\Phi$) &  $2.20 \pm 1.45 \times 10^{-4}$ & $1.71 \pm 2.36 \times 10^{-6}$ & $3.28 \pm 3.22\times10^{-4}$& \\
Density ($\rho$) & $3.00 \pm 3.34 \times 10^{-4}$ & $7.73 \pm 13.1 \times 10^{-6}$ & $6.37 \pm 5.51 \times 10^{-4}$ & \\\bottomrule
\end{tabular}
\end{adjustbox}
\caption{MSE on mininmum output length (meaning 5 timesteps) for each rescaled dataset variable averaged across different starting points for all dataset trajectories for different resolutions of STORM.}
\label{table:initial_results:error_table_diff_res_storm}
\end{table}

\section{MJOREK heatflux}

\begin{figure}
    \includegraphics[width=0.5\textwidth]{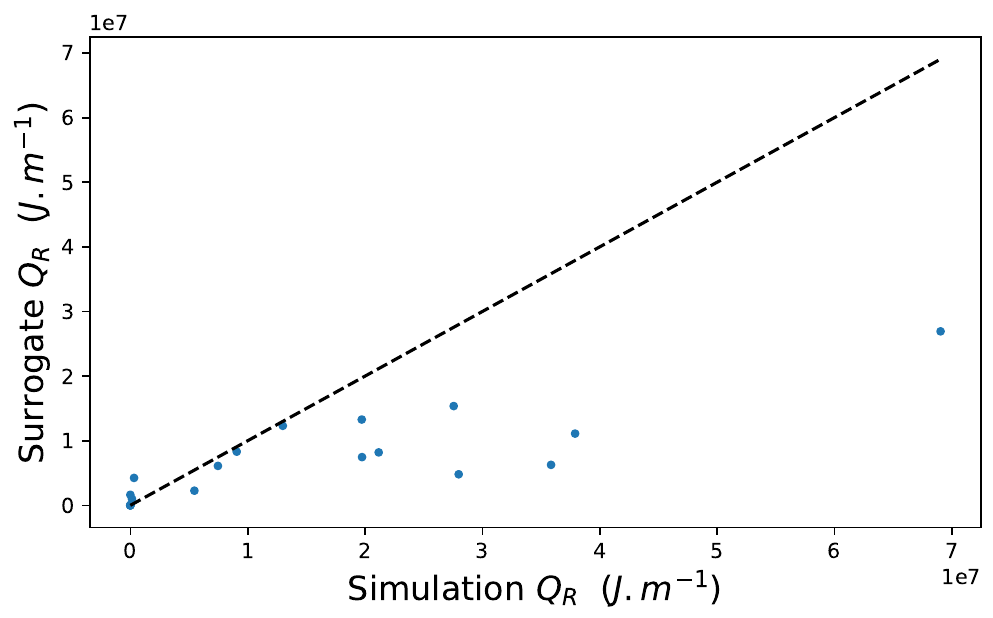}
    \caption{Heatflux at boundary integrated over poloidal direction (y-axis) and time comparing surrogate and simulation. Each datapoint is trajectory. Black lines are reference identity lines $(0,0)$, $(10^{7},0)$ and $(0,10^{7})$}
    \label{fig:heatflux_scatter_mjorek}
\end{figure}

Similar to section \ref{sec:heatflux}, we plotted an example scatter plot in Figure \ref{fig:heatflux_scatter_mjorek} which compares the total heat flux at this boundary, integrated over time and the poloidal direction and shows similar results.

\section{Cross code transfer learning additional figures} \label{cross_code_transfer_additional}

\begin{figure}
    \centering
    \subfigure[Case 1 Transfer learning from STORM to electrostatic JOREK]{
        \includegraphics[width=1\textwidth]{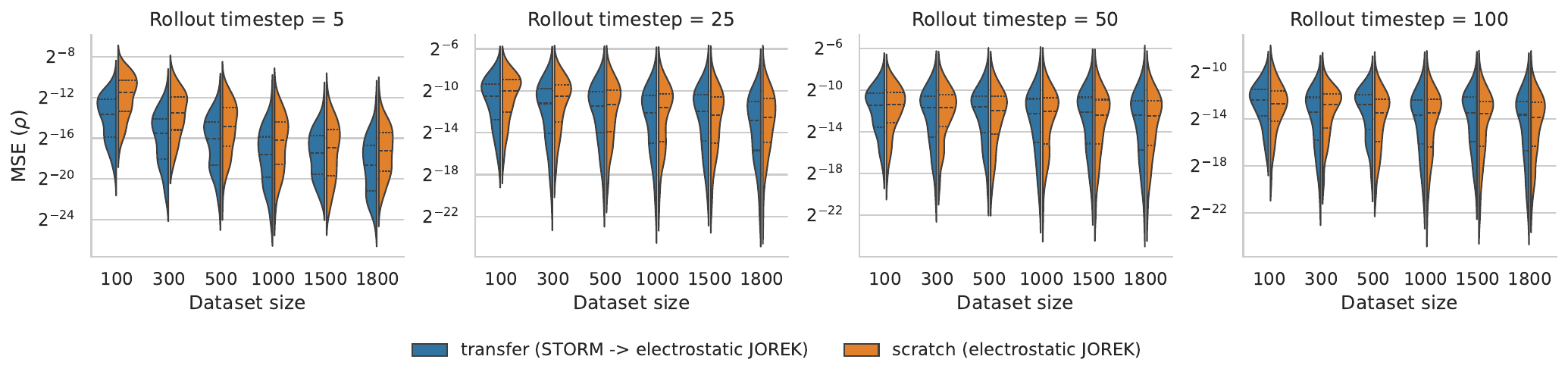}
    }
    \subfigure[Case 2 Transfer learning from electrostatic JOREK to STORM]{
        \includegraphics[width=1\textwidth]{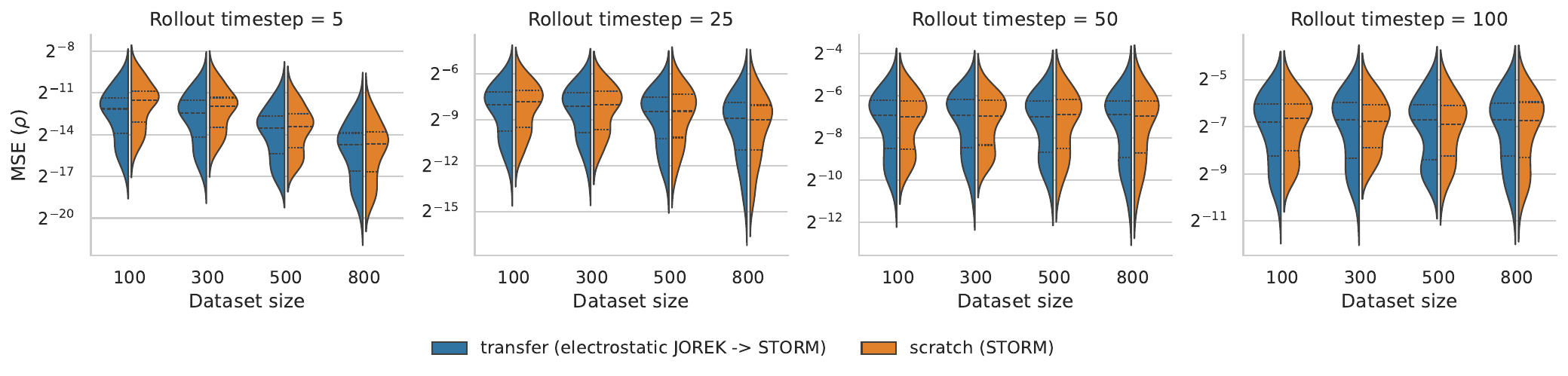}
    }
    \caption{Scratch and transfer model error for density at different timesteps and different dataset sizes. The line is the medium error and the error bars correspond to the 16th and 84th percentile.}
    \label{fig:transfer_code_dataset_size_density}
\end{figure}

Additional figures (figure \ref{fig:transfer_code_dataset_size_density})  for the other field (electrostatic JOREK) discussed in section \ref{sec:cross_code_transfer}. 

\section{Changing t-out} \label{sec:increasing t_out}

The experiment explored larger output steps ($T_{out}>5$ revealing that models trained to predict more $T_{out}$ steps performed similar or worse than those trained with smaller $T_{out}$ steps rolled out iteratively to the same total time length for both short and long rollouts. Larger $T_{out}$ models also consistently lost finer spatial details, aligning with the findings from \cite{lippe2023pderefiner}, which suggested that a common weakness of long rollouts was the inability to learn higher frequency components which seemed to be exacerbated when training on large rollouts.

\begin{figure}
    \centering
    \includegraphics[width=\textwidth]{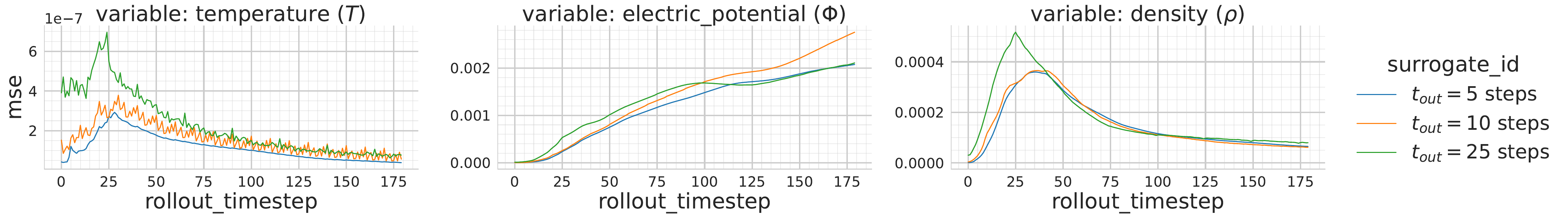}
    \caption{Electrostatic JOREK model rollout error (starting at $t=20$ trained with different number of $t_{out}$ steps. }
    \label{fig:diff_t_out_electrostatic JOREK}
\end{figure}

\begin{figure}
    \centering
    \includegraphics[width=0.8\textwidth]{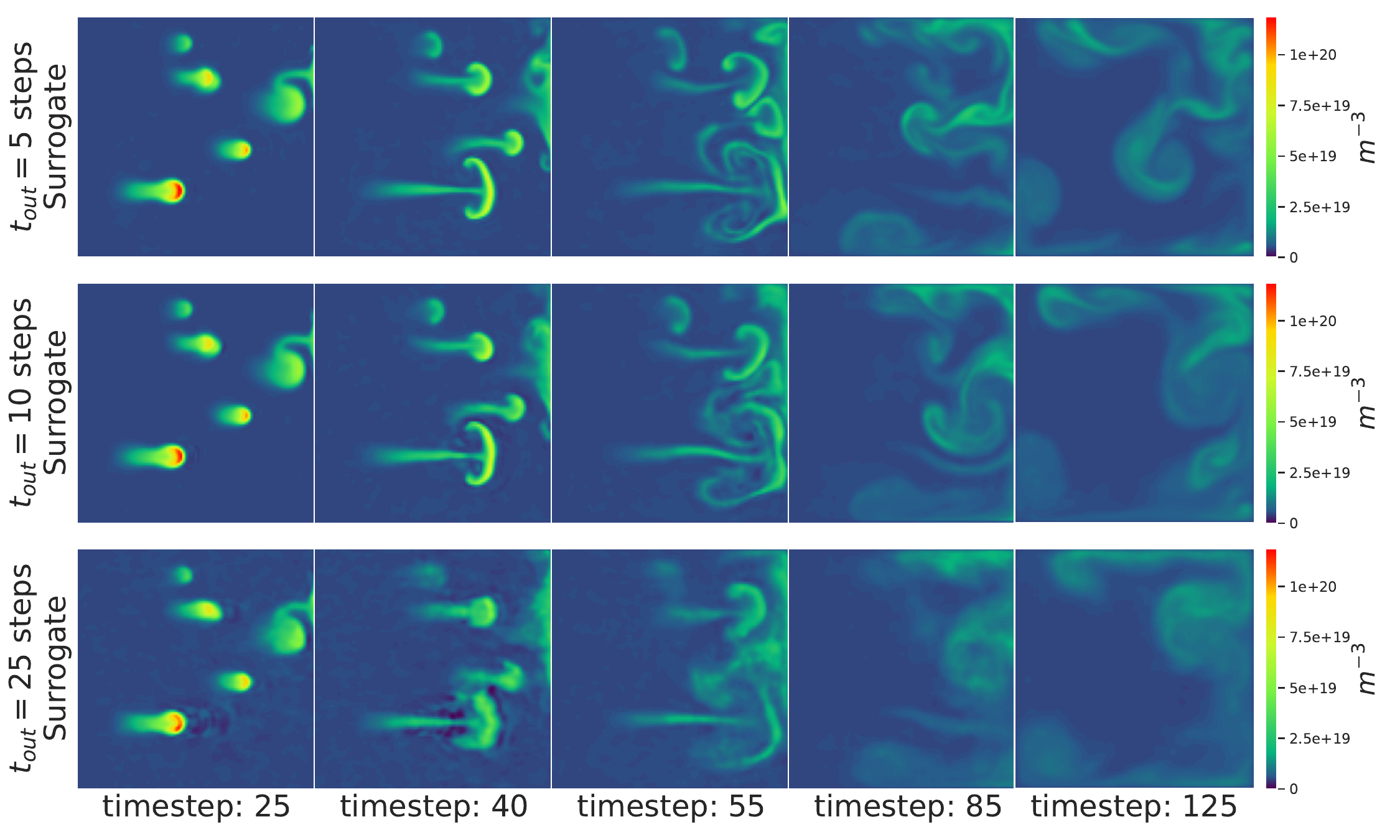}
    \caption{Electrostatic JOREK model rollout error trained with different number of $t_{out}$ steps}
    \label{fig:diff_t_out_electrostatic JOREK_example}
\end{figure}

The results in this appendix (see figures \ref{fig:diff_t_out_electrostatic JOREK} and \ref{fig:diff_t_out_electrostatic JOREK_example}) show that increasing the output step size ($T_{out}>5$) led to models that performed similarly or worse than those trained with smaller $T_{out}$ rolled out iteratively to reach the same time horizon. Notably, models trained with larger $T_{out}$ also lost fine spatial structures, consistent with \cite{lippe2023pderefiner}, which highlighted that long rollouts struggle to capture high-frequency details. These findings suggest that training on large temporal steps may amplify this limitation, emphasizing the trade-off between rollout length and spatial fidelity.

\section{Varying sampling of hump region (electrostatic JOREK)} \label{sec:sampling_ejorek}

\begin{figure}[h!]
    \centering
    \includegraphics[width=\textwidth]{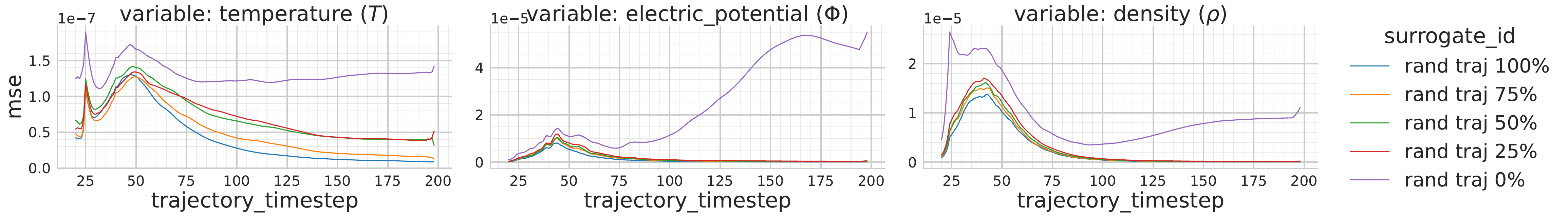}
    \caption{Increased sampling of region between t=30-70 by ensuring that 1-X\% of the training sampling was within this region, X being the percentage of samples being completely random. Rollout error with only ground truth simulation data as input (in other words, not running the surrogate in rollout set-up, but just looking at single, short-term predictions). }
    \label{fig:changing_sampling_hump_region_jorek}
\end{figure}

Resampling around the error peaks (as shown as in figure \ref{fig:changing_sampling_hump_region_jorek}) to improve representation of these dynamics did not enhance performance, suggesting that the issue is not due to under-representation in the training dataset.

\section{Blob number (electrostatic JOREK)} \label{sec:split_blob_num_ejorek}

\begin{figure}[h!]
    \centering
    \includegraphics[width=\textwidth]{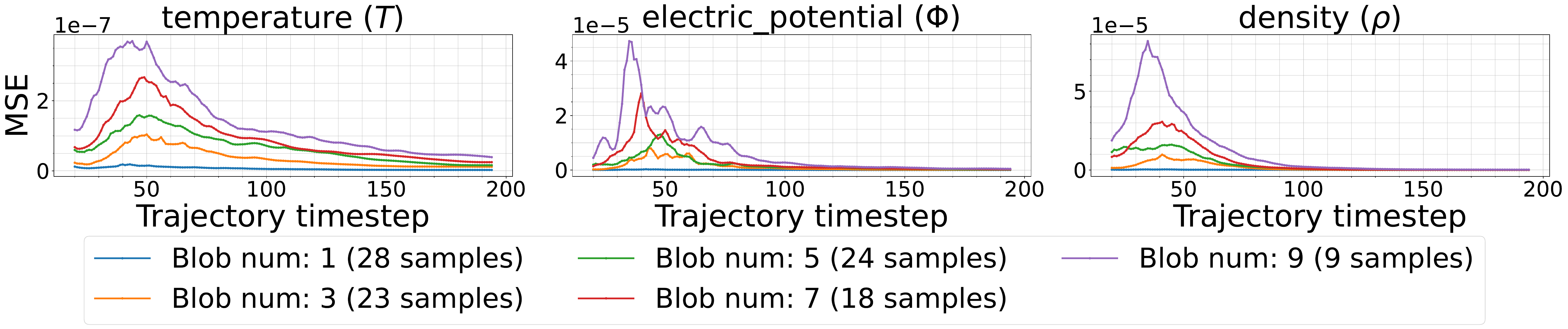}
    \caption{FNO model trained on electrostatic JOREK split according to number of blobs in simulation. Rollout error with only ground truth simulation data as input (in other words, not running the surrogate in rollout set-up, but just looking at single, short-term predictions). } \label{fig:training_dataset}
\end{figure}

The magnitude of error increased with the number of blobs in a trajectory (see figure \ref{fig:training_dataset}), reduced-MHD JOREK validation dataset was too small with only 2 samples per blob number on average), the timing and frequency of error spikes were not correlated with blob count. 

\section{Pointwise error (JOREK)} \label{sec:pointwise_error_jorek}

\begin{figure}[h!]
    \centering
    \includegraphics[width=0.7\textwidth]{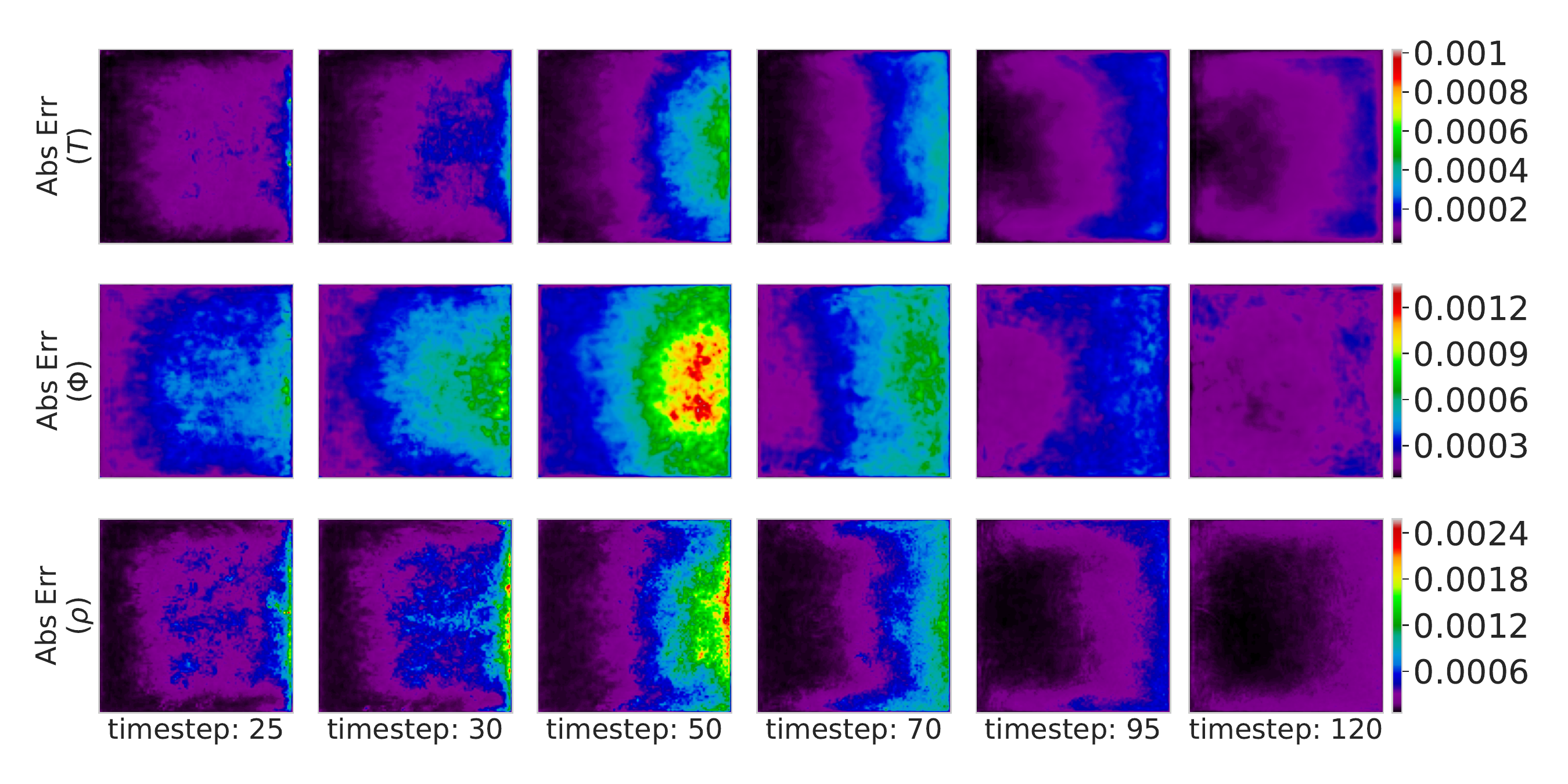}
    \caption{This figure illustrates the pointwise model error for the electrostatic JOREK dataset at specific time steps, averaged across the validation dataset. The results indicate that the error is predominantly concentrated near the boundaries, both during periods of peak error and at other times. This suggests that boundary conditions significantly influence model performance, contributing to localized error accumulation over time. All errors are calculated from the rescaled fields.}
    \label{fig:boundary_cond:pointwise_err_ejorek}
\end{figure}

\begin{figure}[h!]
    \centering
    \includegraphics[width=0.7\textwidth]{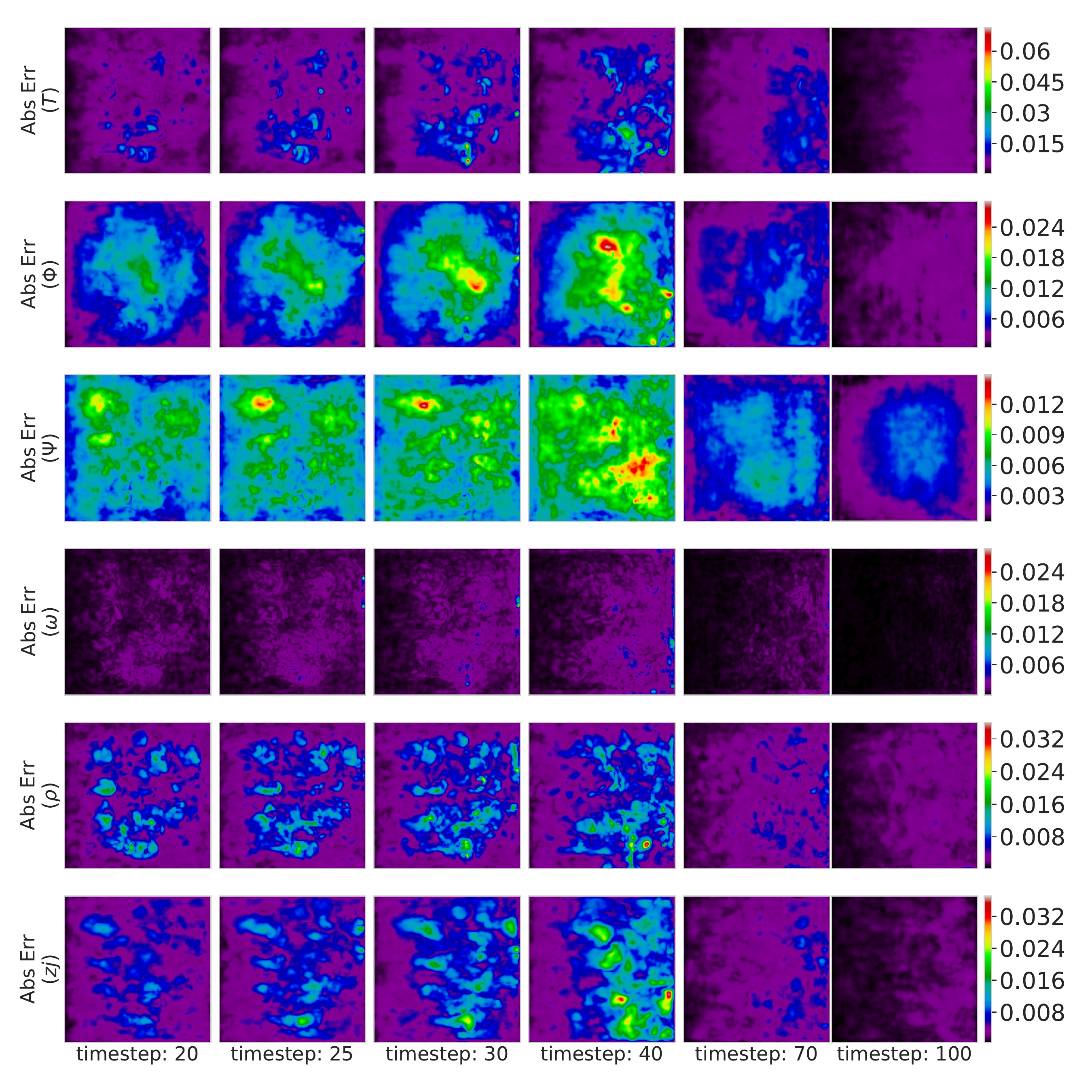}
    \caption{This figure illustrates the pointwise model error for the reduced-MHD JOREK dataset at specific time steps, averaged across the validation dataset. The results indicate that the error is not concentrated near the boundaries unlike electrostatic JOREK. All errors are calculated from the rescaled fields.}
    \label{fig:boundary_cond:pointwise_err_mjorek}
\end{figure}

Using the model trained on the electrostatic JOREK dataset, analysis of the pointwise error (figure \ref{fig:boundary_cond:pointwise_err_ejorek} and \ref{fig:boundary_cond:pointwise_err_mjorek}) revealed distinct spatial patterns. Pointwise error involves computing the error at each spatial point for a given time step, then averaging this error across the validation dataset to assess the model's accuracy across space and time. Errors were predominantly localized near the boundaries, particularly the right-hand side wall during early timesteps. The error peaks at $t=50$, coinciding with the point at which blobs collided with the wall (fig \ref{fig:boundary_cond:snapshots_ejorek}). Over time, these localized errors expanded to encompass other boundaries. This spatial distribution suggests that boundary interactions present a challenge for the neural operator. It is important to note that within the neural operator, there is no explicit consideration of boundary conditions, other than the observed behavior at the edges of the spatial dimensions for the simulation in the dataset.

However, this pattern was not observed in the reduced-MHD JOREK dataset, where blobs took a much longer time to travel to the right boundary and diffused or dissipated before hitting it before reaching the boundary (see fig \ref{fig:initial_results:example_mjorek_traj} for an example), and the peak error at $t=30$ occurred earlier in the trajectory well before the blobs have hit the right wall (see fig \ref{fig:boundary_cond:pointwise_err_mjorek}). This divergence suggests that boundary effects alone cannot fully explain the observed error spikes, as other factors must be contributing to the dynamics.

\section{NMSE (JOREK)} \label{sec:nmse_jorek}

\begin{figure}[h!]
    \centering
    \subfigure[Timewise (avg) normalization]{
        \includegraphics[width=0.9\textwidth]{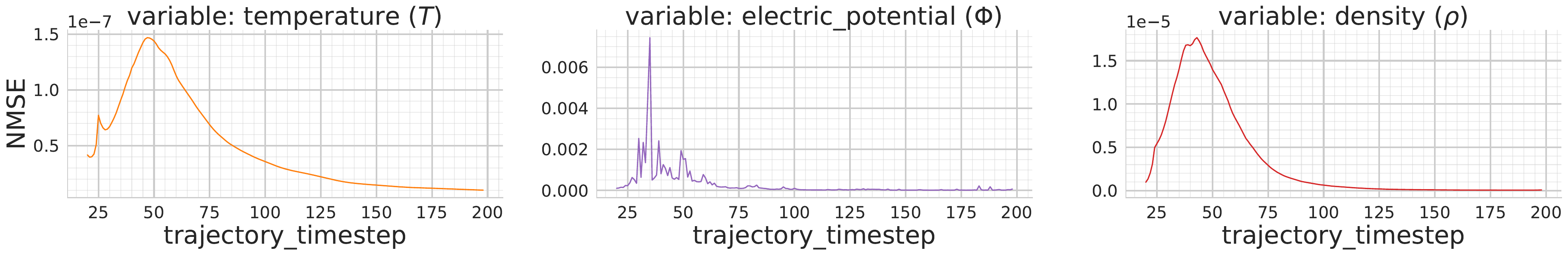}
    }
    \subfigure[Pointwise normalization]{
        \includegraphics[width=0.9\textwidth]{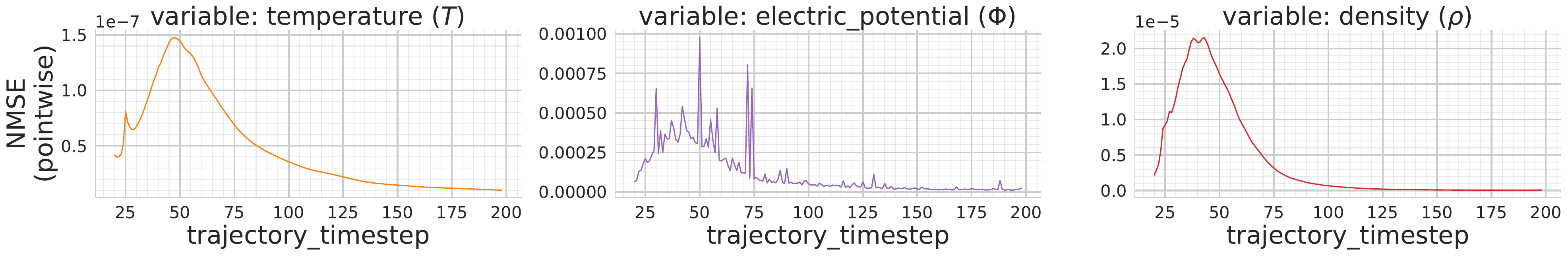}
    }
    \caption{No accumulating error (only simulation input) for FNO model trained on electrostatic. Error is normalized either pointwise, or timewise (avg).}
    \label{fig:norm_ejorek}
\end{figure}

\begin{figure}[h!]
    \centering
    \subfigure[Timewise (avg) normalization]{
        \includegraphics[width=0.9\textwidth]{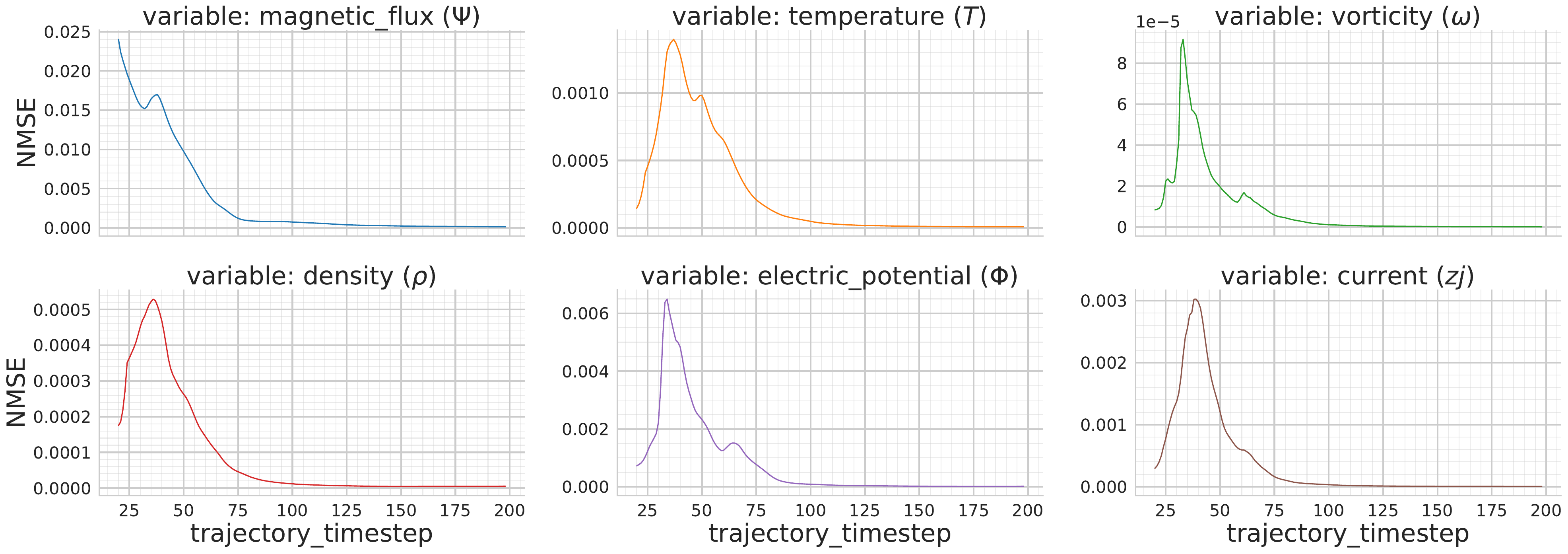}
    }
    \subfigure[Pointwise normalization]{
        \includegraphics[width=0.9\textwidth]{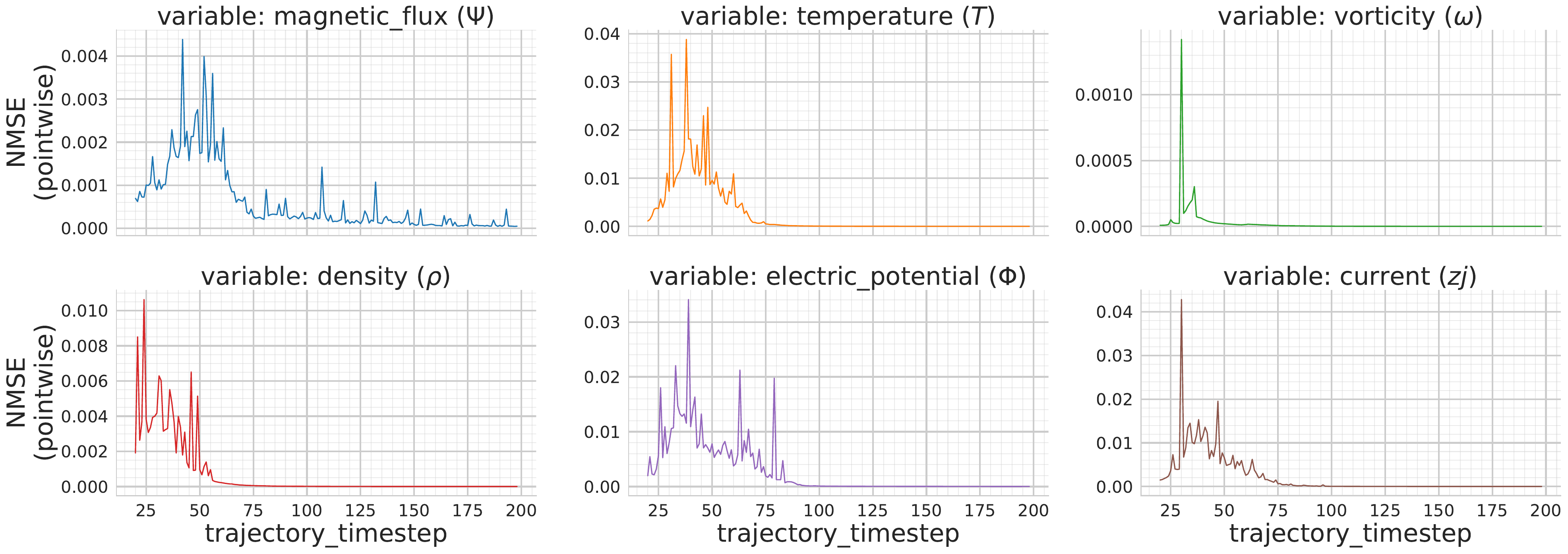}
    }
    \caption{No accumulating error (only simulation input) for FNO model trained on reduced-MHD JOREK split}
    \label{fig:norm_mjorek}
\end{figure}

The lower MSE observed later in the rollout might initially appear to be a consequence of decreasing overall solution magnitudes. This interpretation was further examined using normalized error metrics. Specifically, both pointwise and timewise normalized errors were computed to account for changes in scale and smoothness of the solution over time. These metrics normalize the prediction error by scaling up the error relative to either the pointwise target value or the spatially averaged target at each time step. Importantly, the observed peaks and trends in error remained consistent under these alternative metrics, confirming that the surrogate performance is influenced not solely by solution magnitude as seen in figures \ref{fig:norm_ejorek} and \ref{fig:norm_mjorek}.

\begin{itemize}
    \item Pointwise normalization error $:= MSE/target$ calculated pointwise. This scales error up according to the size of the target value at each point.
    \item Timewise (avg) normalization error $:= MSE/avgTarget$ calculated pointwise where avgTarget is the target averaged across spartial dimensions, but keeping the time dimension. This scales error up according to the average taget value at each time.
\end{itemize}
These are all calculated on the rescaled datasets.

\end{document}